\documentclass[preprint,3p,twocolumn]{elsarticle}

\usepackage{graphicx}
\usepackage{amsmath}
\usepackage{amssymb}
\usepackage{natbib}
\usepackage{hyperref}
\hypersetup{
    colorlinks = true,
    urlcolor   = blue,
    citecolor  = black,
}

\usepackage{xcolor}

\graphicspath{{figures/}} 
\usepackage{picture}

\begin{document}

\begin{frontmatter}

\author[1]{Giovanni Soligo\corref{cor1}}
\ead{soligo.giovanni@oist.jp}
\author[1]{Marco Edoardo Rosti\corref{cor2}} 
\ead{marco.rosti@oist.jp}
\affiliation[1]{organization={Complex Fluids and Flows unit, Okinawa Institute of Science and Technology Graduate University}, 
                    addressline={Tancha 1919-1},
                    postcode={904-0495}, city={Onna}, country={Okinawa, Japan}}

\title{Non-Newtonian turbulent jets at low-Reynolds number}

\begin{abstract}

We perform direct numerical simulations of planar jets of non-Newtonian fluids at low Reynolds number, in typical laminar conditions for a Newtonian fluid. We select three different non-Newtonian fluid models mainly characterized by shear-thinning (Carreau), viscoelasticity (Oldroyd-B) and shear-thinning and viscoelasticity together (Giesekus), and perform a thorough analysis of the resulting flow statistics. We characterize the fluids using the parameter $\gamma$, defined as the ratio of the relevant non-Newtonian time scale over a flow time scale. We observe that, as $\gamma$ is increased, the jet transitions from a laminar flow at low $\gamma$, to a turbulent flow at high $\gamma$. We show that the different non-Newtonian features and their combination give rise to rather different flowing regimes, originating from the competition of viscous, elastic and inertial effects. We observe that both viscoelasticity and shear-thinning can develop the instability and the consequent transition to a turbulent flowing regime; however, the purely viscoelastic Oldroyd-B fluid exhibits the onset of disordered fluid motions at a lower value of $\gamma$ than what observed for the purely shear-thinning Carreau fluid. When the two effects are both present, an intermediate condition is found, suggesting that, in this case, the shear-thinning feature is acting against the fluid elasticity. Despite the qualitative differences observed in the flowing regime, the bulk statistics, namely the centerline velocity and jet thickness, follow almost the same power-law scalings obtained for laminar and turbulent Newtonian planar jets. The simulations reported here are, to the best of our knowledge, the first direct numerical simulations showing the appearance of turbulence at low Reynolds number in jets, with the turbulent motions fully induced by the non-Newtonian properties of the fluid, since the Newtonian case at the same Reynolds number is characterized by steady, laminar flow.

\end{abstract}


\end{frontmatter}

\section{Introduction}
Non-Newtonian fluids possess unique features such as shear-dependent viscosity, fluid elasticity or yield stress, which lead to complex flowing dynamics. These fluids are more common than one would think: examples include biological fluids, such as blood and mucus, foods and cosmetics, such as ketchup or toothpaste, and can be found in oil, chemical processing and pharmaceutical industrial applications.

Fluids laden with long-chain polymers have been extensively studied in recent years: adding a relatively small quantity of polymers to a Newtonian solvent makes it a non-Newtonian fluid. The additional contribution from fluid elasticity is able to develop flow instabilities \citep{sanchez2022understanding}; these were first observed in experiments on flows with curved streamlines \citep{PadkelM_1996,QinSHA_2019,Shaqfeh_1996}, such as flows between cylinders or plates \citep{groisman1998mechanism, larson1990purely, mckinley1996rheological, muller1989purely}, in curved channels \citep{groisman2001efficient, groisman2004elastic} or around obstacles \citep{arora2002experimental, browne2020pore, KawaleMZKMRB_2017, MachadoBBCC_2016,  mitchell2016viscoelastic}. More recently it has been shown that elastic instabilities can develop also in parallel flows \citep{PanMWA_2013, QinSHA_2019}. 
Elastic instabilities manifest at low or vanishing Reynolds numbers, and the resulting disordered fluid motion shows features similar to the classical Newtonian turbulence, as for instance a power-law scaling in the velocity spectra \citep{qin2017characterizing, QinSHA_2019}. The instability, however, is driven and sustained by the elastic stresses \citep{warholic1999influence} rather than inertial terms, as occurs instead for Newtonian turbulence.

Depending on the flow parameter and on the fluid characteristics, very different flowing regimes can be observed. Recent experiments performed by \citet{yamani2021spectral} on a round jet of viscoelastic fluid entering a bath of water showed the onset of different flowing regimes: at low values of the Reynolds number and of the fluid elasticity, a laminar-like regime is observed; as the contribution of viscoelasticity was increased, the formation of filaments of highly-stretched fluid was observed around the core of the jet, indicating the transition towards elastic turbulence. 

It is clear that the flowing of non-Newtonian fluids exhibits richer and more complex dynamics than Newtonian fluids: the competition among elastic, inertial and viscous contributions can lead to a non-trivial flowing pattern. These peculiarities have caught the attention of many researchers, as proven by the considerable amount of archival literature on the topic. There are however much fewer works on jets, despite them being a typical example of free shear flows and among the canonical problems of fluid mechanics. \citet{parvar2020local} performed direct numerical simulations of laminar planar jets of a FENE-P fluid; the analysis was however limited to low values of the Weissenberg number, ratio of the elastic relaxation time scale over a flow time scale. Semi-analytic solutions for the governing equations were derived; an excellent agreement was obtained for the velocity field, while the accurate computation of the conformation tensor -- which is an indicator of the averaged local deformation of the polymers -- proved to be more challenging, with results strongly depending on the Weissenberg number. 
The conformation tensor is an indicator of the micro-structural state of the polymers, and from a qualitative standpoint it can be thought as the average value of the end-to-end distance of the polymers.
It was also reported that the solution shows self-similarity both at different distances from the inlet and at different Weissenberg numbers, provided that a suitable normalization is adopted for the rescaled profiles. Linear stability analysis on planar jets of FENE-P fluid at moderate Reynolds numbers showed the importance of viscoelasticity in the onset of the instability \citep{ray2015absolute}. A similar result was obtained also for planar mixing layers \citep{ray2014absolute}. The authors pointed out that stronger effects from the non-Newtonian component are observed at transitional Reynolds numbers, as also hinted by the flow instability reported in the experimental results by \citet{fruman1984swelling}. It was also observed that reducing the Reynolds number promotes the onset of the viscoelastic instability \citep{ray2015absolute}.
Conversely, at high Reynolds numbers linear stability analysis showed that fluid elasticity contributes to the stabilization of the jet \citep{rallison1995instability}, with hoop stresses stabilizing the sinuous mode and partially stabilizing the varicose mode.

Direct numerical simulations of turbulent planar jets of FENE-P fluid at high Reynolds number showed that the non-Newtonian jet has a slower spreading rate and decay rate with respect to the equivalent Newtonian jet \citep{guimaraes2020direct}. The classical jet statistics however still follow the power-law scalings computed for the Newtonian case at the same flowing conditions, although these scalings are recovered further away from the inlet, thus resulting in the observed lower spreading and decay rate. Differences from the Newtonian case are minimal for low Weissenberg numbers ($Wi\ll1$, ratio of the polymer relaxation time over the flow time scale), while they become clear for larger values of the Weissenberg number, when a larger fraction of turbulent kinetic energy is transferred into elastic energy.

The presence of polymers alters the energy cascade from large to small scales: in turbulent Newtonian flows the non-linear energy cascade transfers energy from the larger eddies to progressively smaller and smaller eddies, till it is dissipated at the smallest scales. When polymers are added to the Newtonian fluid, part of the energy from the larger eddies is first transferred to the polymers \citep{valente2014effect,valente2016,zhang2021experimental}, and then transferred to the small eddies or dissipated by the polymers. At high values of the Weissenberg number, ratio of the polymer relaxation time over the flow time-scale, the energy cascade via the polymers competes with the non-linear energy cascade from the fluid \citep{valente2014effect}. This effect is reflected in the turbulent kinetic energy power spectrum: the energy decay still follows a power-law scaling, however with a different exponent from the $-5/3$ obtained for Newtonian turbulence \citep{lee2003turbulent}. However, there is yet no clear agreement on the value of the power-law exponent of the turbulent kinetic energy spectrum in viscoelastic fluids, with the theory predicting an exponent lower or equal to $-3$ \citep{balkovsky2001turbulence,fouxon2003spectra}, and experiments and direct numerical simulations reporting a value equal to $-3$ \citep{valente2016,vonlanthen2013grid}.

In this paper we report numerical simulations of planar jets at low Reynolds number using three different non-Newtonian models: Carreau (shear-thinning), Giesekus (shear-thinning and viscoelastic) and Oldroyd-B (viscoelastic). We extend the work of \citet{parvar2020local} to different types of non-Newtonian fluids -- namely shear-thinning and viscoelastic fluids -- and to about one order of magnitude larger values of $\gamma$, the ratio of the relevant non-Newtonian time scale (relaxation time or consistency index) over a flow time scale. By varying $\gamma$, we trace a map of the transition from a laminar to a turbulent flowing regime, and assess the differences among the fluids with different non-Newtonian features.

The manuscript is organised as follows. In section~\ref{sec: num}, we start by reporting the mathematical model and the details on its numerical discretization; then, we introduce the selected computational setup, together with the relevant parameters, in section~\ref{sec: compset}. Section~\ref{sec: res} is dedicated to the analysis and presentation of the results obtained from the numerical simulations. In particular, we consider the turbulent energy spectra and structure functions in the core of the jet, the bulk statistics of the jet computed along the stream-wise direction, the profile of the flow velocity and of the non-Newtonian stresses along the jet-normal direction, and the map of the transition from a laminar to a turbulent flowing regime. Lastly, conclusions are drawn in section~\ref{sec: concl}.

\section{Mathematical model and numerical method}
\label{sec: num}
We study jets of non-Newtonian fluids using numerical simulations; their dynamics are governed by the conservation of momentum and the incompressibility constraint, coupled with the transport equation for the extra-stress tensor describing the non-Newtonian effects,
\begin{equation}
\rho \left( \frac{\partial \mathbf{u}}{\partial t}+\mathbf{u}\cdot\nabla\mathbf{u} \right)=-\nabla p+\nabla \cdot \left[ \eta_s\left( \nabla\mathbf{u} +\nabla\mathbf{u}^T \right) \right]+\nabla\cdot\tau,
\label{eq: ns}
\end{equation}
\begin{equation}
\nabla \cdot \mathbf{u}=0,
\label{eq: cont}
\end{equation}
\begin{equation}
\lambda \overset{\nabla}{\tau}+\tau +\frac{\alpha \lambda}{\eta_p} \tau\cdot\tau =\eta_p\left( \nabla\mathbf{u} +\nabla\mathbf{u}^T \right).
\label{eq: extras}
\end{equation}
In the previous system of equations, $\mathbf{u}$ denotes the velocity field, $p$ the pressure, and $\tau$ the non-Newtonian extra-stress-tensor, with $\rho$ being the fluid density (homogeneous in space and constant in time), $\eta_s$ and $\eta_p$ the solvent and polymeric dynamic viscosity, $\lambda$ the polymer relaxation time scale, and $\alpha$ the mobility parameter of the Giesekus fluid (equal to zero for the Oldroyd-B fluid model). $\overset{\nabla}{\tau}$ indicates the upper-convected derivative defined as
\begin{equation}
\overset{\nabla}{\tau}=\frac{\partial \tau}{\partial t}+\mathbf{u}\cdot\nabla\tau- \left( \nabla\mathbf{u}^T\cdot \tau+\tau\cdot \nabla\mathbf{u}\right).
\end{equation}
When considering the inelastic Carreau fluid model, the extra-stress transport equation is not solved, whereas the solvent viscosity $\eta_s$ is a function of the local applied shear rate $\dot \gamma$ as
\begin{equation}
\eta_s=\eta_\infty+(\eta_0-\eta_\infty)\left[ 1+\left(\lambda \dot\gamma \right)^2\right]^{\frac{n-1}{2}}.
\label{eq: sh_thinn}
\end{equation}
Here, $\eta_0$ and $\eta_\infty$ are the zero-shear viscosity and the viscosity for $\dot \gamma\to\infty$, $\lambda$ is the fluid consistency index, and $n$ is the power-law index defining the non-Newtonian fluid behaviour.

The numerical solution of the extra-stress transport equation, Eq.~\ref{eq: extras}, is cumbersome due to numerical instabilities arising for high values of the Weissenberg number (ratio of the elastic relaxation time scale of the polymers over a flow time scale), when disturbances amplifying over time \citep{dupret1986loss,min2001effect,sureshkumar1995effect} can cause the loss of positive-definiteness of the conformation tensor. This issue is known as the high-Weissenberg number problem \citep{keunings1986high} and different methods have been proposed to tackle it; in this work the matrix-logarithm formulation \citep{fattal2004constitutive,hulsen2005flow} of the transport equation of the non-Newtonian extra-stress is adopted. The transport equation for the extra-stress is first recast in a transport equation for the conformation tensor; then the matrix logarithm of this latter equation is taken. The matrix-logarithm formulation is exact, as no approximation is introduced, and solves the problem by implicitly enforcing the positive-definiteness of the conformation tensor. However, this formulation adds a significant overhead to the computational requirements of numerical simulations, since taking the matrix-logarithm of the conformation tensor requires computing eigenvalues and eigenvectors of the tensor itself at every grid point. 

The systems of equations is discretized on a staggered, uniform, Cartesian grid; most variables -- namely pressure, density, viscosity and extra-stresses -- are stored at the cell centre, while velocity data is stored at the cell faces. All spatial derivatives are approximated using a second-order finite difference scheme, the only exception being the advection term of the extra-stress transport equation, which is discretized using a fifth-order WENO scheme \citep{shu2009high,sugiyama2011full}. Time discretization is performed using a second-order, explicit Adams-Bashforth scheme. The pressure coupling is solved by using the fractional-step method \citep{kim1985application}, with the resulting Poisson equation solved using a fast pressure solver. 
The in-house code \textbf{\textit{Fujin}} is used to solve the coupled systems of equations; the very same code has been extensively used and validated in the past on a variety of problems \citep{abdelgawad2023scaling,brizzolara2021fiber, cannon2021effect, mazzino2021unraveling, olivieri2020dispersed, rosti2020increase, rosti2021turbulence, rosti2023large, rosti2021shear}. The full list of validation cases is available at \url{https://groups.oist.jp/cffu/code}.

\section{Computational setup}
\label{sec: compset}
\begin{table}
\caption{List of the simulations performed and of the corresponding parameters. All simulations were performed at a constant inlet Reynolds number, $Re=20$. The non-dimensional number $\gamma$ corresponds to the Weissenberg number $Wi$ for the Giesekus and Oldroyd-B fluid, and to the Carreau number $Cu$ for the Carreau fluid.}
\label{tab: params}
\centering
\begin{tabular}{lccccc}
Fluid   & $\gamma$ & $\eta_p/\eta_0$ & $\eta_\infty/\eta_0$ & $n$ & $\alpha$ \\ \hline
Newtonian  & $-$ & $-$ & $-$ & $-$ & $-$ \\
Carreau  & 1 & $-$ & 0.02 & 0.2 & $-$ \\
Carreau  & 10 & $-$ & 0.02 & 0.2 & $-$ \\
Carreau  & 100 & $-$ & 0.02 & 0.2 & $-$ \\
Giesekus & 1 & 0.98 & $-$ & $-$& 0.125 \\
Giesekus &10 & 0.98 & $-$ & $-$& 0.125 \\
Giesekus  & 100 & 0.98 & $-$& $-$ &0.125 \\
Oldroyd-B & 1 & 0.98 & $-$ & $-$& $-$ \\
Oldroyd-B & 10 & 0.98 & $-$ & $-$& $-$ \\
Oldroyd-B & 100 & 0.98 & $-$ & $-$ & $-$ \\
\end{tabular}
\end{table}

\begin{figure}
\setlength{\unitlength}{\columnwidth}
\begin{picture}(1,1.85)
\put(0.,1.29){\includegraphics[width=\columnwidth, keepaspectratio]{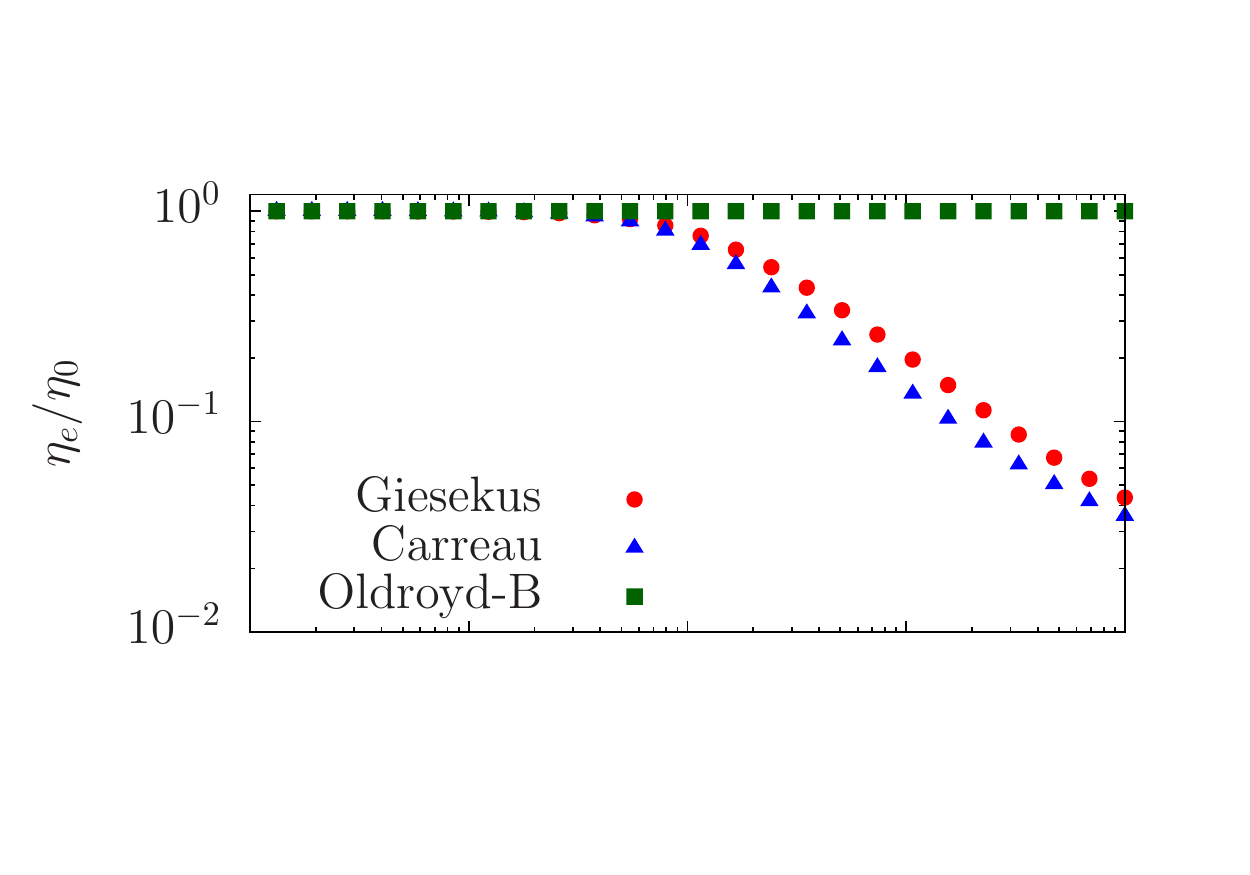}}
\put(0.,0.87){\includegraphics[width=\columnwidth, keepaspectratio]{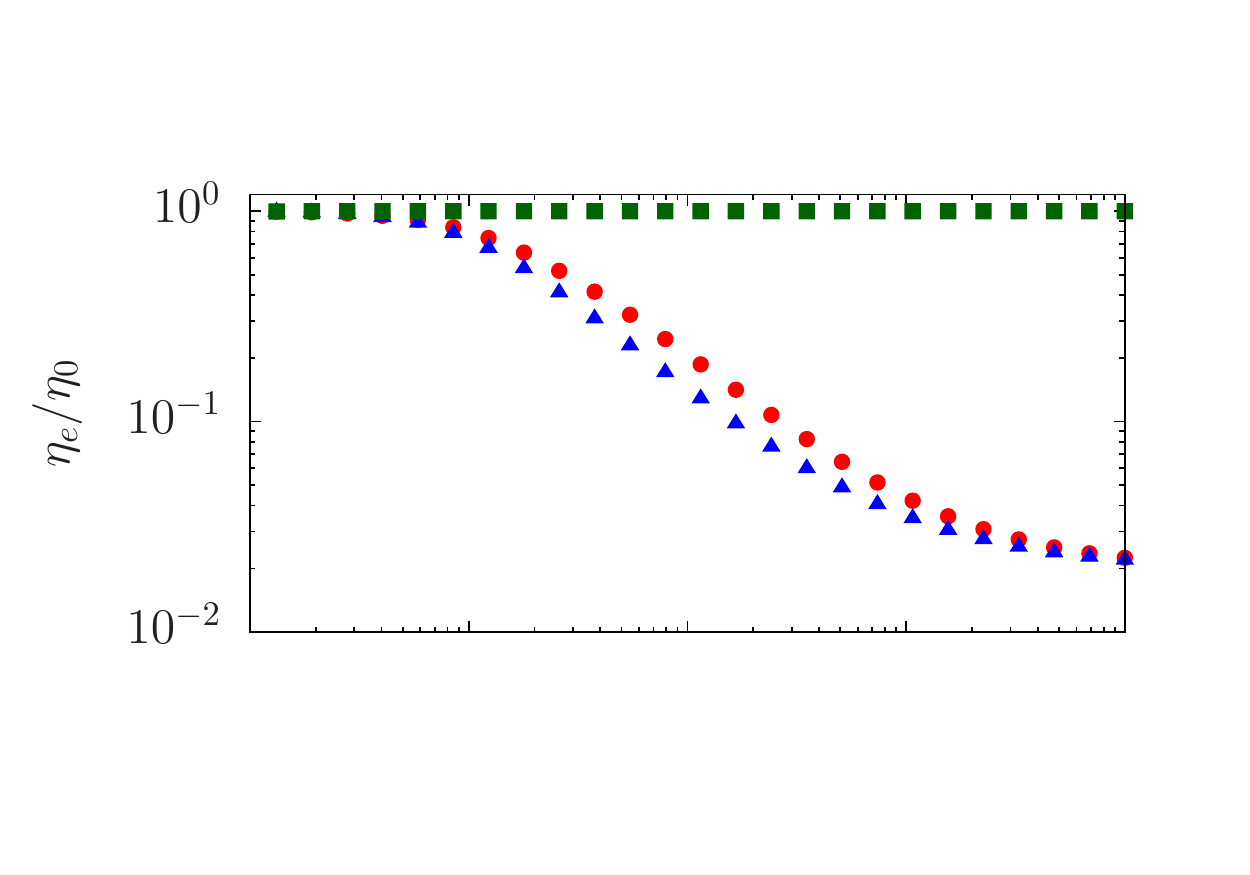}}
\put(0.,0.45){\includegraphics[width=\columnwidth, keepaspectratio]{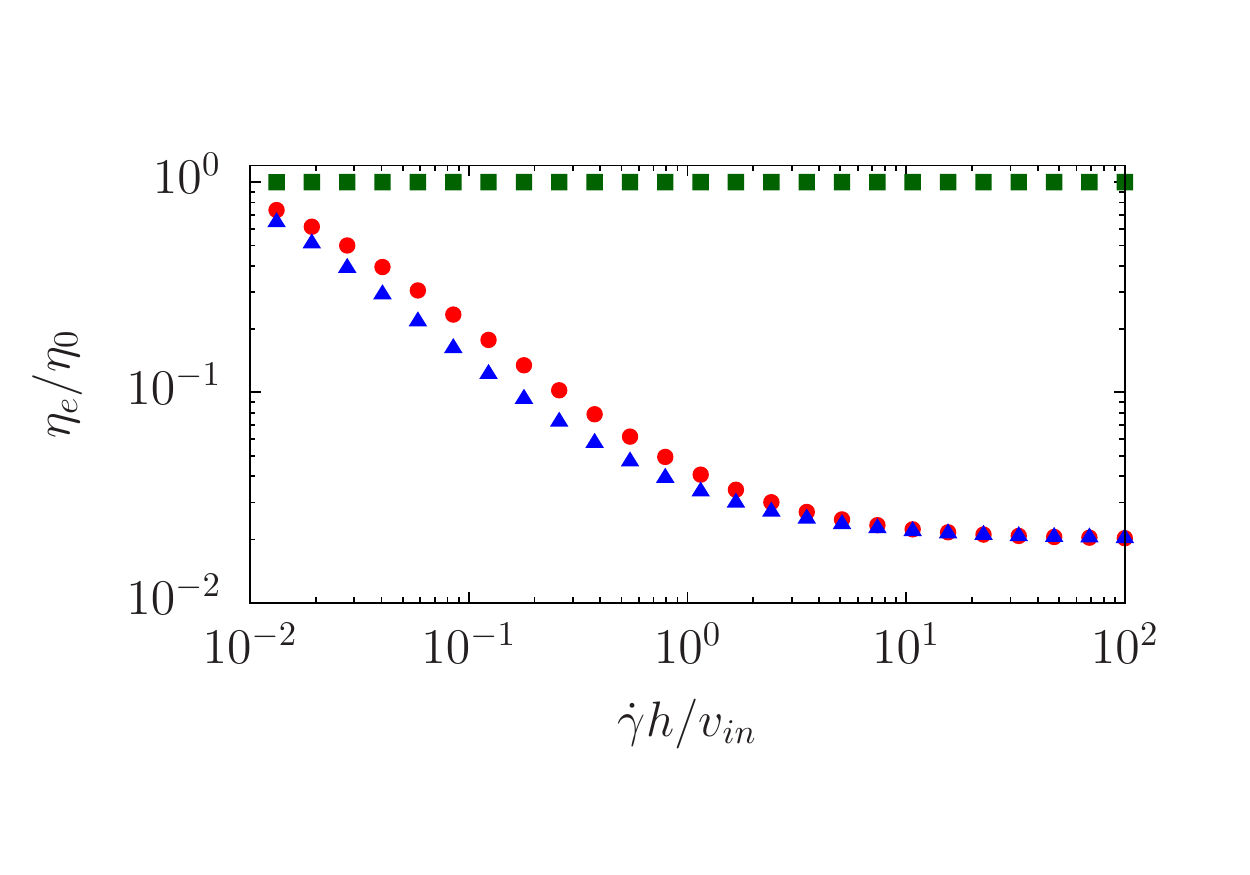}}
\put(0.,-0.1){\includegraphics[width=\columnwidth, keepaspectratio]{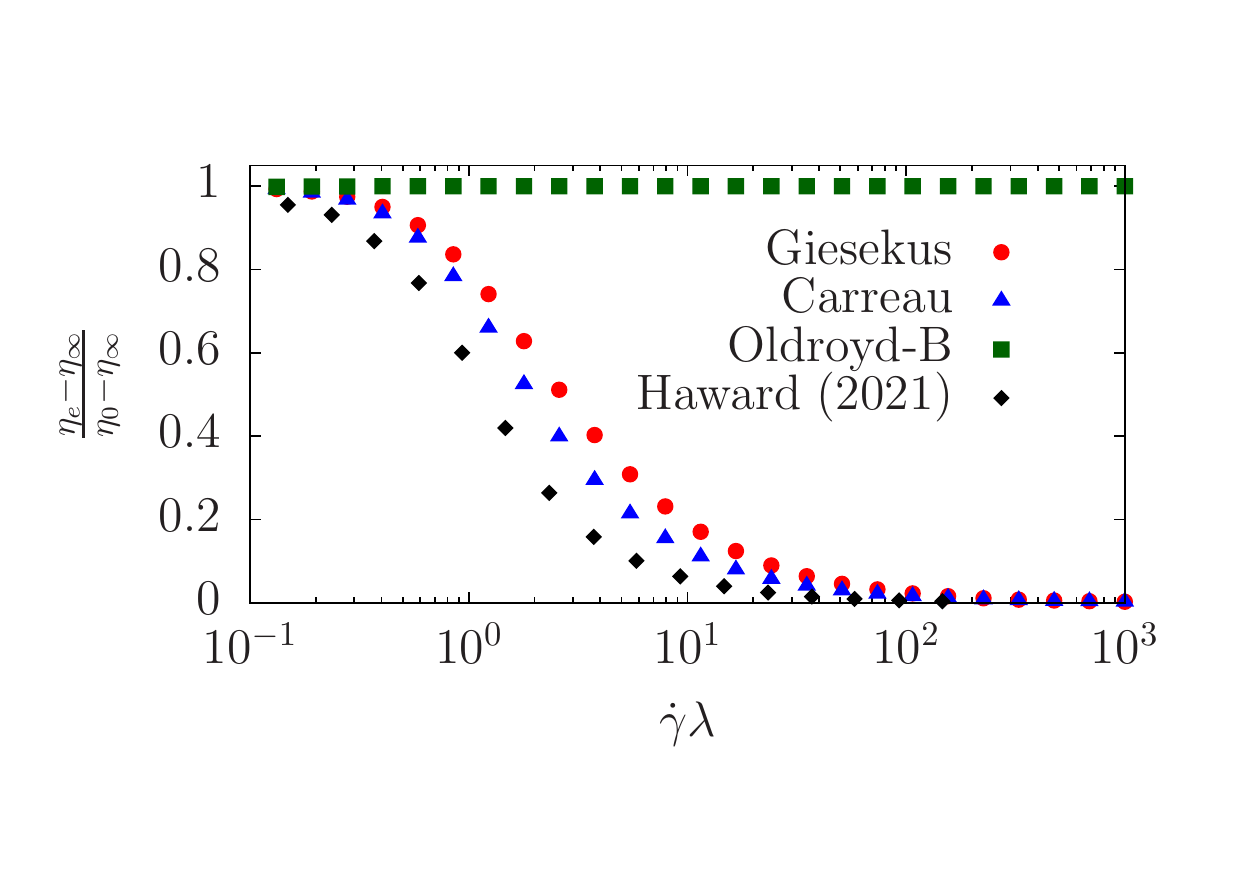}}
\put(0.04,1.8){(a)}
\put(0.04,1.38){(b)}
\put(0.04,0.96){(c)}
\put(0.04,0.41){(d)}
\end{picture}
\caption{Panels (a)-(c) report the fluid rheology for three values of $\gamma$: $\gamma=1$, $10$,  and $100$ from top to bottom.  Panel (d) compares the rheology of the fluids used in our simulations with that of a 100:60 mM CPyCl:NaSal wormlike micelle solution in deionized water used in the experiments by \citet{haward2021stagnation}.} 
\label{fig: rheology}
\end{figure}
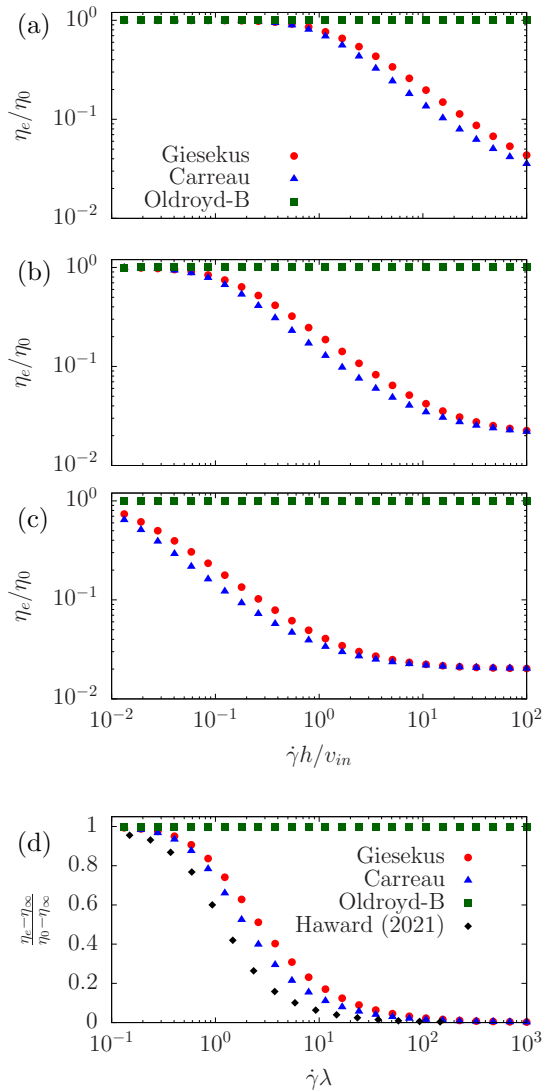

We perform numerical simulations of planar jets for Newtonian and non-Newtonian fluids. The flow is injected from a slit with height $2h$ on the left side of the domain with constant velocity $v_{in}$; the polymers are at rest (zero extra-stresses) when injected in the domain. The fluid injected from the slit is the same fluid found in the rest of the domain. The velocity profile at the inlet is a plug flow; we tested also different initial conditions, namely a parabolic velocity profile having the same flow-rate. Details on the effect of the different inlet boundary conditions can be found in appendix \ref{sec: inlet}.

The computational domain is a box of size $160h\times240h\times13.3h$ (stream-wise $\times$ jet-normal $\times$ span-wise directions) and it is discretized using $1536\times2304\times128$ grid points in the corresponding directions. Note that, when the flow is fully laminar, simulations have been performed in two-dimensional (2D) domains. No-slip and no-penetration boundary conditions are imposed at the left wall, exception made for the inlet slit. Free-slip boundary conditions are enforced at the top and bottom boundaries, and an outflow non-reflective boundary condition \citep{orlanski1976simple} is implemented at the right boundary. Periodic boundary conditions are enforced in the span-wise direction.
The resulting computational cost for two-dimensional simulations is up to about 250'000 core hours, while for three-dimensional simulations is up to about 3'000'000 core hours.

We tested the independence of the obtained results on both the domain size and the grid resolution. In particular, the stream-wise size of the domain is large enough for the jet to fully develop: the region where inlet effects are still dominant, up to about $y=40h$ for some of the non-Newtonian cases, is followed by an approximately self-similar region. Most of the statistics are taken within the self-similar region, unless stated otherwise. 
We also tested different domain sizes in the jet-normal direction and we ensured that the jet evolves without being affected by the presence of the top and bottom boundaries. For the three-dimensional simulations we tested several domain sizes in the span-wise direction in order to verify that the width of the domain does not affect our results. Finally, the grid resolution was tuned to properly resolve all the relevant scales of the flow. We performed an additional simulation for the Oldroyd-B case at $\gamma=100$ using a finer grid resolution to verify the independence of the results on the computational grid, observing no difference in the bulk jet statistics between the two cases at different grid resolutions. The ratio between the grid spacing used in the current setup, $\Delta$, over the grid spacing of the finer grid tested, $\Delta_f$, is $\Delta/\Delta_f=1.5$.

We limit our analysis to low Reynolds number flows: the Reynolds number, computed using the velocity at the inlet $v_{in}$, the half-height of the slit through which the fluid is injected $h$ and the zero-shear viscosity $\eta_0$ is $Re=\rho h v_{in}/ \eta_0=20$. Newtonian jets at this value of the Reynolds number are laminar, thus any unstable behavior is to be attributed to the non-Newtonian property of the fluid. Apart from the reference Newtonian simulation, three other non-Newtonian fluid models are considered: the Carreau, Oldroyd-B and Giesekus fluid models. The Carreau fluid model describes an inelastic shear-thinning fluid, whose viscosity depends on the instantaneous shear rate according to the power-law relation defined in Eq.~\ref{eq: sh_thinn}, the Oldroyd-B model represents a purely elastic fluid with constant viscosity, and the Giesekus model describes an elastic fluid with shear-dependent viscosity. Note that, the differences among the Giesekus and Oldroyd-B fluid models are not limited to the presence or absence of shear-thinning, but are characterized by additional different behaviors of the extra-stress components, which can be exactly computed in simpler flow configurations. 
The full analysis of the components of the non-Newtonian extra-stress tensor and on the normal stress differences are thus included in the discussion of the results.

The rheology of the three non-Newtonian fluids is chosen to provide a reasonable comparison among fluid with different properties. All the fluids have the same zero-shear viscosity $\eta_0$,  which is equal to the solvent viscosity $\eta_s$ in the Newtonian simulation, and to the sum of the solvent and polymeric viscosity in the Oldroyd-B and Giesekus models $\eta_0=\eta_s+\eta_p$, with the ratio of the polymeric to total viscosity being set equal to $\eta_p/\eta_0=0.98$. We consider a strong shear-thinning effect, provided by a power index equal to $n=0.2$ for the Carreau model; the same level of shear-thinning is achieved in the Giesekus model with the mobility coefficient $\alpha$ equal to $0.125$. Figure~\ref{fig: rheology} reports the effective viscosity of each non-Newtonian fluid for three different values of $\lambda$, with $\lambda$ being the polymer relaxation time (Giesekus and Oldroyd-B) or the fluid consistency index (Carreau). The fluid parameters we selected for the Carreau and Giesekus fluids yield to a shear rheology qualitatively similar to that of a 100:60 mM CPyCl:NaSal wormlike micelle solution in deionized water that was used in the experiments by \citet{haward2021stagnation}. The normalized fluid rheology for the three fluid models used in this work and that for the 100:60 mM CPyCl:NaSal wormlike micelle solution is compared in panel (d) of figure~\ref{fig: rheology}.

Throughout the rest of the paper, we will discuss the effect of the ratio $\gamma$ between a non-Newtonian time scale and a flow time scale, $h/v_{in}$, based on the inlet velocity and dimension. For fluids characterized by fluid elasticity, namely the Giesekus and Oldroyd-B fluid models, we use the polymer relaxation time as the non-Newtonian time scale, hence the parameter $\gamma$ corresponds to the Weissenberg number, $Wi$. For the Carreau fluid model we use instead the fluid consistency index as the non-Newtonian time scale; in this case $\gamma$ is usually called the Carreau number, $Cu$. Table~\ref{tab: params} reports a list of the simulations performed and all the main parameters. 

\section{Results}
\label{sec: res}
\begin{figure*}
\centering
\includegraphics[width=0.9\textwidth, keepaspectratio]{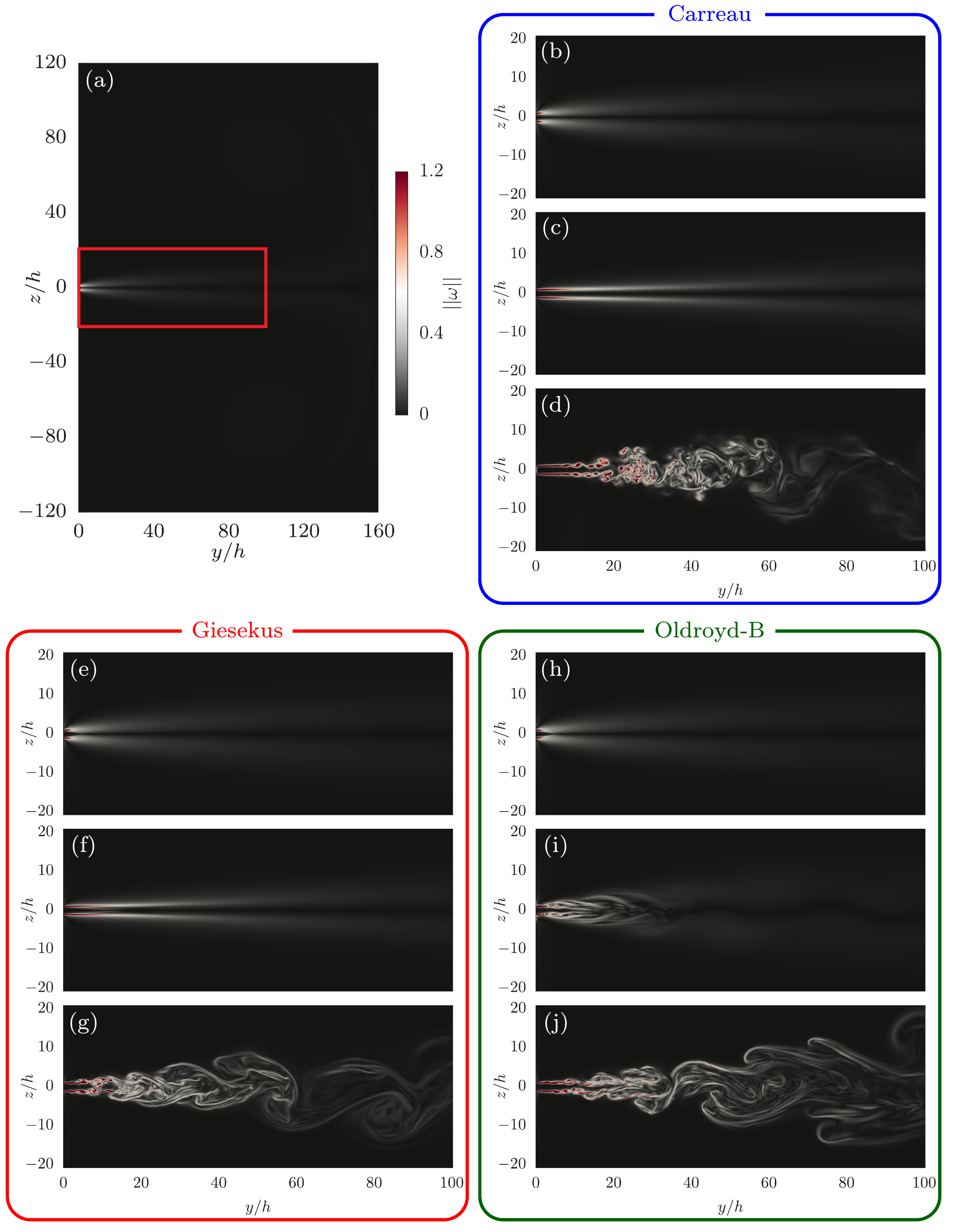}
\caption{Vorticity magnitude for the fully-developed jet flow. Panel (a) reports the Newtonian case on the full domain; for all the non-Newtonian cases, panels (b)-(j), only a subset of the domain is shown, represented by the red box in panel (a). Panels (b)-(j): each colored box groups data at different Weissenberg numbers for the same non-Newtonian fluid. Each row of the boxes corresponds to a different $\gamma$: panels (b), (e), (h) $\gamma=1$, panels (c), (f), (i) $\gamma=10$ and panels (d), (g), (j) $\gamma=100$. }
\label{fig: quali}
\end{figure*}
We start our analysis of the non-Newtonian jets by showing an instantaneous snapshot of the vorticity magnitude, reported in figure~\ref{fig: quali} for all the simulated cases once the flow becomes stationary over time. Panel (a) shows the magnitude of the vorticity for the Newtonian reference case over the full computational domain, whereas panels (b)-(j) report the same quantity in a subset of the domain, identified by the red box in panel (a). Each box groups a different fluid model -- Carreau, Giesekus and Oldroyd-B -- and, within each group, each row corresponds to a different $\gamma$ -- from top to bottom: $\gamma=1$, $\gamma=10$ and $\gamma=100$.  It is clear that at the lowest $\gamma$ the flow is steady and laminar (panels (b), (e), (h)), similarly to what observed for a Newtonian fluid (panel (a)). The effect of the non-Newtonian features -- viscoelasticity or shear-thinning --  is limited and does not produce any clear change in the flowing regime with respect to the Newtonian case. Moving to $\gamma=10$, non-Newtonian effects become clearer: the Carreau and Giesekus models (panels (c) and (f)) generate a high-speed region at the core of the jet near the inlet, originated by the strong shear-thinning effect. The Oldroyd-B fluid shows instead the onset of a transition towards a turbulent-like regime (panel (i)). Note that the observed flow is clearly different from the classical turbulence experienced at high-Reynolds number; however here we use the term turbulence in a broader sense to denote disordered fluid motions. From the same figure we can also observe that only the region close to the inlet, up to $y\simeq40h$ is characterized by a clear turbulent-like regime, which then transitions towards a more regular flow, resembling the laminar regime, further downstream.  At the highest $\gamma$ considered here (panels (d), (g), (j)), a clear turbulent-like regime is observed for all cases. However, the disordered flowing pattern is qualitatively very different among them.

\begin{figure}
\setlength{\unitlength}{\columnwidth}
\centering
\includegraphics[width=1.\columnwidth, keepaspectratio]{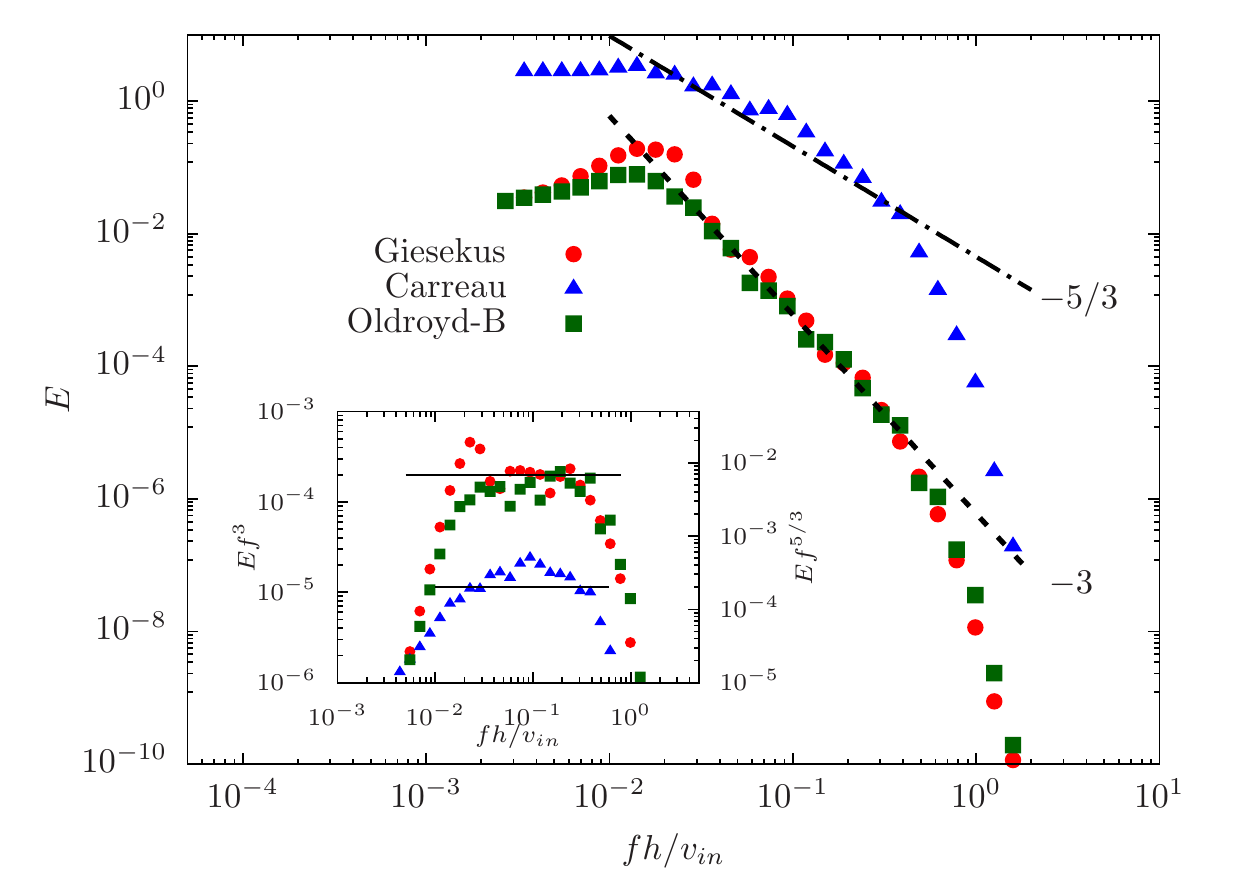}
\caption{Power spectra of the turbulent kinetic energy for the three fluid models considered at  $\gamma=100$. Data for the Carreau fluid is shifted upwards by a factor 100. The inset shows the compensated power spectra, where the Giesekus and Oldroyd-B cases are compensated by $f^{3}$ (left axis) and the Carreau case by $f^{5/3}$ (right axis).} 
\label{fig: PS}
\end{figure}

To provide a quantitative description of the resulting flowing regimes observed at the highest $\gamma$, we compute the power spectrum of the turbulent kinetic energy, with velocity data extracted from a probe at the jet centerline at $y=40h$; the power spectrum is shown in figure~\ref{fig: PS}. \citet{lumley1973drag} theorized the existence of two different regimes in viscoelastic turbulent flows: at larger scales inertial effect are dominant, while at smaller scales elastic effects from the polymers dominate \citep{de2005homogeneous,lumley1973drag}. 
In our simulations, at the highest value of $\gamma$, the turbulent kinetic energy power spectrum is characterized by the presence of a single regime, either inertial or elastic, extending over more than one decade; fluid elasticity appears to be the relevant factor in determining the turbulent regime observed. For the inelastic Carreau fluid, we observe the power-law scaling $f^{-5/3}$, similarly to what found in classical Newtonian turbulence, while for both the Giesekus and Oldroyd-B fluids we observe instead a clear $f^{-3}$ power-law scaling. The inset shows the compensated power spectrum of the turbulent kinetic energy for the three cases; we compensate by $f^3$ the Giesekus and Oldroyd-B results (left axis), and by $f^{5/3}$ the Carreau one (right axis). The compensated power spectra further prove the existence of the elastic (Giesekus and Oldroyd-B fluid) or inertial (Carreau fluid) regimes.
Recent experimental results of non-Newtonian jets \citep{vonlanthen2013grid,yamani2021spectral} reported a $-3$ spectra,  which was also suggested to be universal \citep{yamani2021spectral}. This value is also confirmed by theoretical arguments, which predict a scaling exponent equal to $-3$ or steeper \citep{balkovsky2001turbulence, fouxon2003spectra}. Indeed, our results from figure~\ref{fig: PS} clearly show that, even at our very low Reynolds number, in the presence of elasticity the $-3$ power-law scaling persists. It is interesting to note that the Giesekus fluid, which is also characterized by a strong shear-thinning much alike the Carreau fluid, shows only the elastic $-3$ power-law scaling and no trace of a $-5/3$ scaling, observed instead for the inelastic Carreau fluid. 

\begin{figure}
\setlength{\unitlength}{\columnwidth}
\centering
\includegraphics[width=1.\columnwidth, keepaspectratio]{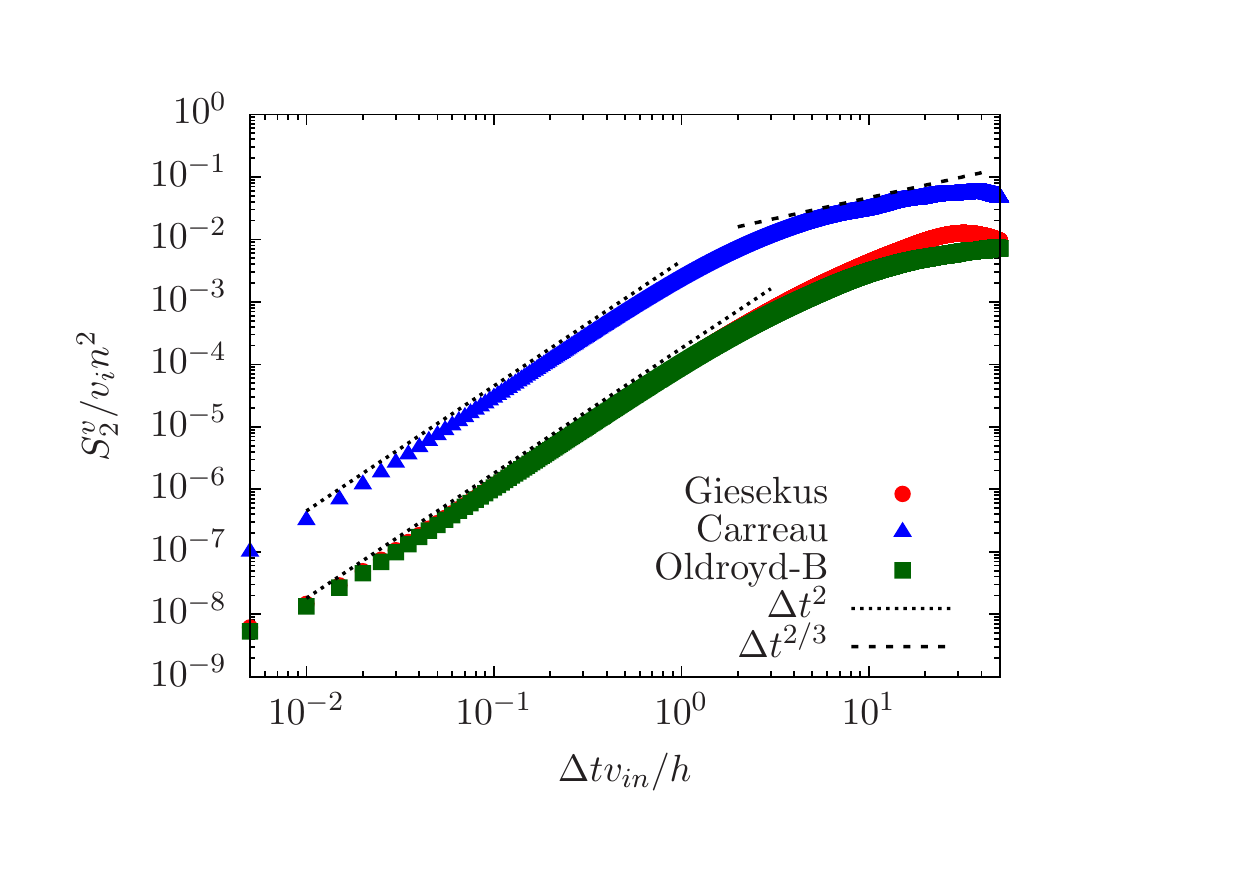}
\caption{Second-order structure function of the velocity fluctuations as a function of the separation time $\Delta t$ for all the turbulent cases at the highest value of $\gamma$. The inertial scaling for the velocity structure function $r^{2/3}$ is reported with a dashed line,  the dissipative regime scaling for the velocity structure function $r^2$ with a dotted line. Data for the Carreau fluid has been shifted upwards by a factor 10 for better readability. } 
\label{fig: strucfun}
\end{figure}

We observe that the power spectrum of the turbulent kinetic energy scales as $f^{-q}$, with $q=5/3$ for the Carreau case and $q=3$ for the Giesekus and Oldroyd-B cases. 
For high-Reynolds Newtonian turbulence, the exponent $q$ of the turbulent kinetic energy spectrum (in the wavenumber space) can be linked to the exponent $p$ of the second-order velocity structure function through the relation $q=p+1$, under the hypothesis of sufficient homogeneity and isotropy. 
We compute the second-order structure function in time over separation times $\Delta t$ as $S_2^v(\Delta t)=\langle \left[v'(t+\Delta t)-v'(t) \right] ^2\rangle$, where the angle brackets denote averaging over time. The structure function, shown in figure~\ref{fig: strucfun}, is computed from the velocity data recorded over time at the jet centerline at $y=40h$, as done for the turbulent kinetic energy power spectrum. For the Carreau fluid we observe a dissipative regime for small time separations characterized by a $\Delta t^2$ power-law scaling, which is then replaced by the $\Delta t^{2/3}$ at larger separation times. The latter power law scaling is characteristic of the inertial range and is in agreement with the inertial range exponent found in the turbulent kinetic energy power spectrum. The inertial range is not found for the Giesekus and Oldroyd-B cases: here we have the $\Delta t^2$ power-law scaling that further extends towards higher time separation, up to about one decade more than what observed for the Carreau case. The power-law scaling of the second order structure function is undistinguished between the dissipative regime at low separation time scales and the elastic regime at larger separation time scales. The power-law exponent for the elastic regime of the second-order structure function $p=2$ is coherent with the exponent $q=p+1=3$ measured in the turbulent kinetic energy power spectrum.

\begin{figure*}
\setlength{\unitlength}{\columnwidth}
\begin{picture}(2,0.6)
\put(0.,-0.1){\includegraphics[width=1.1\columnwidth, keepaspectratio]{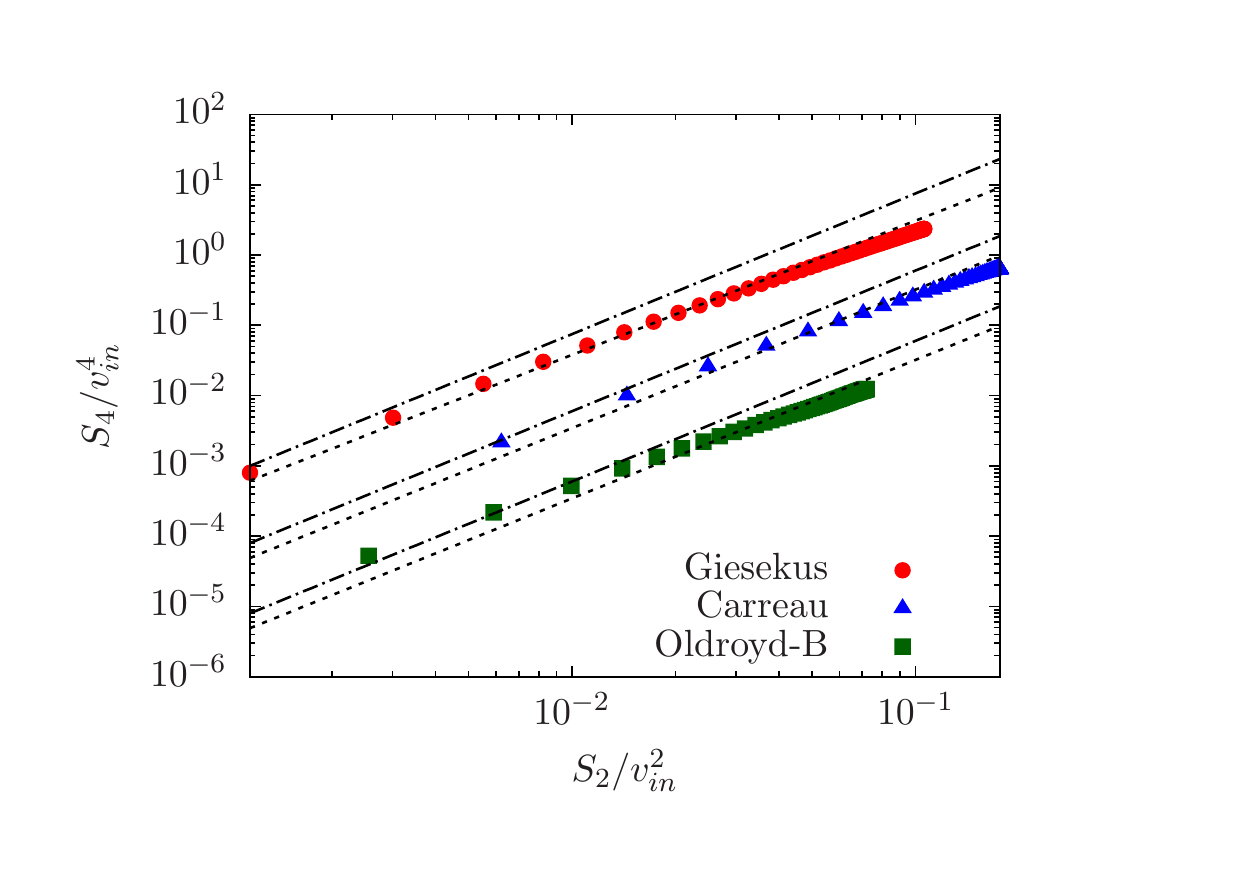}}
\put(0.95,-0.1){\includegraphics[width=1.1\columnwidth, keepaspectratio]{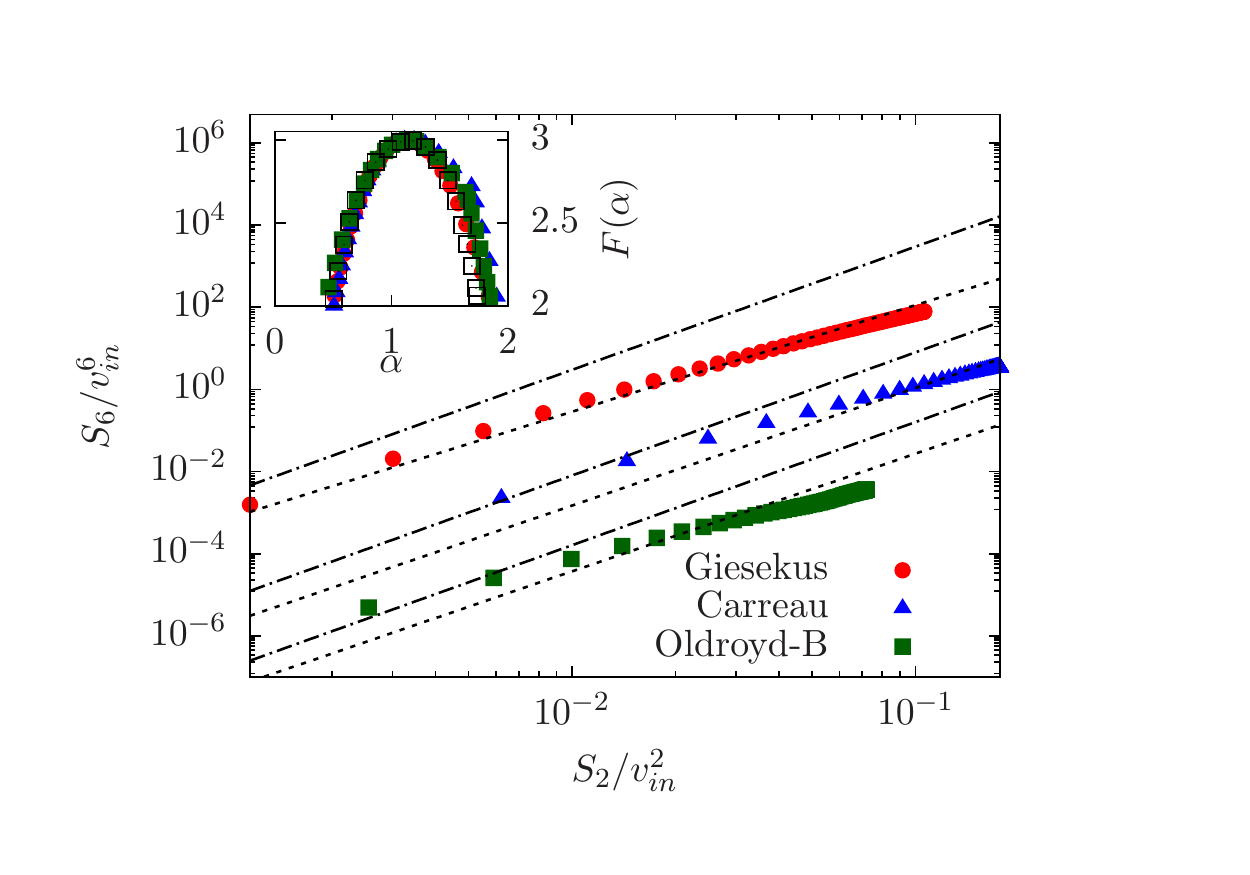}}
\put(0.05,0.5){(\textit{a})}
\put(1.,0.5){(\textit{b})}
\end{picture}
\caption{Extended self-similarity: (a) $S_2-S_4$, (b) $S_2-S_6$ for all the turbulent cases at the highest value of $\gamma$. The dash-dotted lines show the K41 scaling, the dotted lines the K41 scaling with intermittency corrections (obtained at high Reynolds number). The various structure functions are shifted upwards for clarity. Inset: the multifractal spectrum computed from our numerical simulations is compared with experimental data (black empty squares) from \citet{SreenevasanM_1988}.} 
\label{fig: intermittency}
\end{figure*}

We now investigate the presence of intermittency for the cases at the highest $\gamma$. To do so, we consider the higher order velocity structure functions, namely the fourth- and sixth-order structure functions. To investigate the presence of intermittency, the structure functions are computed in space, within a cylinder with the axis at $y=40h$ and with radius equal to one jet thickness ($\delta_{0.5}(40h)$, defined in the following section). The velocity differences are computed among all points within this cylinder and averaged in space and time. For Newtonian turbulence, Kolmogorov theory (K41) predicts that the $p$-th order structure function scales as $p/3$ \citep{kolmogorov1941local}. The K41 theory does not however account for presence of intermittency, which manifests as a deviation from the expected scalings that becomes more evident for the high-order structure functions. Corrections to the exponents of the fourth- and sixth-order structure function exponents have been measured experimentally and numerically in the past \citep{watanabe2004statistics}: the fourth- and sixth-order structure function power-law exponents are reduced by respectively $\chi_4=0.0602$ and $\chi_6=0.2674$. The presence of intermittency has been shown for moderate- and high-Reynolds number Newtonian turbulence; in the following we will show how intermittency characterizes as well low-Reynolds non-Newtonian turbulence. We report the fourth- and sixth-order structure functions against the second-order structure function, figure~\ref{fig: intermittency}, in the so-called extended self-similarity \citep{benzi1995scaling,benzi1993extended,benzi1993extendedb}: in the absence of intermittency, the $p$-order structure function scales as $p/2$, shown with a dash-dotted line, while the dashed line show the scaling corrected for intermittency. It is clear that data from our numerical simulations deviates from the K41 scaling and follows instead the K41 scaling with intermittency correction. While this may be expected for the Carreau case, where the turbulent fluid motions originate from local inertial effects, this is less obvious in the Giesekus and Oldroyd-B cases characterized by elastic turbulence. While the current data does not allow to provide a definitive estimate for the correction coefficients due to intermittency in elastic turbulence, the values obtained for high-Reynolds Newtonian turbulence seems to provide already a good fit. The presence of intermittency is further confirmed by the Cram\'{e}r function $F(\alpha)$, reported in the inset in figure~\ref{fig: intermittency}(b). The Cram\'{e}r function, or multifractal spectrum, uses the coarse graining of the local dissipation to show the presence of intermittency: eddies of progressively smaller sizes are less and less space-filling and a fractal dimension can be associated to each size; the Cram\'{e}r function represents the dimension of the fractal set where power-law scalings with exponent $\alpha$ hold. The multifractal spectrum extracted from our numerical simulations follows fairly well the values measured in the atmospheric surface layer \citep{SreenevasanM_1988}, and no clear difference can be observed among the different non-Newtonian fluids.

\subsection{Bulk statistics}
\begin{figure}
\setlength{\unitlength}{\columnwidth}
\begin{picture}(1,1.20)
\put(0.,0.71){\includegraphics[width=\columnwidth, keepaspectratio]{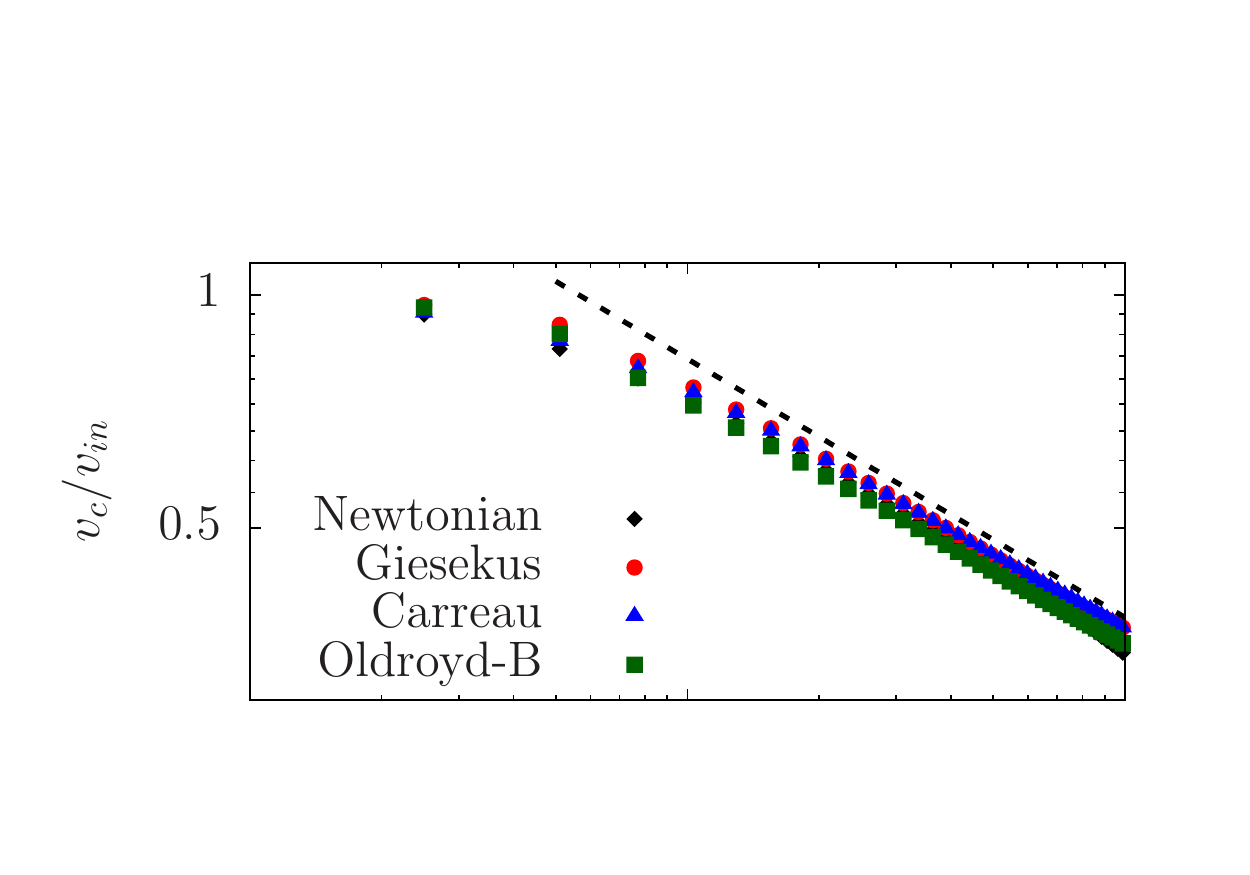}}
\put(0.,0.33){\includegraphics[width=\columnwidth, keepaspectratio]{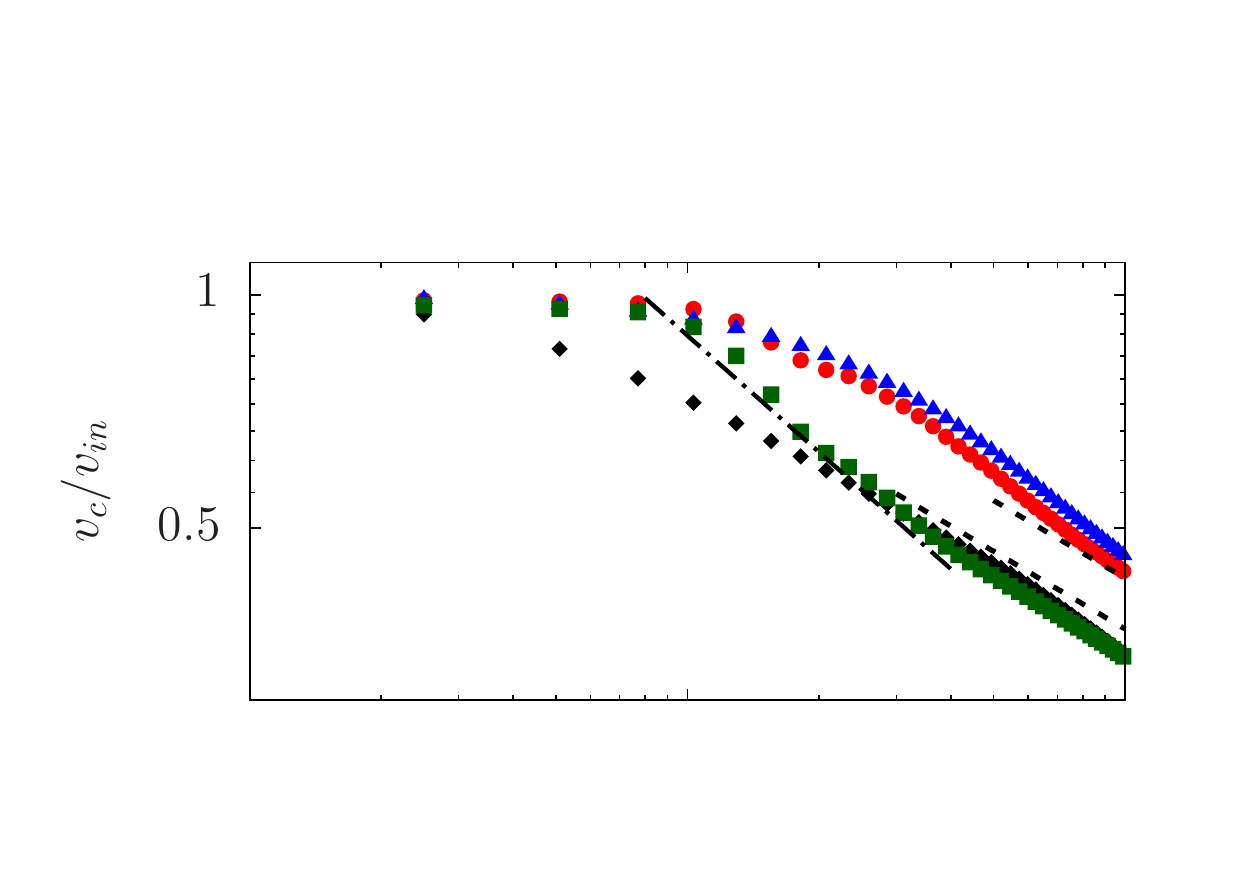}}
\put(0.,-0.05){\includegraphics[width=\columnwidth, keepaspectratio]{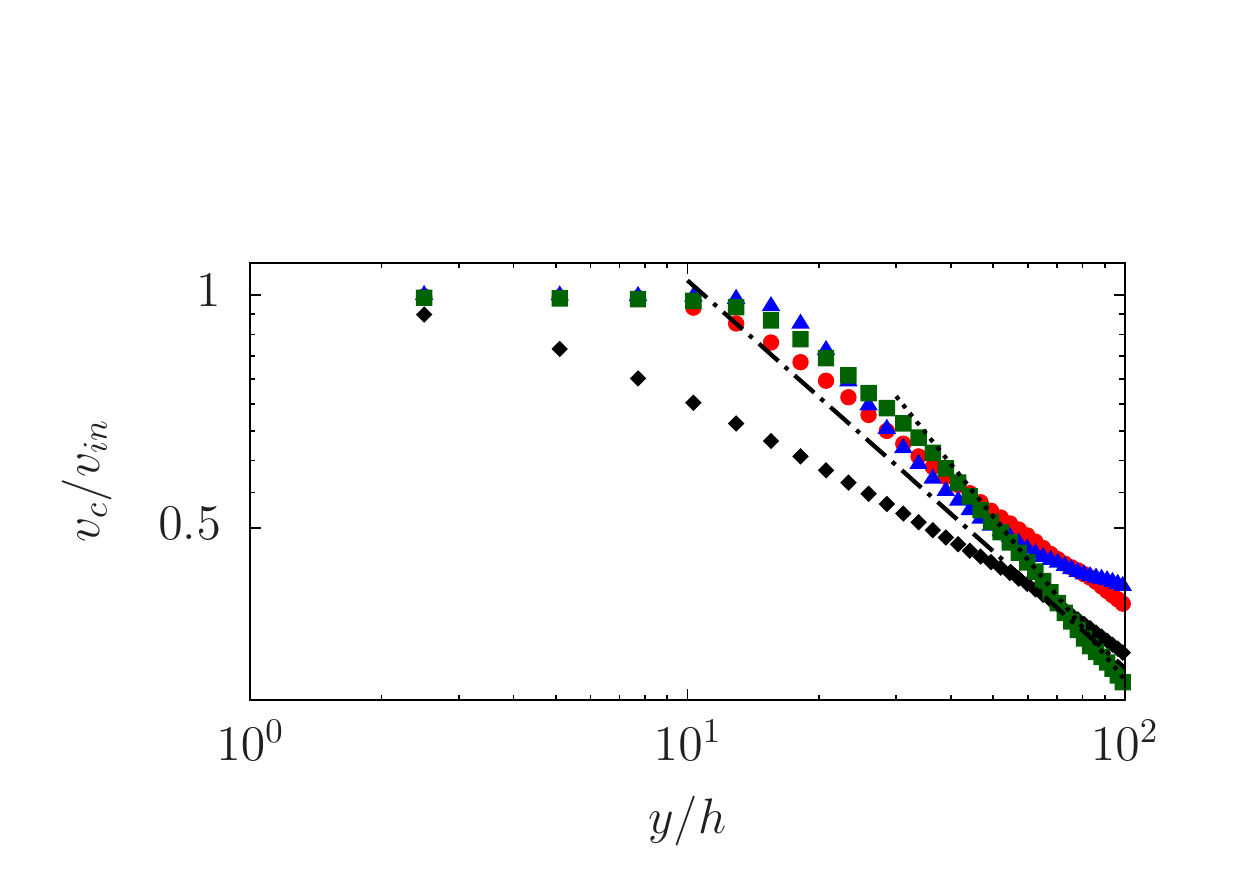}}
\put(0.04,1.16){(a)}
\put(0.04,0.78){(b)}
\put(0.04,0.4){(c)}
\end{picture}
\caption{Centerline velocity for the three non-Newtonian fluid models, with the data for the Newtonian case also reported for reference. The expected laminar, $v_c\propto y^{-1/3}$, and inertial turbulent, $v_c\propto y^{-1/2}$, power-law scalings are shown with dashed and dash-dotted lines respectively. The fit for the Oldroyd-B case at high $\gamma$, $v_c\propto y^{-0.7}$, is also reported with a dotted line. The three panels correspond to different $\gamma$: (a) $\gamma=1$, (b) $\gamma=10$ and (c) $\gamma=100$.} 
\label{fig: vcent}
\end{figure}
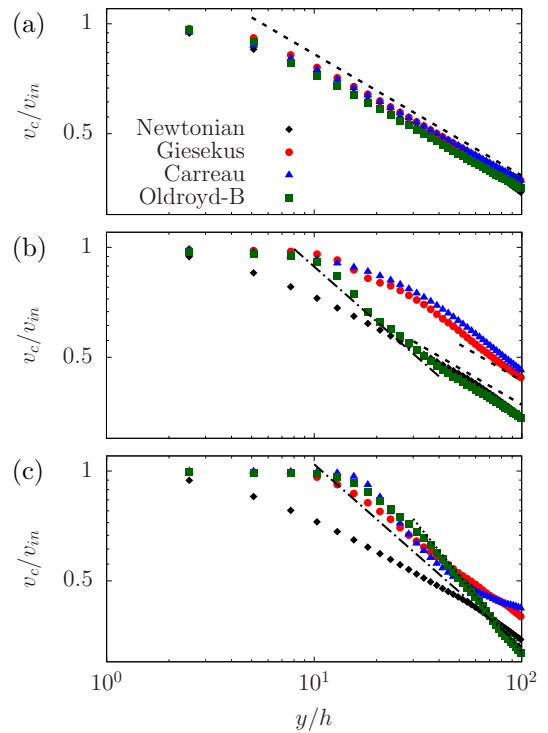
In this section we compute the classical bulk jet statistics and we compare them with analytical scalings obtained for Newtonian planar jets. 

We start by analyzing the stream-wise velocity at the centerline of the jet, $v_c$; scaling laws for the centerline velocity of both laminar \citep{Mei_2001} and turbulent \citep{POPE2000} Newtonian planar jets can be obtained by applying the conservation of momentum and introducing similarity solutions. The following power-law scalings have been obtained for Newtonian planar jets: the centerline velocity decays as $v_c\propto y^{-1/3}$ (laminar case) and as $v_c\propto y^{-1/2}$ (inertial turbulent case), with $y$ being the stream-wise coordinate measuring the distance from the jet inlet. For non-Newtonian laminar planar jets the fluid rheology has to be taken into account when deriving the scaling law for the decay of the centerline velocity; \citet{parvar2020local} however showed that there is only a weak dependence on the fluid rheology, and a fair agreement with the Newtonian scaling is obtained. For this reason we will use the Newtonian power-law scaling in this work, thus neglecting the weak dependence on the fluid rheology on the power-law scaling in the laminar regime. Conversely, \citet{guimaraes2020direct} showed that the same scaling for the Newtonian case still holds for non-Newtonian planar turbulent jets: no hypothesis on the fluid model is needed in the derivation of the decay law for the jet centerline velocity, thus the scaling obtained for Newtonian jets still holds true.

The centerline velocity obtained for all non-Newtonian fluid models and $\gamma$ considered is reported in figure~\ref{fig: vcent}, together with the laminar and inertial turbulent power-law scalings. 
At $\gamma=1$, figure~\ref{fig: vcent}(\textit{a}), there is no appreciable difference among the different fluid models,  with the centerline velocity for all non-Newtonian cases falling on top of the Newtonian case. This result is not surprising, as away from the inlet effects from the non-Newtonian component (shear-thinning, viscoelasticity or both combined) are weaker, especially at this low value of $\gamma$. When  $\gamma$ is increased to $\gamma=10$ and $100$, figures~\ref{fig: vcent}(\textit{b}) and (\textit{c}), the effect of the non-Newtonian component strengthens and we can immediately notice that the memory of the inlet conditions persists over a longer distance: while the Newtonian jet recovers the laminar scaling at about $5h$ from the inlet, for the non-Newtonian cases the transition occurs much later, at distances larger than $10h$ from the inlet. Before this, a region with almost constant velocity can be found. This result confirms the lower decay rate of the centerline velocity and the reduced spreading of the jet reported by \citet{guimaraes2020direct}. It is also interesting to note that, although $\gamma$ is increased tenfold between panels (\textit{b}) and (\textit{c}), the extension of this initial region does not change appreciably. In particular, at the intermediate $\gamma=10$, figure~\ref{fig: vcent}(\textit{b}), the transition towards the laminar scaling is delayed further downstream for the Carreau and Giesekus models,  with the observed flowing regime being always laminar-like and the centerline velocity approaching the laminar scaling only beyond $70h$, although with some minor difference in the observed slope. For the Oldroyd-B model instead, fluid elasticity plays a fundamental role in destabilizing the flow: a transition to an early turbulent-like regime is observed at about $10h$ from the inlet, figure~\ref{fig: quali}(\textit{i}). Further downstream fluid elasticity is not sufficient to sustain this turbulent-like regime and the flow reverts to the laminar regime at around $40h$. Consistently, we can observe a first region, between $10h$ and $40h$, where the centerline velocity follows the inertial turbulent scaling, followed by a region where the flow approaches the laminar scaling. Finally, at the highest value of $\gamma$ considered, $\gamma=100$ in figure~\ref{fig: vcent}(c), we observe a lengthening of the region where memory of inlet conditions persists for the Oldroyd-B model: this region now extends up to about $15h$ from the inlet and is of similar length for all non-Newtonian cases. An intermediate region where the centerline velocity for all three non-Newtonian fluid models decays according to the inertial turbulent scaling, $v_c^{-1/2}$, follows the initial region where inlet effects are still persistent. Further away from the inlet, beyond about $60h$, we observe a substantial reduction in the decay rate from the inertial turbulent scaling for the two fluids with shear-thinning,  the Carreau and Giesekus fluids, while a faster decay is observed for the purely Oldroyd-B fluid. We quantified the decay rate of the centerline velocity of the Oldroyd-B at $\gamma=100$ and found a good agreement with a power-law decay $v_c \propto y^{-0.7}$, with an exponent about 40\% larger than the expected scaling.

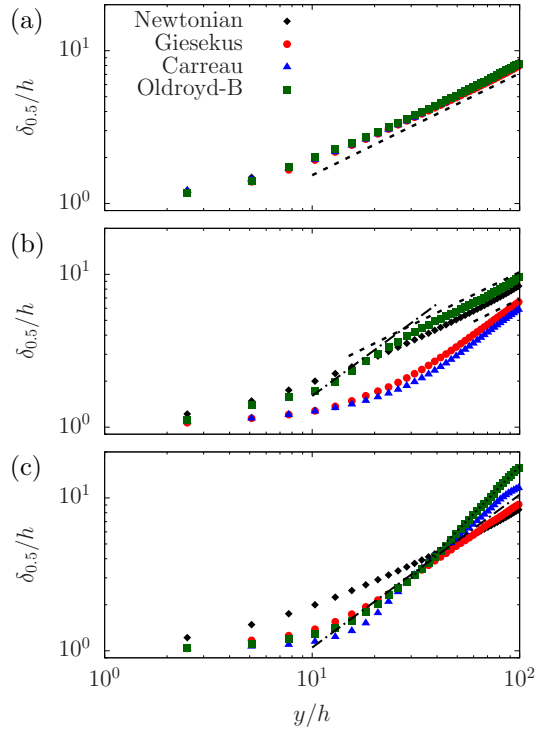
\begin{figure}
\setlength{\unitlength}{\columnwidth}
\begin{picture}(1,1.20)
\put(0.,0.71){\includegraphics[width=\columnwidth, keepaspectratio]{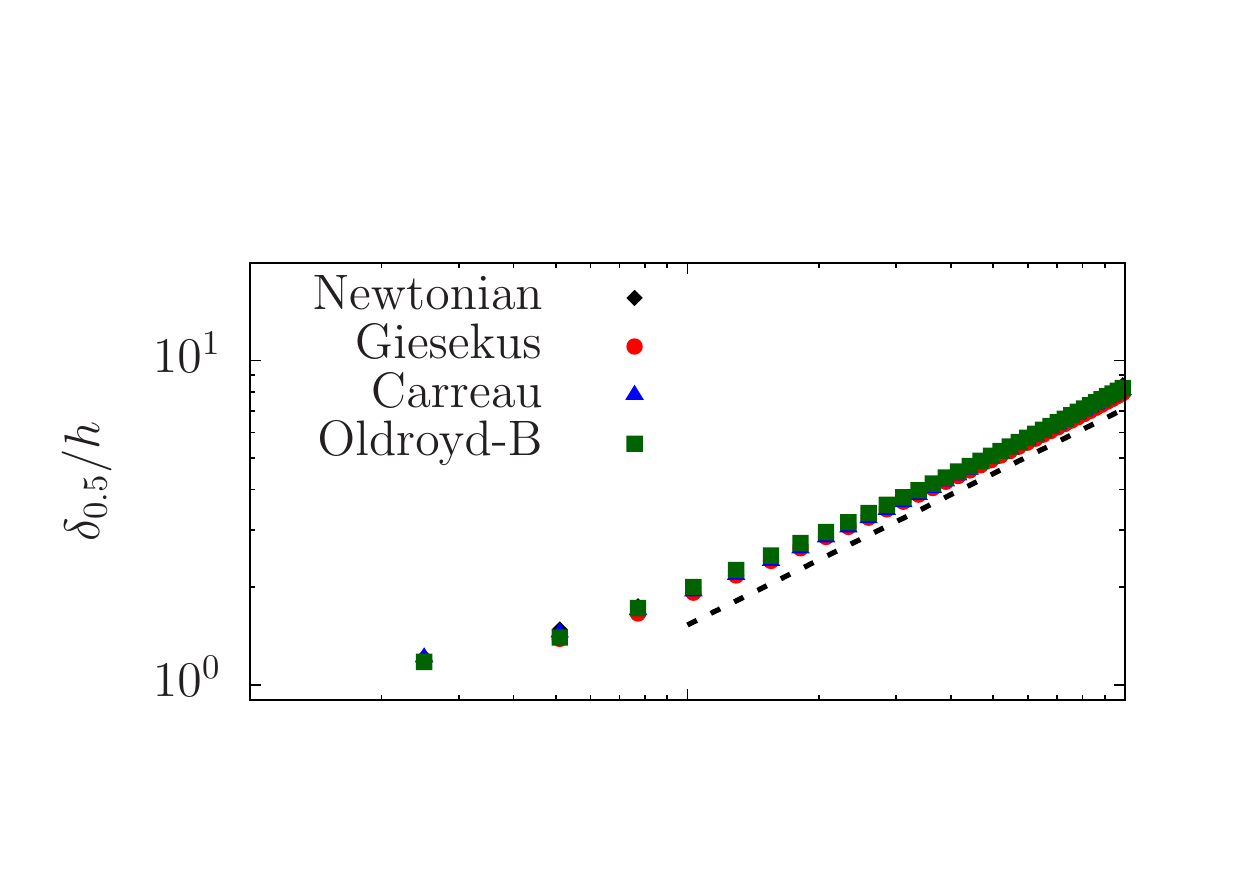}}
\put(0.,0.33){\includegraphics[width=\columnwidth, keepaspectratio]{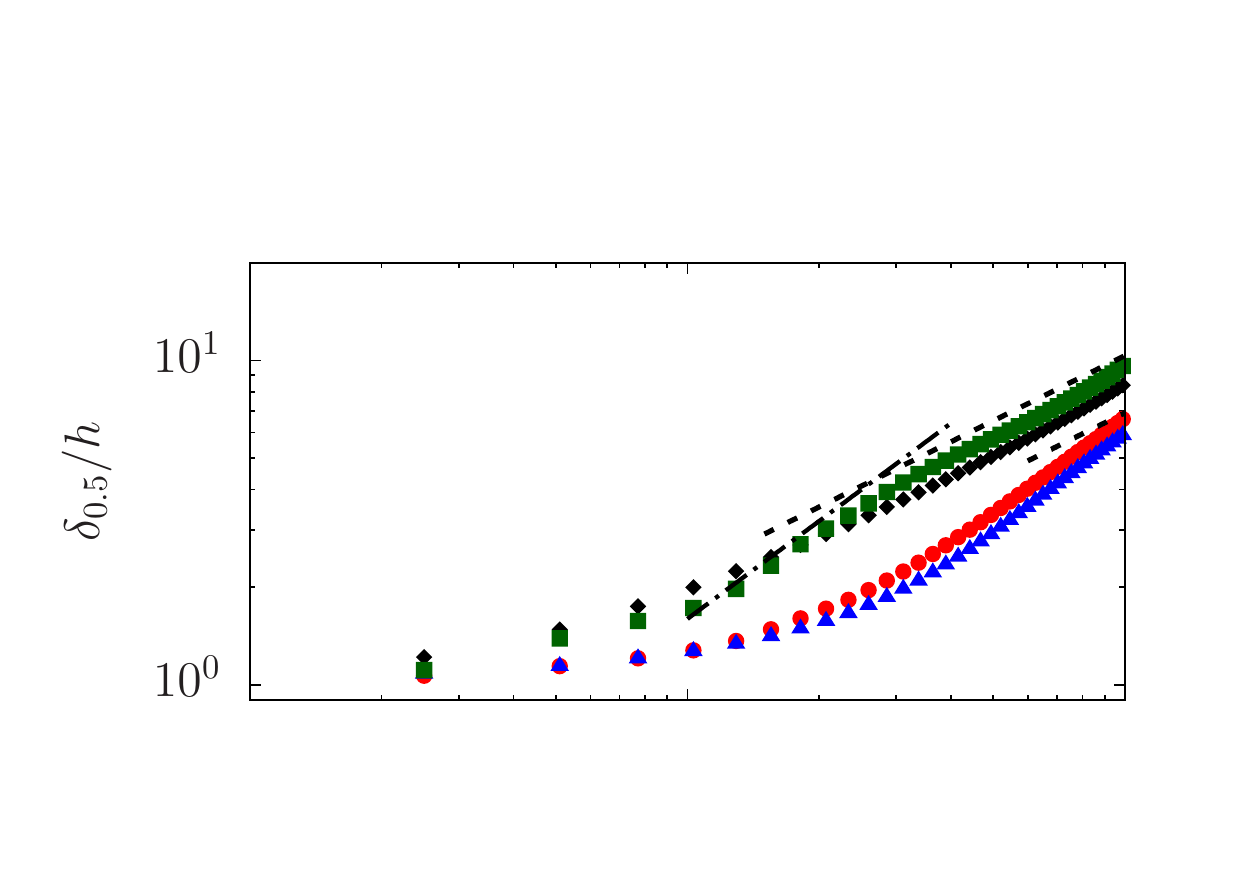}}
\put(0.,-0.05){\includegraphics[width=\columnwidth, keepaspectratio]{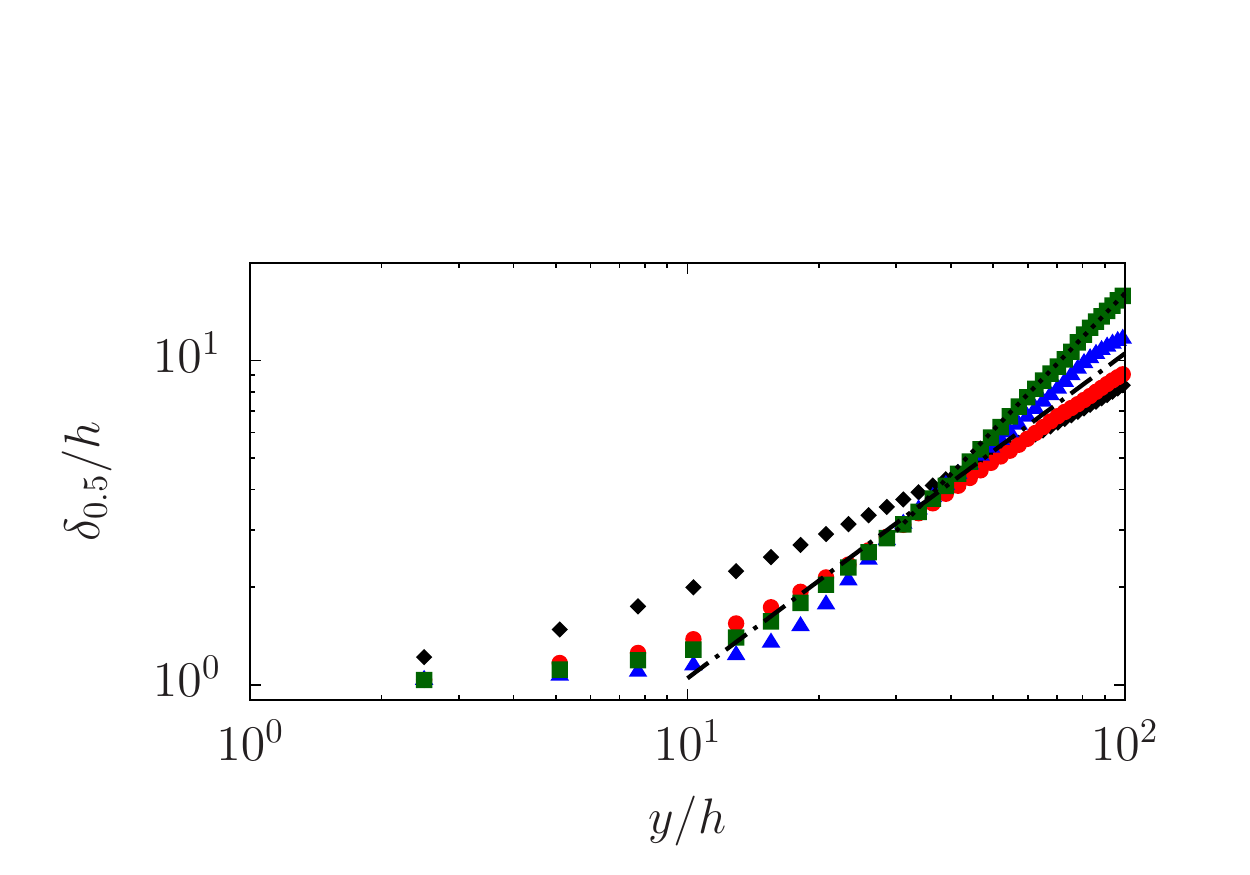}}
\put(0.04,1.16){(a)}
\put(0.04,0.78){(b)}
\put(0.04,0.4){(c)}
\end{picture}
\caption{Jet thickness for the three non-Newtonian fluid models, with the data for the Newtonian case also reported for reference. The expected laminar, $\delta_{0.5}\propto y^{2/3}$, and inertial turbulent, $\delta_{0.5}\propto y$, power-law scalings are shown with dashed and dash-dotted lines respectively. The fit for the Oldroyd-B case at high $\gamma$, $\delta_{0.5}\propto y^{1.4}$, is also reported with a dotted line. The three panels correspond to the different $\gamma$: (a) $\gamma=1$, (b) $\gamma=10$ and (c) $\gamma=100$.} 
\label{fig: delta}
\end{figure}

We now move to consider the jet thickness; various definitions can be chosen, such as the shear-layer thickness, corresponding to the thickness of an inviscid, shear-less jet flow with velocity equal to the centerline velocity and carrying the same volumetric flow-rate, or the distance from the centerline at which the stream-wise velocity equals a certain percentage of the centerline velocity at the same stream-wise location \citep{POPE2000}. Here, we chose the latter and define the jet thickness $\delta_{0.5}$ as the distance from the centerline at which the stream-wise velocity equals half of the centerline velocity. Scaling laws for the jet thickness of a Newtonian planar jet have been obtained for both laminar \citep{Mei_2001,parvar2020local} and inertial turbulent \citep{POPE2000} conditions: $\delta_{0.5}\propto y^{2/3}$ (laminar) and $\delta_{0.5}\propto y$ (inertial turbulent).

Overall, all the results for the jet thickness reported in figure~\ref{fig: delta} provide a picture similar to what already observed for the jet centerline velocity. At the lowest value of $\gamma$ considered, $\gamma=1$, shown in figure~\ref{fig: delta}(\textit{a}), non-Newtonian effects are negligible and the jet thickness for all the cases falls on top of each other, well in agreement with the laminar scaling. At the intermediate value, $\gamma=10$, the Oldroyd-B fluid shows an initial turbulent-like regime followed by a relaminarization further downstream, where the laminar scaling is recovered. Conversely, the Carreau and Giesekus fluid models show laminar flowing regime throughout the entire domain, with a slightly different slope than the expected one for a laminar Newtonian fluid. Also, we report a lengthening of the region where memory of the inlet conditions still persists, up to about $15h$ for these cases; then, further away from the inlet (beyond about $60h$), the jet growth rate approaches the predicted laminar growth rate, again showing some minore deviations from the expected scaling law. Finally, at the highest $\gamma$ considered, we observe that all non-Newtonian cases follow the jet growth rate predicted by the inertial turbulent scaling between roughly $10h$ and $60h$; beyond $60h$ we observe a substantial increase in the growth rate for the purely elastic Oldroyd-B fluid, with almost no appreciable difference instead for the fluids with shear-thinning, i.e. the Carreau and Giesekus fluids. We provide a fit for the increased growth rate of the Oldroyd-B case at $\gamma=100$ and we report a good agreement with a power-law $\delta_{0.5}\propto y^{1.4}$, resulting in a 40\% increase in the power-law exponent compared to the expected inertial turbulent scaling. It is interesting to note that a similar increase, although in absolute value, was reported for the exponent of the power-law scaling for the centerline velocity of the same case.

The results obtained from the previous figures provide a consistent picture in terms of flow regimes. In particular, we observe that both elasticity and shear-thinning can promote a transition to turbulence, with elasticity triggering the transition at lower values of $\gamma$ than shear-thinning. Also, when both mechanisms are present, we observe a competition between fluid elasticity and shear-thinning in triggering the transition to the turbulent regime: at the intermediate value of the time scale ratio $\gamma$, the purely viscoelastic Oldroyd-B fluid shows an early transition to turbulence, which is not observed for the viscoelastic and shear-thinning Giesekus fluid. Another consistent result among the different cases studied is that the scaling regions for non-Newtonian jets are reached at larger distances than what found in a Newtonian fluid, thus resulting in a longer region where the centerline velocity remains almost constant and the jet-thickness grows slowly. Finally, concerning the validity of the expected scalings, we can conclude that the laminar scaling applies up to moderate values of $\gamma$ only for the non-Newtonian fluids, while the inertial turbulent scaling is achieved only in a limited region up to about $60h$; beyond this we observe a faster growth rate of the jet thickness for the purely elastic Oldroyd-B fluid, and a slower decay-rate of the centerline velocity for the Carreau and Giesekus fluids with shear-thinning.

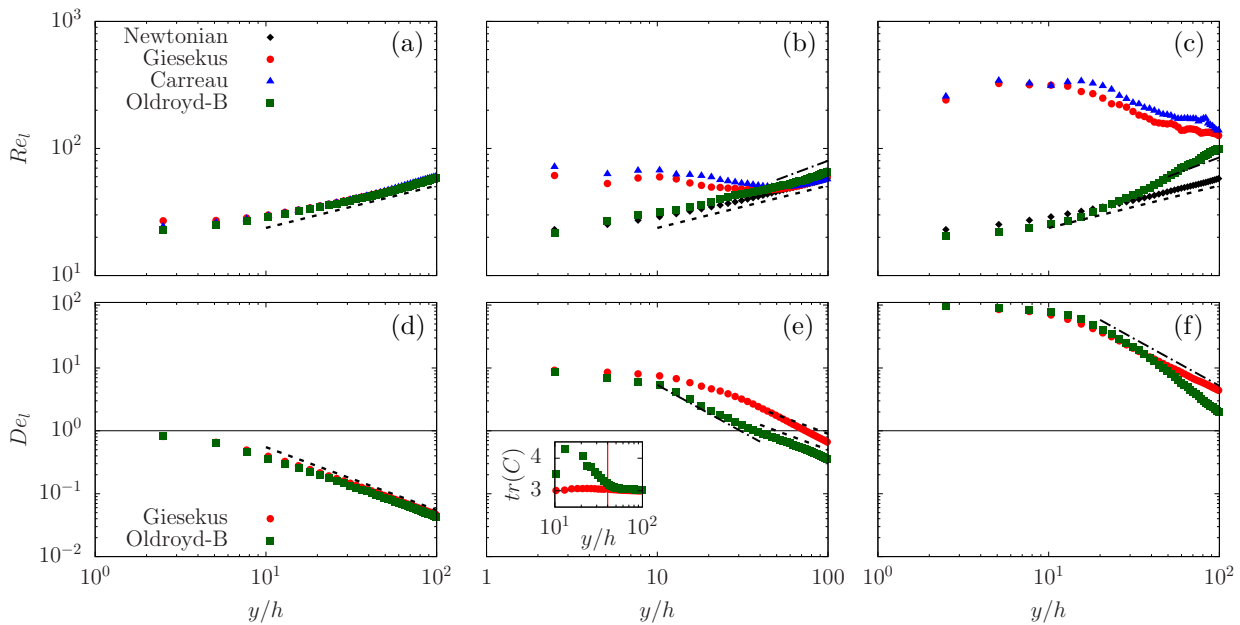
\begin{figure*}
\setlength{\unitlength}{\columnwidth}
\begin{picture}(1,1.0)
\put(-0.05,0.44){\includegraphics[width=0.96\columnwidth, keepaspectratio]{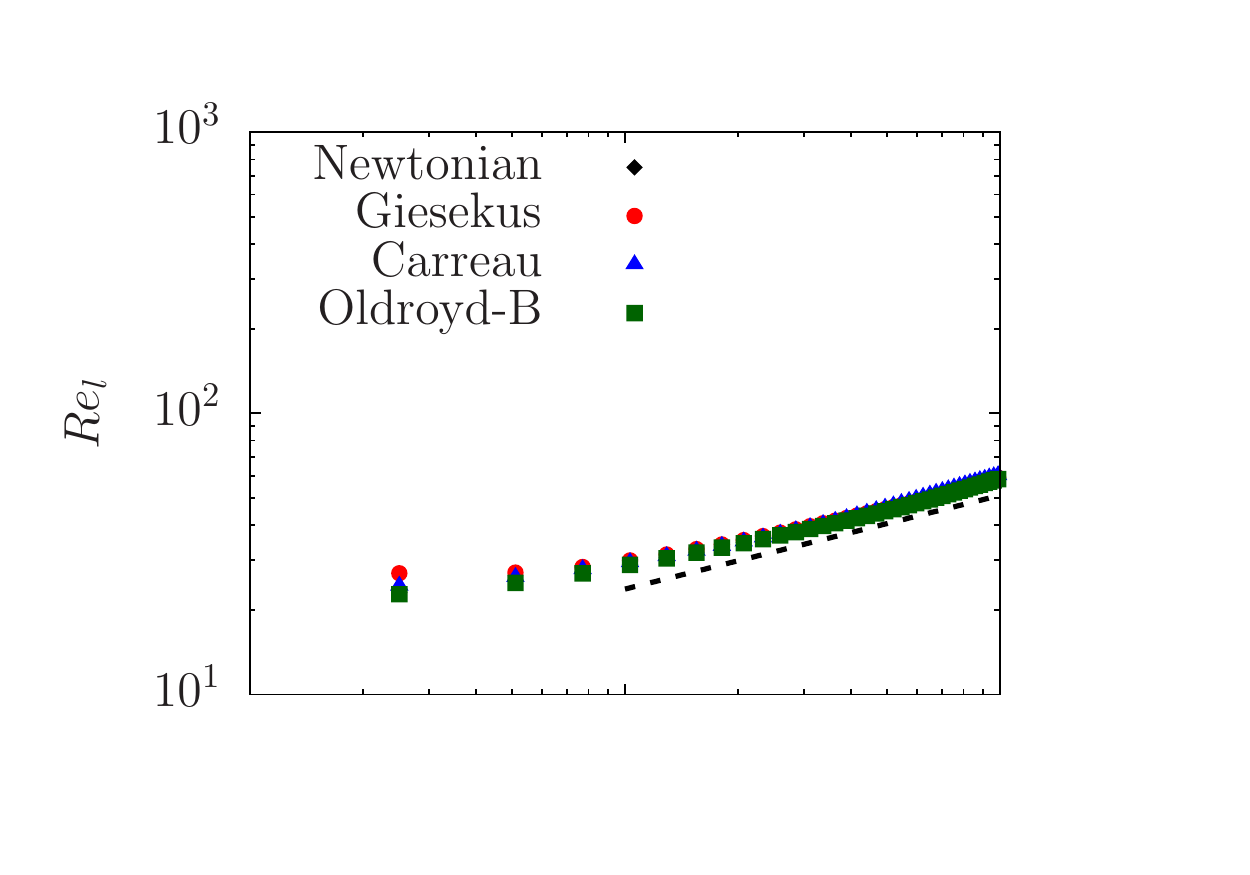}}
\put(0.61,0.44){\includegraphics[width=0.96\columnwidth, keepaspectratio]{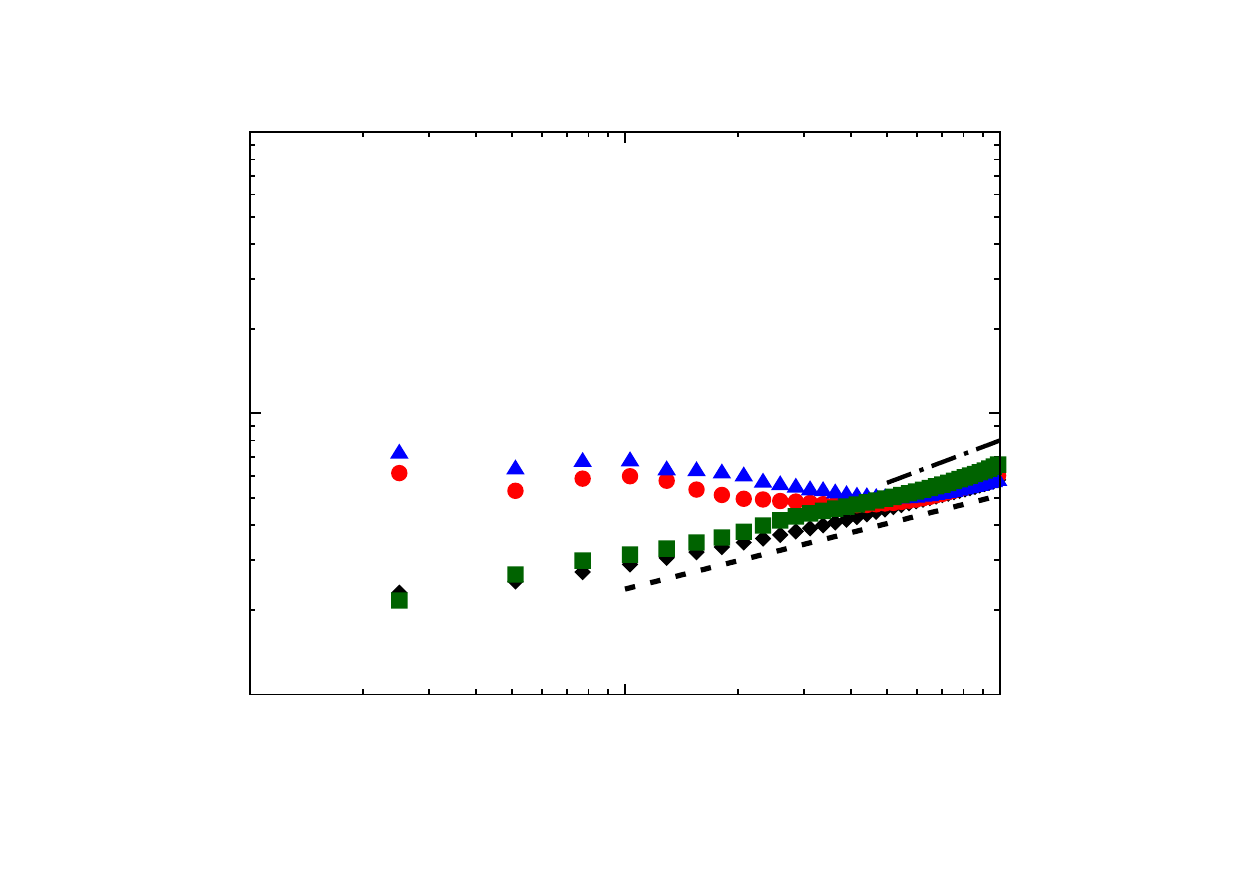}}
\put(1.27,0.44){\includegraphics[width=0.96\columnwidth, keepaspectratio]{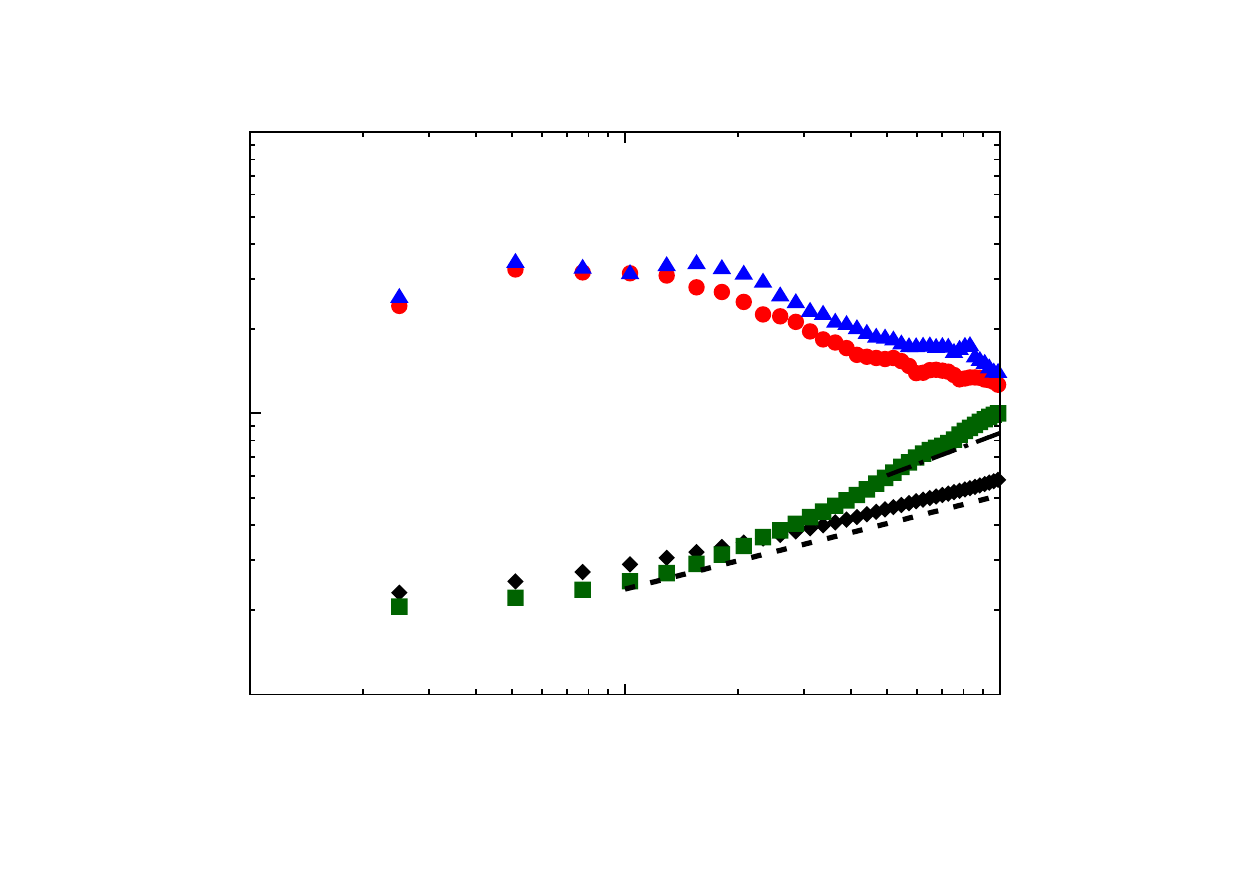}}
\put(-0.05,-0.06){\includegraphics[width=0.96\columnwidth, keepaspectratio]{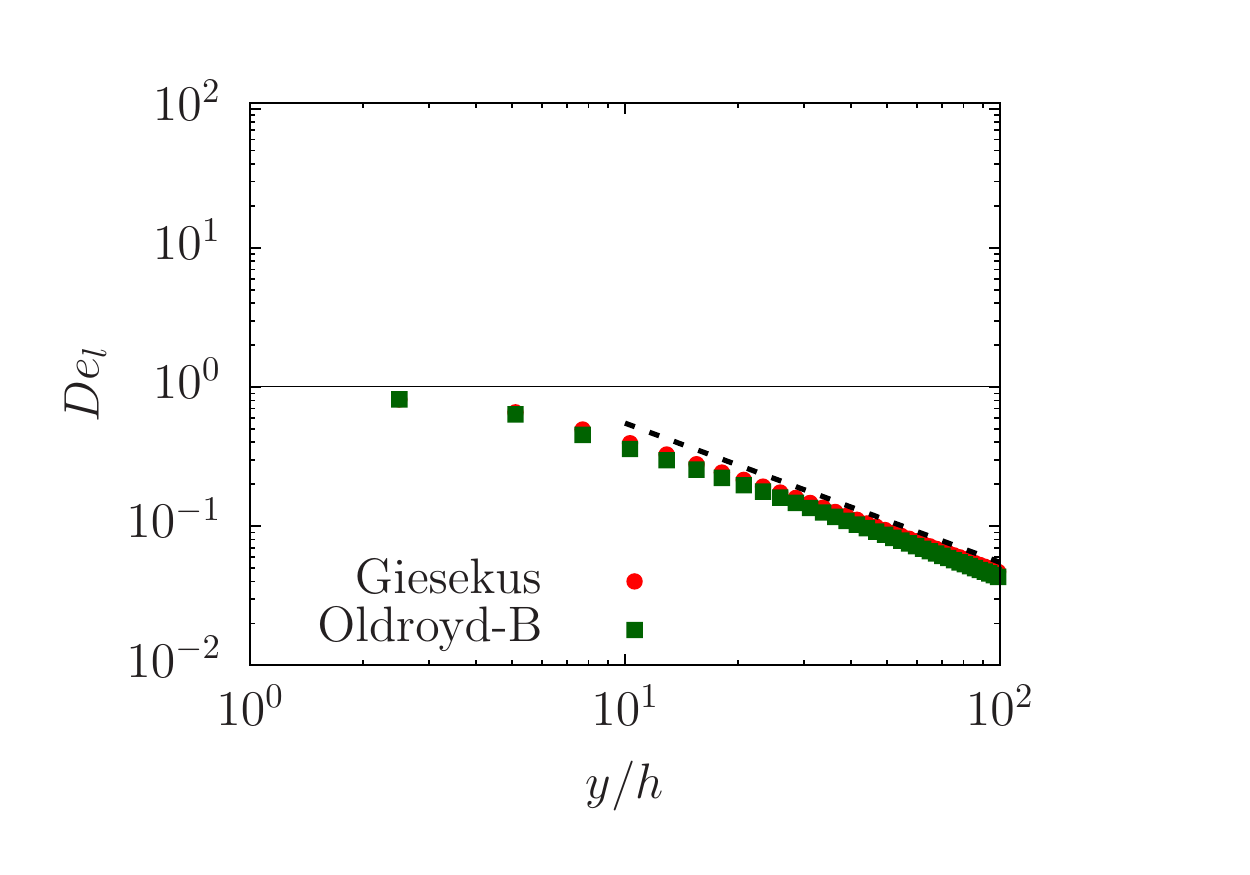}}
\put(0.61,-0.06){\includegraphics[width=0.96\columnwidth, keepaspectratio]{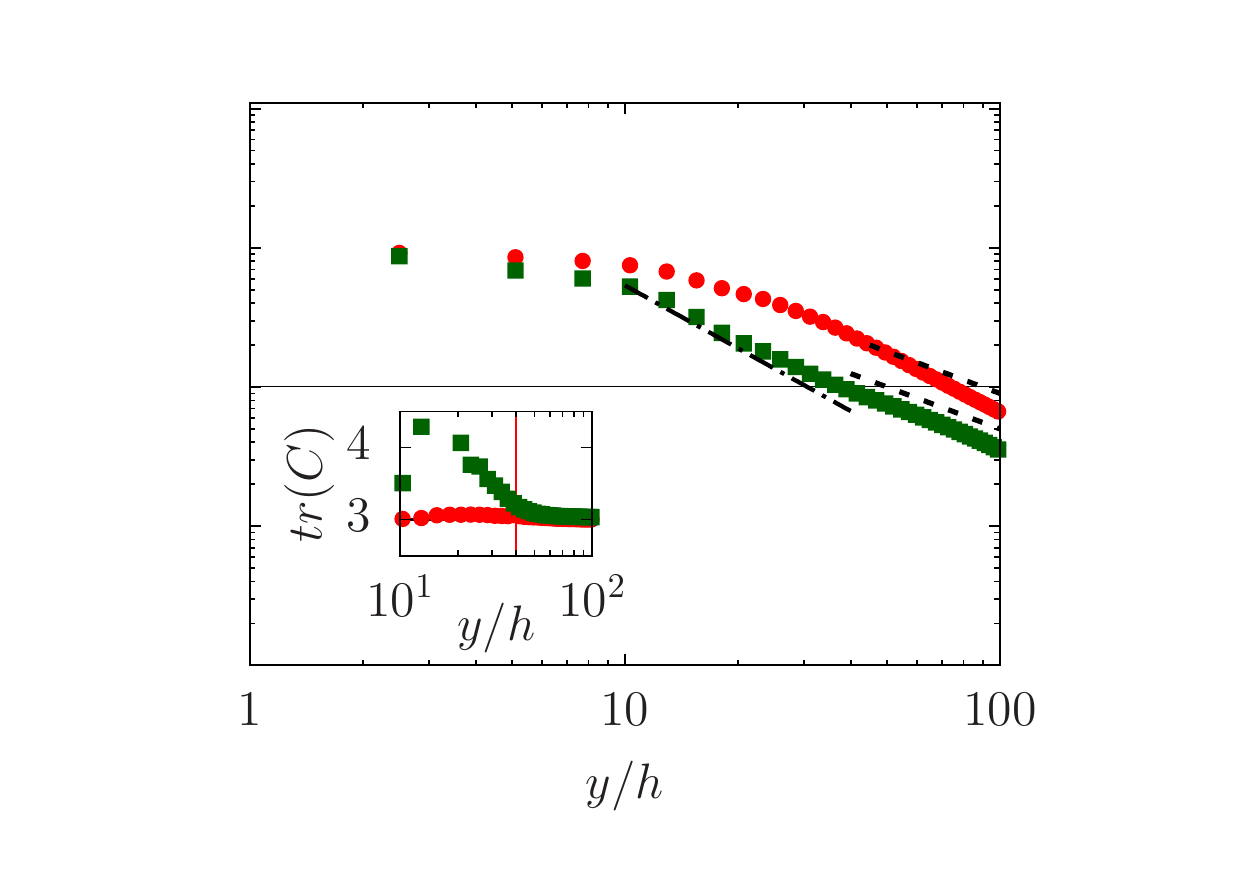}}
\put(1.27,-0.06){\includegraphics[width=0.96\columnwidth, keepaspectratio]{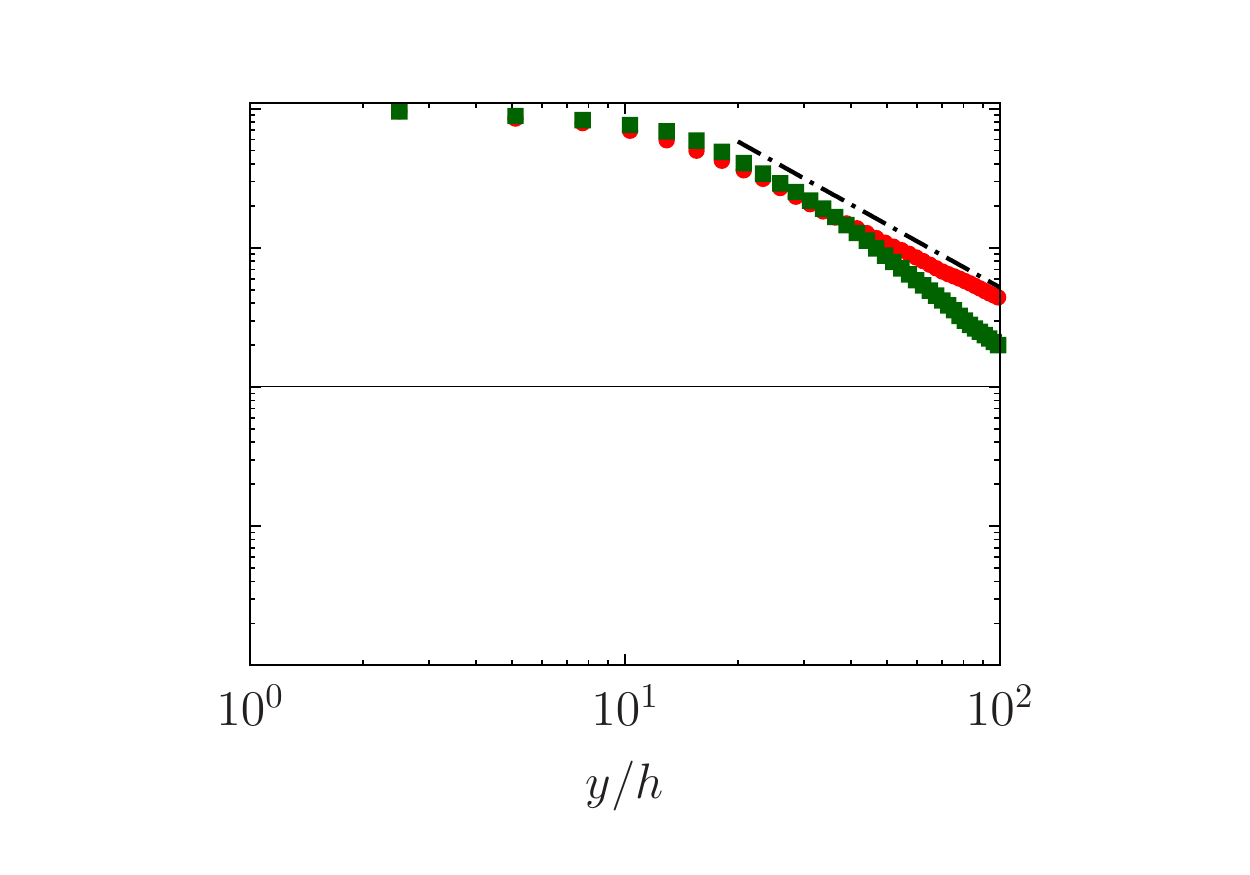}}
\put(0.64,0.96){(a)}
\put(1.30,0.96){(b)}
\put(1.96,0.96){(c)}
\put(0.64,0.48){(d)}
\put(1.30,0.48){(e)}
\put(1.96,0.48){(f)}
\end{picture}
\caption{Local (a-c) Reynolds $Re_l$ and (d-e) Deborah $De_l$ numbers computed using the local jet thickness, centerline velocity and viscosity. The three panels in each row correspond to the different $\gamma$: (a, d) $\gamma=1$, (b, e) $\gamma=10$ and (c, f) $\gamma=100$. Dashed and dash-dotted lines identify the laminar ($Re_l \propto y^{1/3} $, $De_l \propto y^{-1} $) and turbulent ($Re_l \propto y^{1/2} $, $De_l \propto y^{-3/2}$) scalings. The inset in panel (e) shows the trace of the conformation tensor, $tr(C)$, at the centerline.} 
\label{fig: Recent}
\end{figure*}

The information from the centerline velocity and jet thickness can be combined to form two non-dimensional numbers, the local Reynolds and Deborah numbers, and thus study their behavior as a function of the stream-wise distance from the inlet. 
We define a local Reynolds number, $Re_l$, as the product of the centerline velocity $v_c$ and the jet thickness $\delta_{0.5}$, divided by the local viscosity, with the latter taken as the lowest value of the viscosity (corresponding to the highest shear rate) present at each stream-wise location. The local Deborah number, $De_l$,  instead is defined as the ratio between the polymer relaxation time $\lambda$ over a characteristic local flow time-scale; the latter is chosen as the ratio of the jet thickness $\delta_{0.5}$ over the centerline velocity $v_c$. Note that the Deborah number is computed only for the Giesekus and Oldroyd-B fluids, being the only two fluid models characterized by fluid elasticity, and thus having a polymer relaxation time. 
The local Reynolds number is shown in the top row of figure~\ref{fig: Recent}, while the local Deborah number in the bottom row of the same figure. By combining the expected scalings for the centerline velocity and jet thickness, we can derive scalings for the local Reynolds and Deborah numbers. In particular, in the laminar regime the Reynolds number scales as $Re_l \propto y^{1/3}$ and the Deborah number as $De_{l}\propto y^{-1}$; in the inertial turbulent regime the Reynolds number scales as $Re_l \propto y^{1/2}$ and the Deborah number as $De_{l}\propto y^{-3/2}$. Here, when computing the theoretical scalings for the Reynolds number, we do not account for any shear-thinning effects, hence the effect of the local viscosity reduction is not considered.

At the lowest $\gamma$, shown in figure~\ref{fig: Recent}(a, d), not much difference can be noticed from the Newtonian case, as expected since both the centerline velocity and jet thickness for all non-Newtonian models are very close to those for the Newtonian case. 
Small differences can be observed for the fluid models characterized by shear-thinning, namely Carreau and Giesekus, where the Reynolds number is slightly larger than $20$ as a consequence of a reduction in the local viscosity due to the shear-rate dependent viscosity. At this low value of $\gamma$, the Reynolds number is always less than $100$ and the Deborah number always less than $1$, thus no appreciable effect of inertia or elasticity is present. When $\gamma$ is increased to $\gamma=10$, shown in figure~\ref{fig: Recent}(b, e), the shear-thinning effect becomes evident: the local Reynolds number is around four times larger than the Newtonian value for the Carreau and Giesekus models in the region close to the inlet, where the shear is maximum; downstream, the value of the local Reynolds number first slowly decays, as the maximum value of the shear rate reduces, and finally grows again following the Newtonian data. At this intermediate $\gamma$, the Oldroyd-B fluid model does not depart significantly from the Newtonian (laminar) case in terms of the local Reynolds number, while the Deborah number is larger than $1$ for $y\lesssim40h$ and smaller afterwards. This correlates well with the two flowing regimes observed previously for this model, with the injected flow first transitioning to a turbulent behaviour and later relaminarizing. We observe that this transition from the turbulent and the laminar regimes occurs at $De_l \simeq 1$, with $De_l\gtrsim1$ characterizing the turbulent regime and $De_l\lesssim1$ the laminar regime.
On the other hand, the same can not be said when the fluid has also shear-thinning property. Indeed, although the Giesekus fluid also has a local value of the Deborah number larger than unity, actually being even larger than the values observed for the Oldroyd-B fluid, the flow remains laminar. This result thus suggests that the shear-thinning effect is actually interfering with the arising of the elastic instability and having $De_l \gtrsim 1$ is not a sufficient condition for the development of an elastic instability. The inset in figure~\ref{fig: Recent}(e) reports the value of the trace of the conformation tensor at the centerline, $tr(C)$, for the Giesekus and Oldroyd-B fluid models at $\gamma=Wi=10$. The trace of the conformation tensor at the centerline acts well as an indicator of the flowing regime: for both models, in laminar flow its value is equal to $tr(C)=3$, while in the turbulent regime it has much larger values. We observe that, the trace of the conformation tensor for the Oldroyd-B fluid is much larger than 3 till about $y\simeq40h$ and beyond this point reduces to $tr(C)=3$, being a clear indication of the transition between the two regimes. At the highest $\gamma$ ($\gamma=100$), shown in figure~\ref{fig: Recent}(c, f), the above effects from the non-Newtonian features become even stronger: the local Reynolds number for both shear-thinning fluid models grows up to roughly $400$ near the inlet, then slowly decreases as we move towards the outlet, reaching a value of about $150$ at $y=100h$. Experiments on planar jets at increasing Reynolds numbers showed that a jet of Newtonian fluid is in the transitional regime at inlet Reynolds numbers as low as $Re=125$ \citep{suresh2008reynolds}. The exact value strongly depends on the particular experiment, as for example on the inlet aspect ratio, on the type of nozzle used, on the inlet flow regime or on the eventual flow confinement. The transitional regime is characterized by a value of the inlet Reynolds number spanning from $Re\gtrsim1\mathcal{O}(10^2)$ up to about $Re\gtrsim1\mathcal{O}(10^4)$ \citep{deo2008influence,suresh2008reynolds}. Our numerical simulations further confirm this result: at $\gamma=10$ the local Reynolds number at the inlet for the shear-thinning fluids (Carreau and Giesekus) is about $Re_l\approx70$ and increases to above 200 for $\gamma=100$. While at the lower $\gamma$ the flow is steady and laminar, a turbulent flowing regime can be observed when increasing $\gamma$. Compared to Newtonian jets, the presence of shear-thinning is an additional factor that can possibly change the dynamics of the jet, as the local viscosity changes along the jet cross-section and increases with the distance from the inlet. Thus, the higher local Reynolds number computed at the inlet starts to decrease beyond about $y=20h$ for the Carreau and Giesekus cases.
The Oldroyd-B fluid instead continues to exhibit values of local Reynolds number sufficiently small, initially similar to that of the Newtonian case and then departing from it. The value of the local Deborah number for all cases is larger than one. Hence, we can conclude that the observed instability and the following turbulent flow for the Oldroyd-B fluid are dominated by elastic effects, those for the Carreau model are purely inertial, while those for the Giesekus fluid are a combination of inertial and elastic effects.

\subsection{Jet-normal statistics}
\begin{figure}
\setlength{\unitlength}{\columnwidth}
\begin{picture}(1,1.20)
\put(0.,0.71){\includegraphics[width=\columnwidth, keepaspectratio]{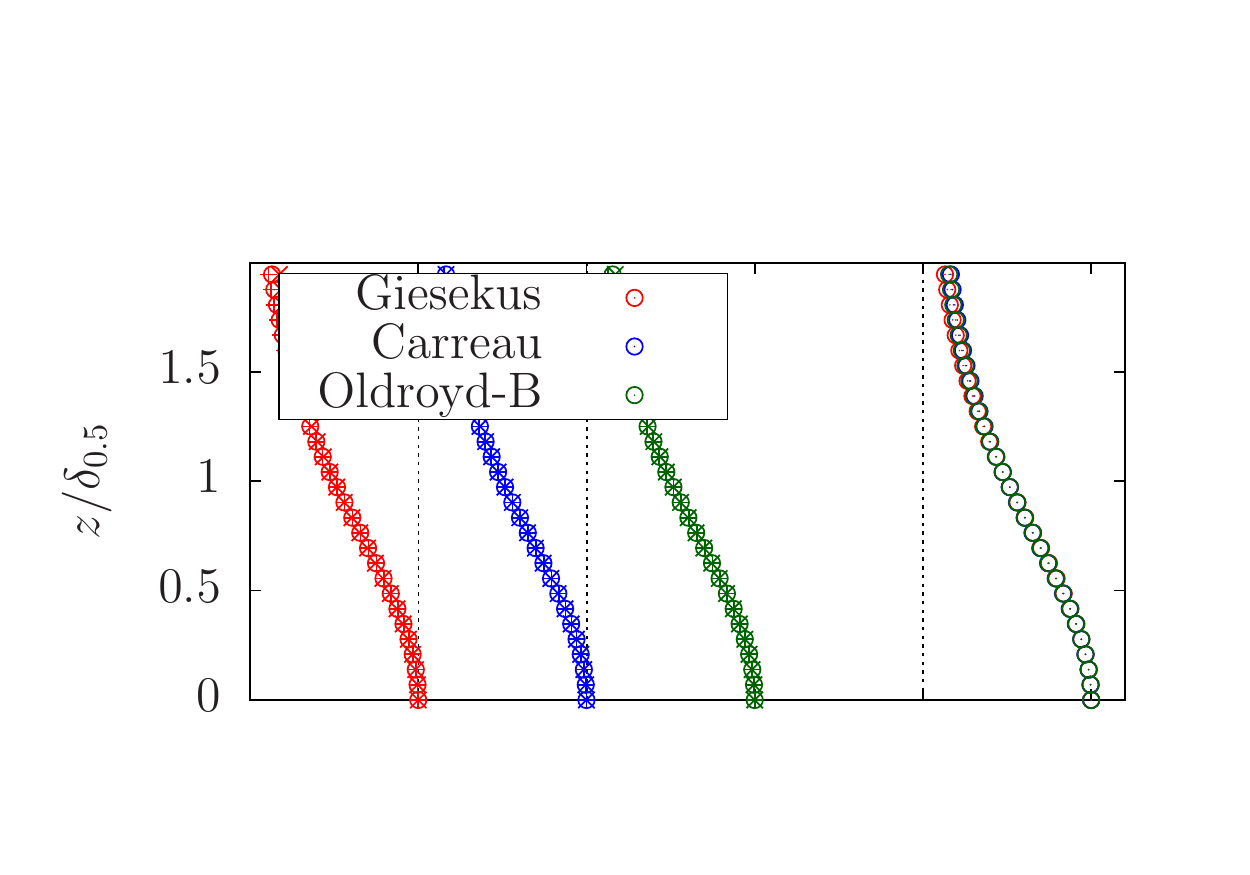}}
\put(0.,0.33){\includegraphics[width=\columnwidth, keepaspectratio]{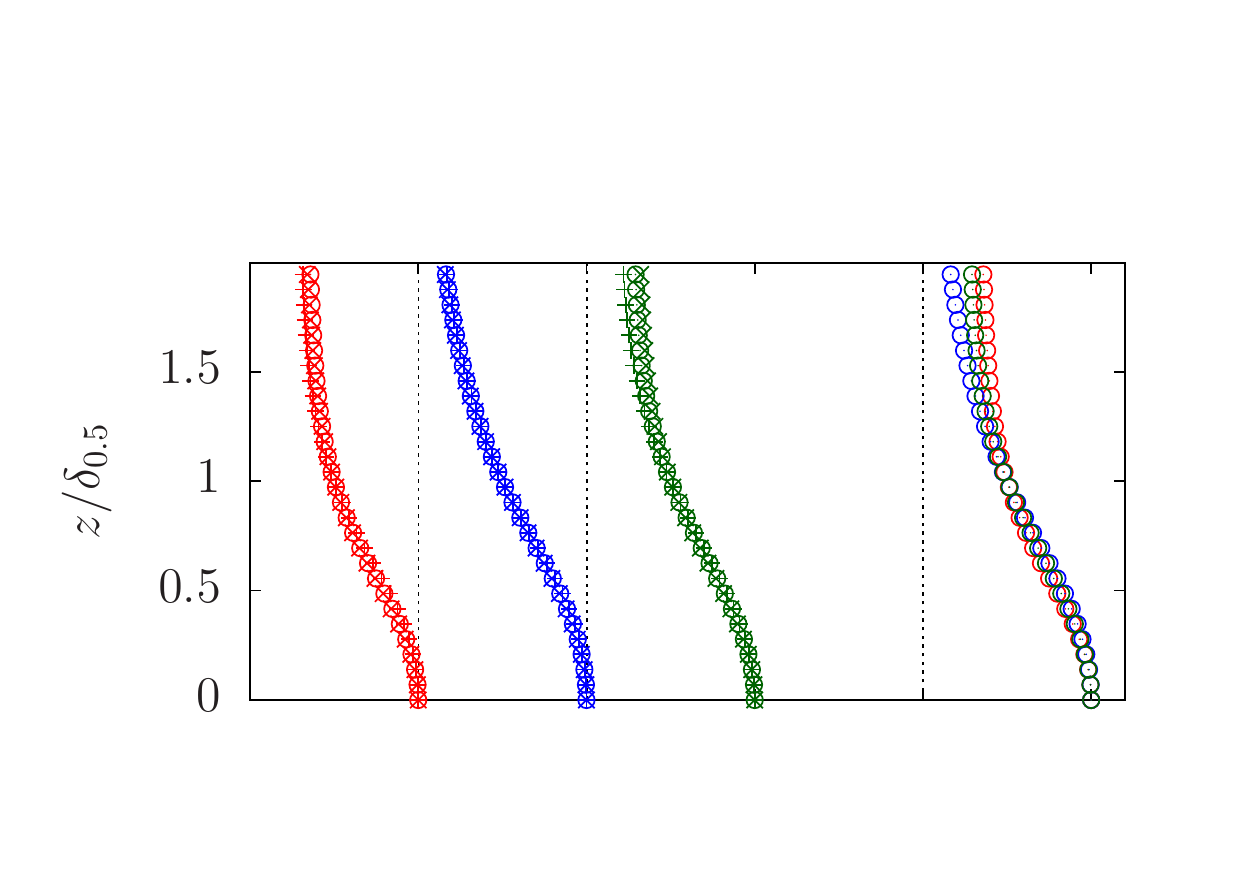}}
\put(0.,-0.05){\includegraphics[width=\columnwidth, keepaspectratio]{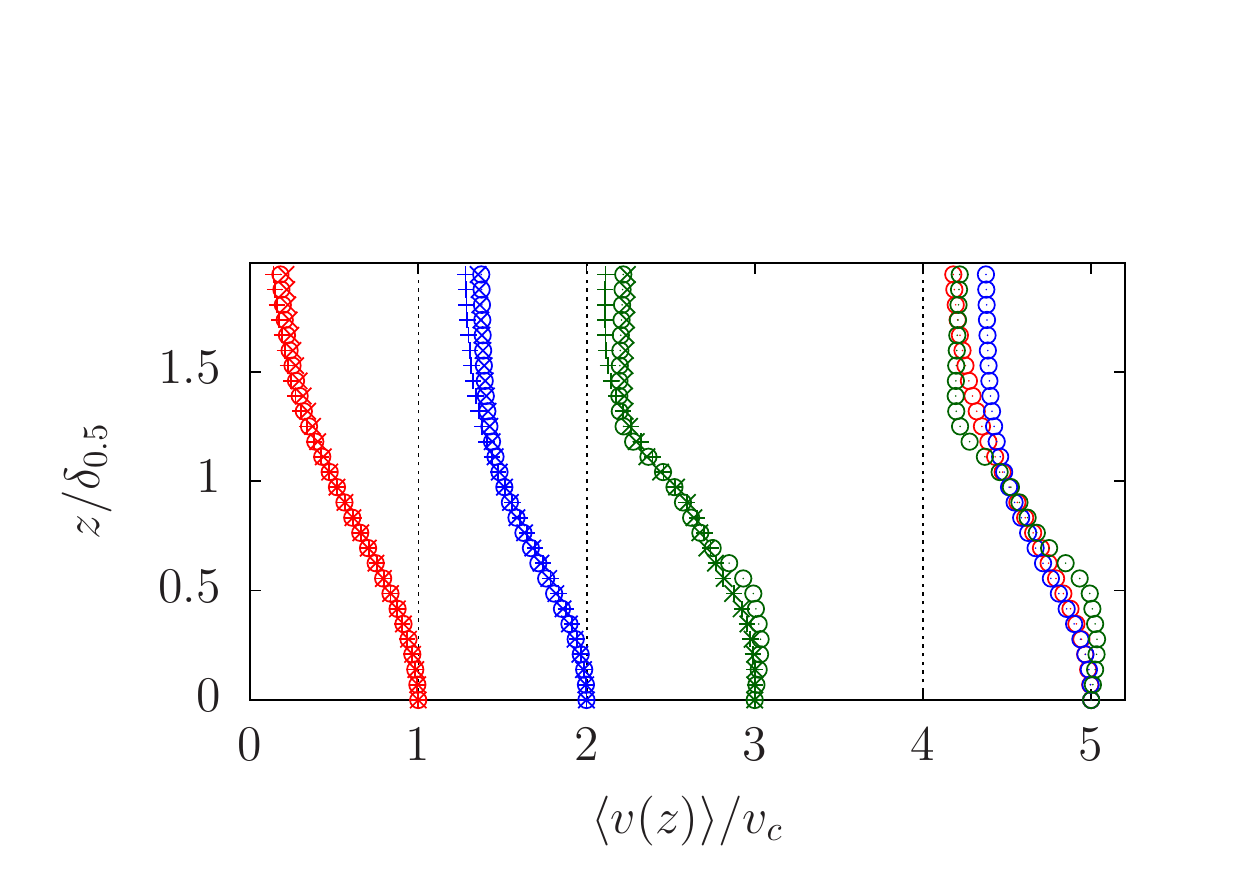}}
\put(0.04,1.16){(a)}
\put(0.04,0.78){(b)}
\put(0.04,0.4){(c)}
\end{picture}
\caption{Stream-wise mean velocity $\langle v \rangle$ at different stream-wise locations: $+$ at $y=60h$, $\circ$ at $y=80h$ and $\times$ at $y=100h$. Each color represents a different model and for clarity the profiles for the different models are shifted; the dashed vertical lines identify the corresponding zero-levels. The rightmost part of each panel compares the mean velocity for different models at the same stream-wise location $y=80h$. The three panels correspond to the different values of $\gamma$: (a) $\gamma=1$, (b) $\gamma=10$ and (c) $\gamma=100$.} 
\label{fig: v_ss}
\end{figure}
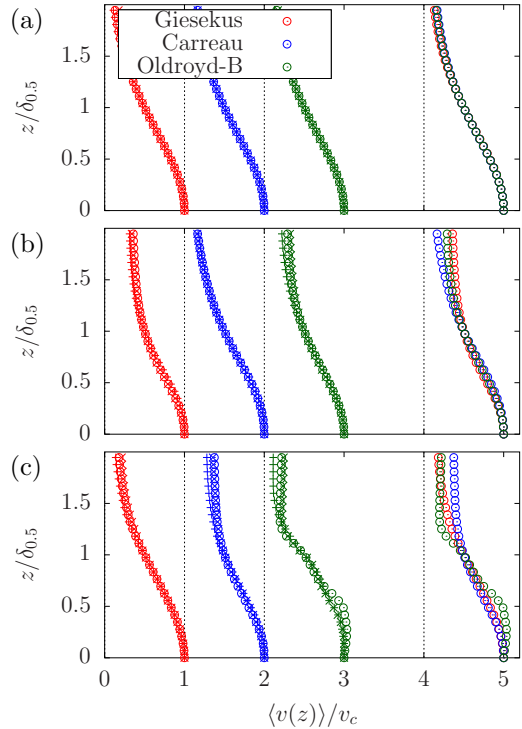

\begin{figure*}
\setlength{\unitlength}{\columnwidth}
\begin{picture}(1,1.1)
\put(-0.05,0.52){\includegraphics[width=0.96\columnwidth, keepaspectratio]{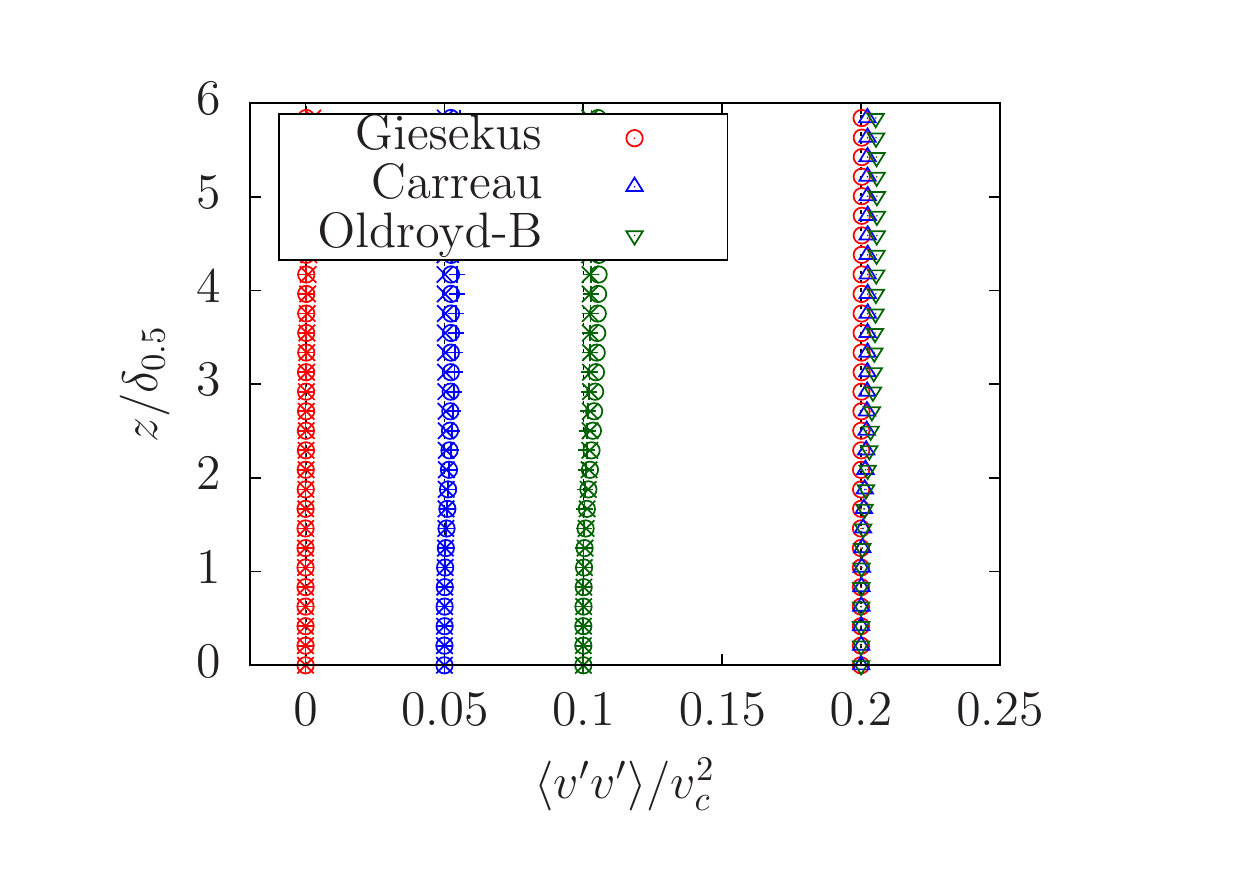}}
\put(0.61,0.52){\includegraphics[width=0.96\columnwidth, keepaspectratio]{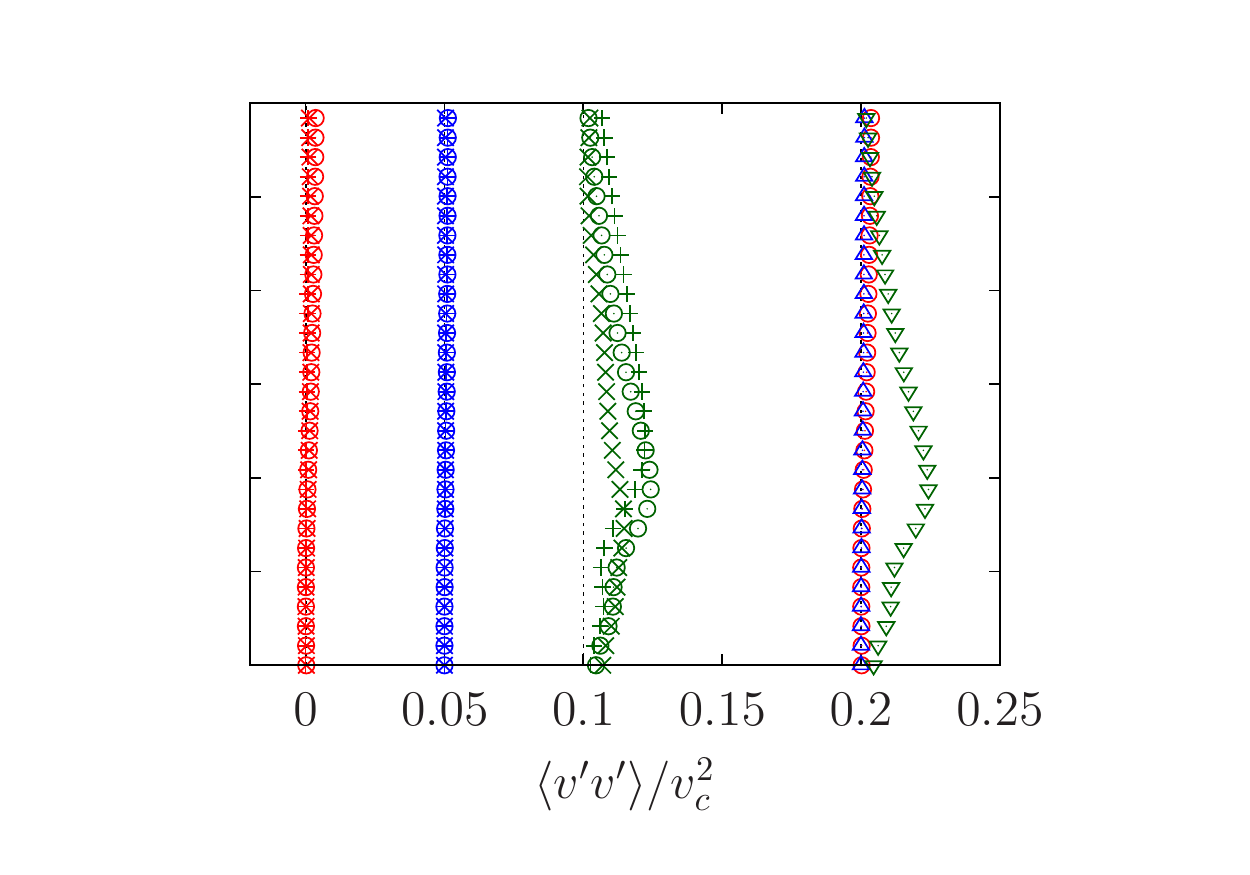}}
\put(1.27,0.52){\includegraphics[width=0.96\columnwidth, keepaspectratio]{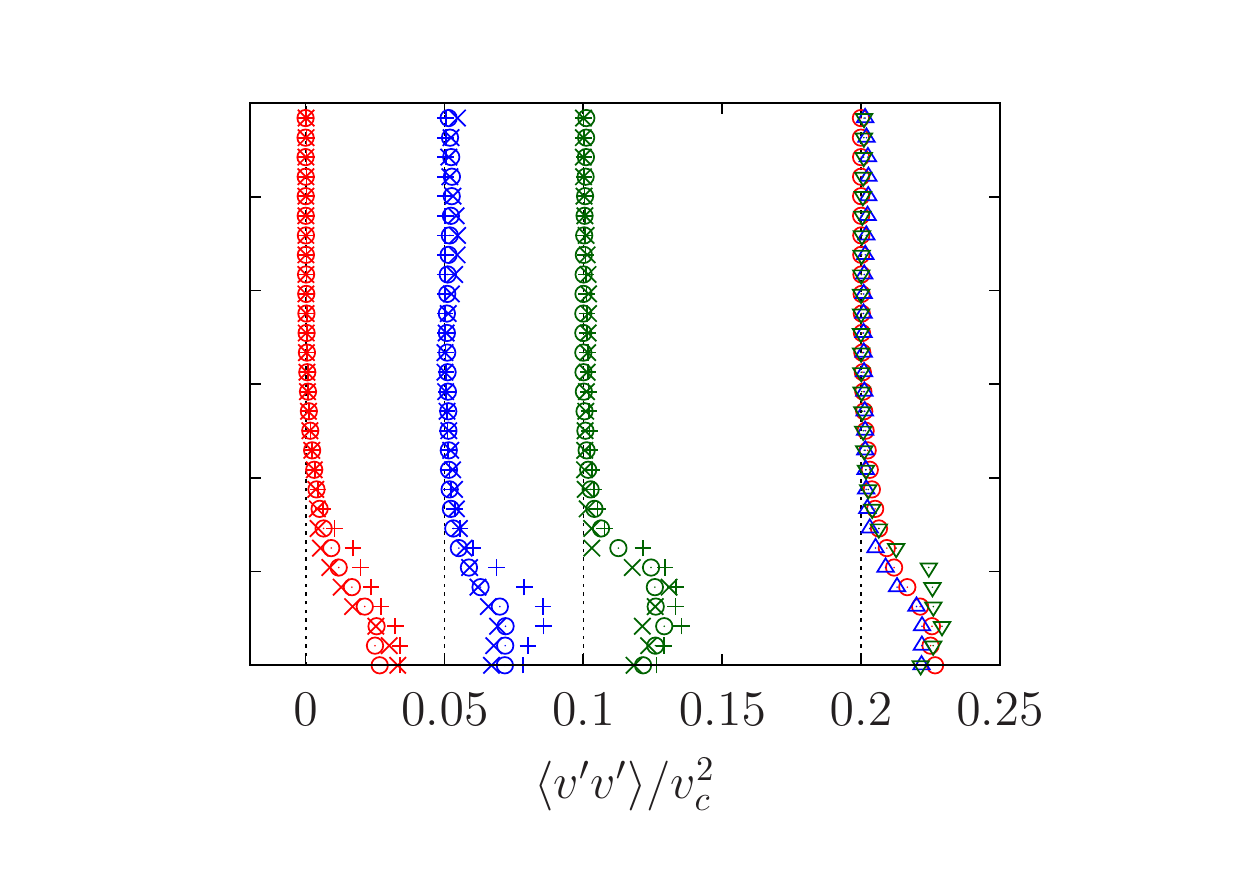}}
\put(-0.05,-0.06){\includegraphics[width=0.96\columnwidth, keepaspectratio]{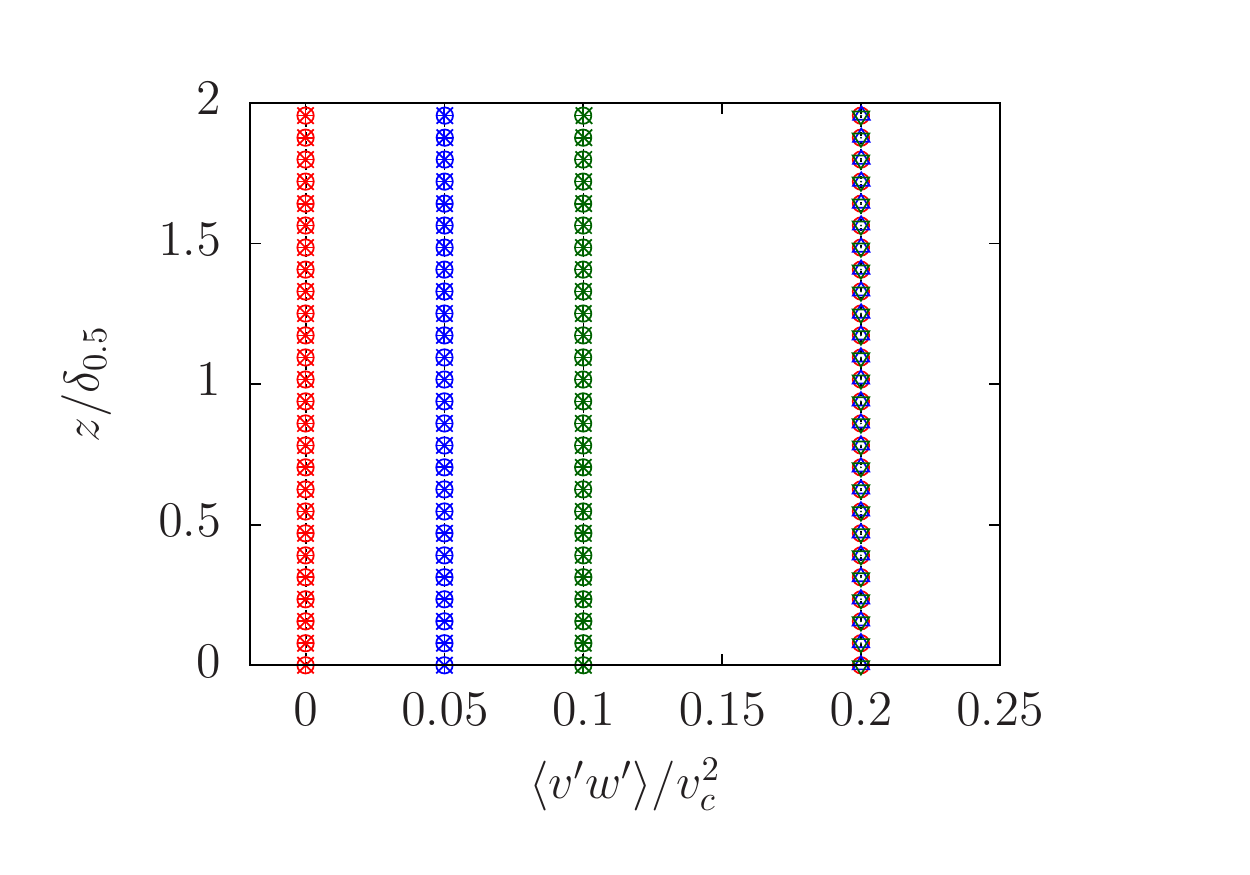}}
\put(0.61,-0.06){\includegraphics[width=0.96\columnwidth, keepaspectratio]{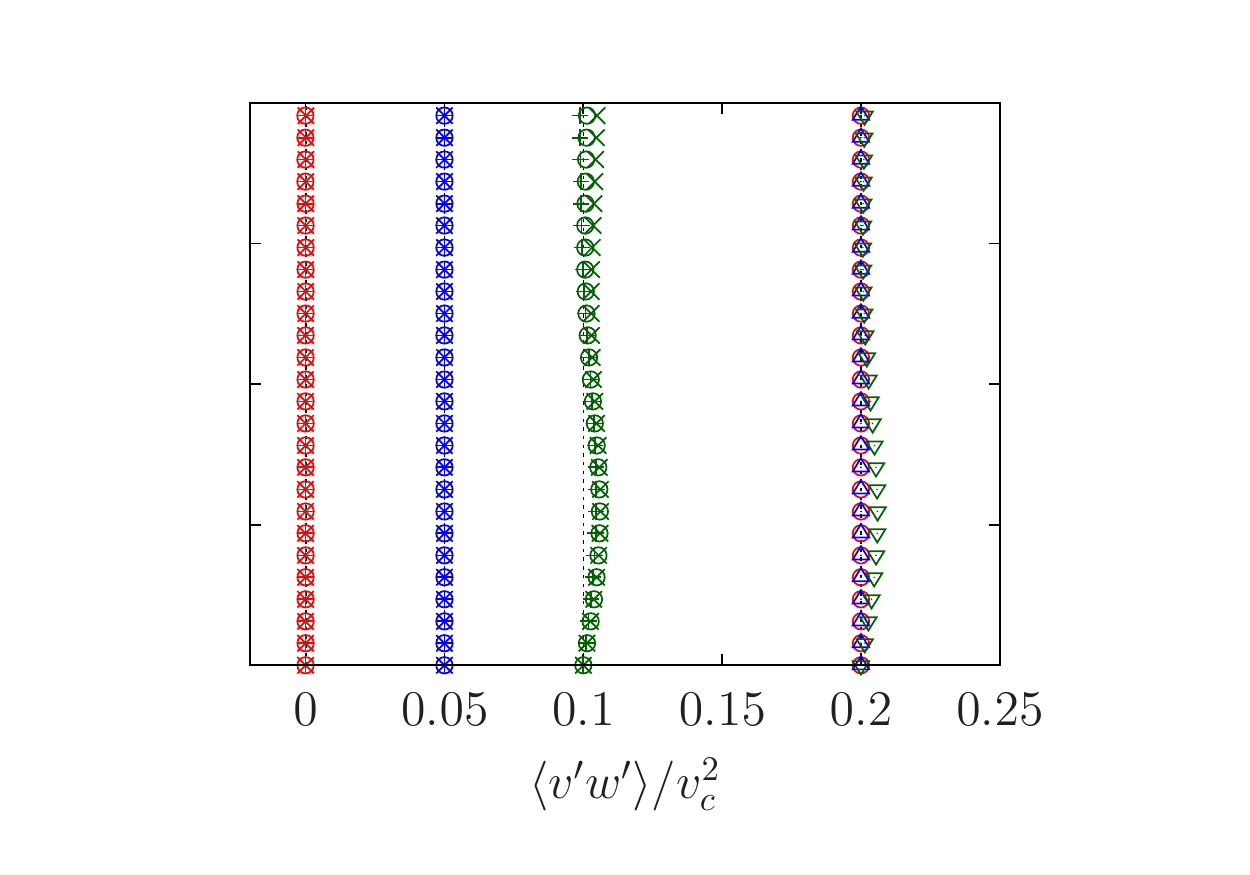}}
\put(1.27,-0.06){\includegraphics[width=0.96\columnwidth, keepaspectratio]{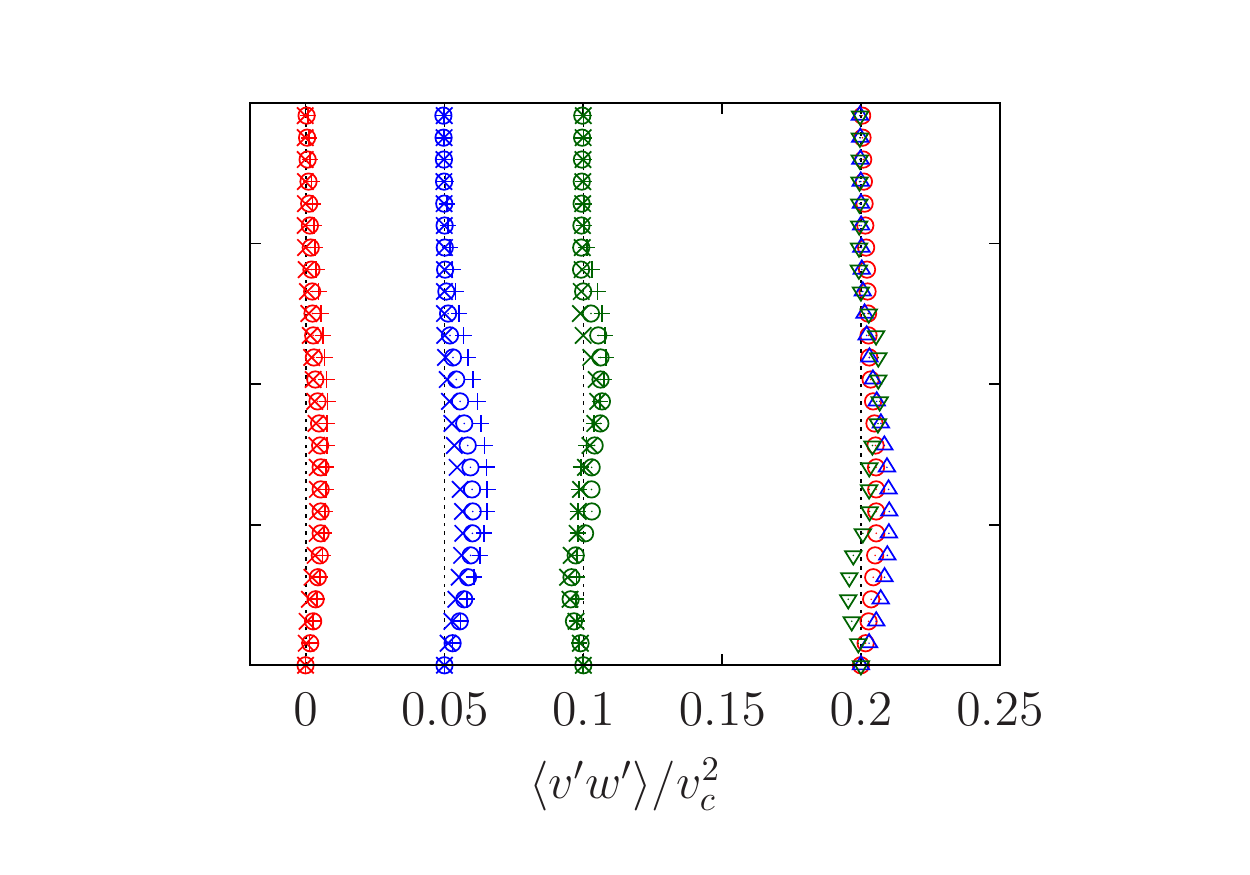}}
\put(0.64,1.04){(a)}
\put(1.30,1.04){(b)}
\put(1.96,1.04){(c)}
\put(0.64,0.48){(d)}
\put(1.30,0.48){(e)}
\put(1.96,0.48){(f)}
\end{picture}
\caption{Reynolds stresses $\langle v'v'\rangle$ and  $\langle v'w'\rangle$ at different stream-wise locations: $+$ at $y=60h$, $\circ$ at $y=80h$ and $\times$ at $y=100h$. Each color represents a different model and for clarity the profiles for the different models are shifted; the dashed vertical lines identify the corresponding zero-levels. The rightmost part of each panel compares the mean Reynolds stresses for different models at the same stream-wise location $y=80h$. The three columns correspond to the different $\gamma$: (a, d) $\gamma=1$, (b, e) $\gamma=10$ and (c, f) $\gamma=100$.} 
\label{fig: Restress_ss}
\end{figure*}
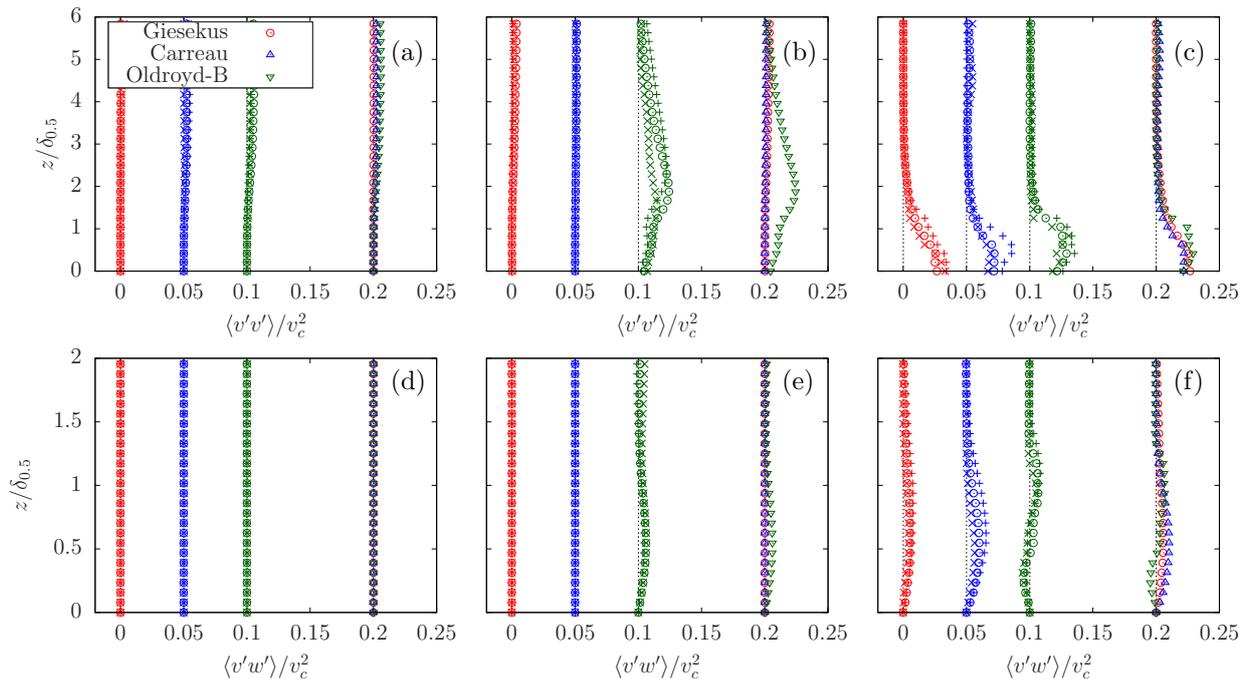

\begin{figure*}
\setlength{\unitlength}{\columnwidth}
\begin{picture}(1,1.5)
\put(-0.05,1.1){\includegraphics[width=0.96\columnwidth, keepaspectratio]{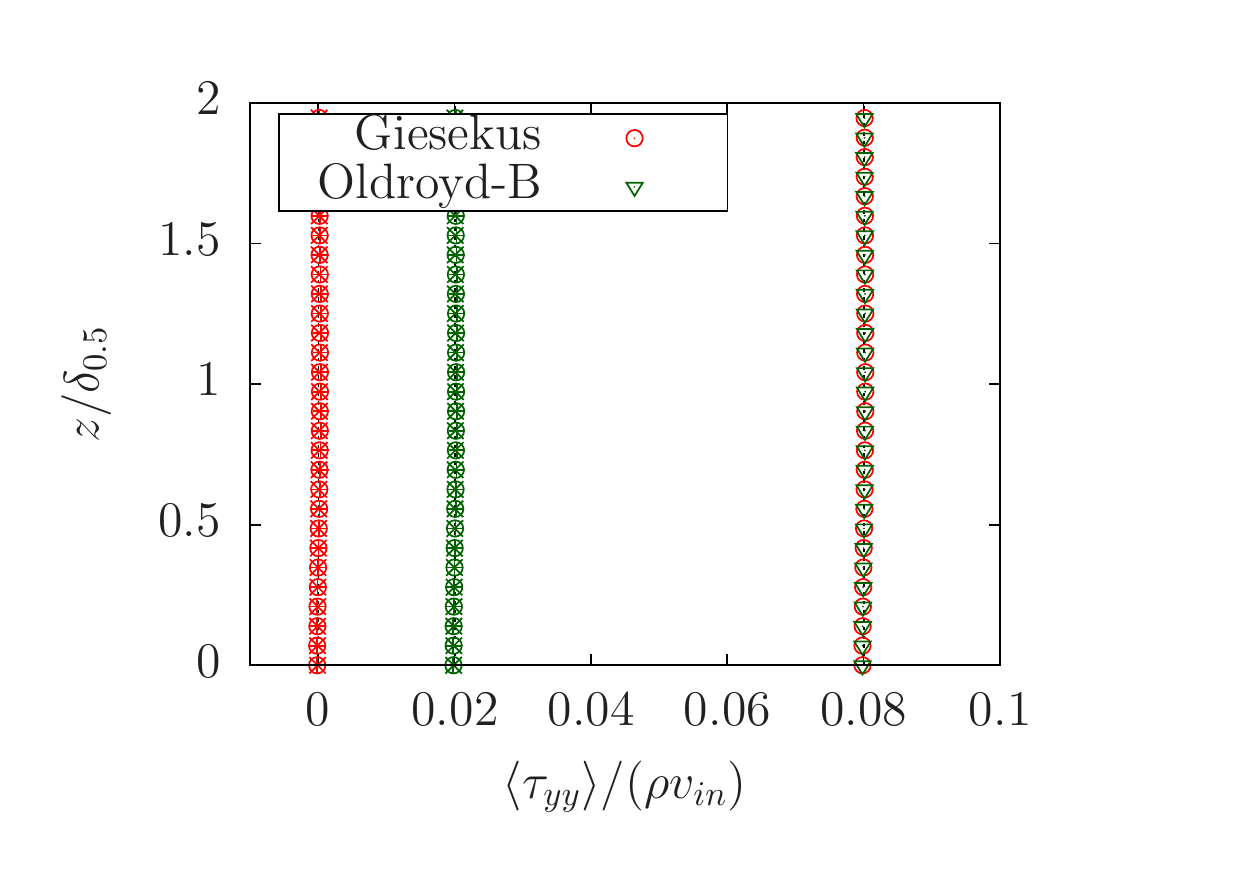}}
\put(0.61,1.1){\includegraphics[width=0.96\columnwidth, keepaspectratio]{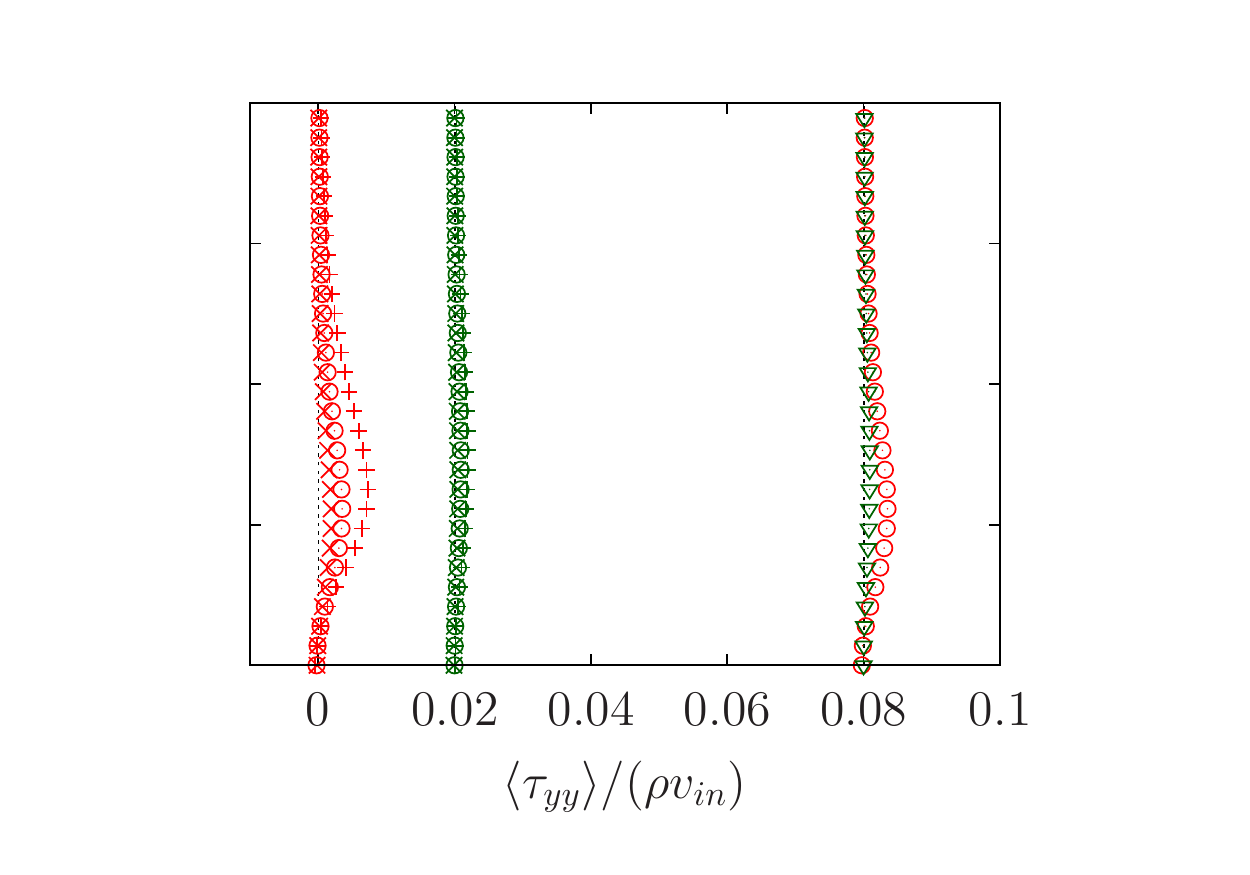}}
\put(1.27,1.1){\includegraphics[width=0.96\columnwidth, keepaspectratio]{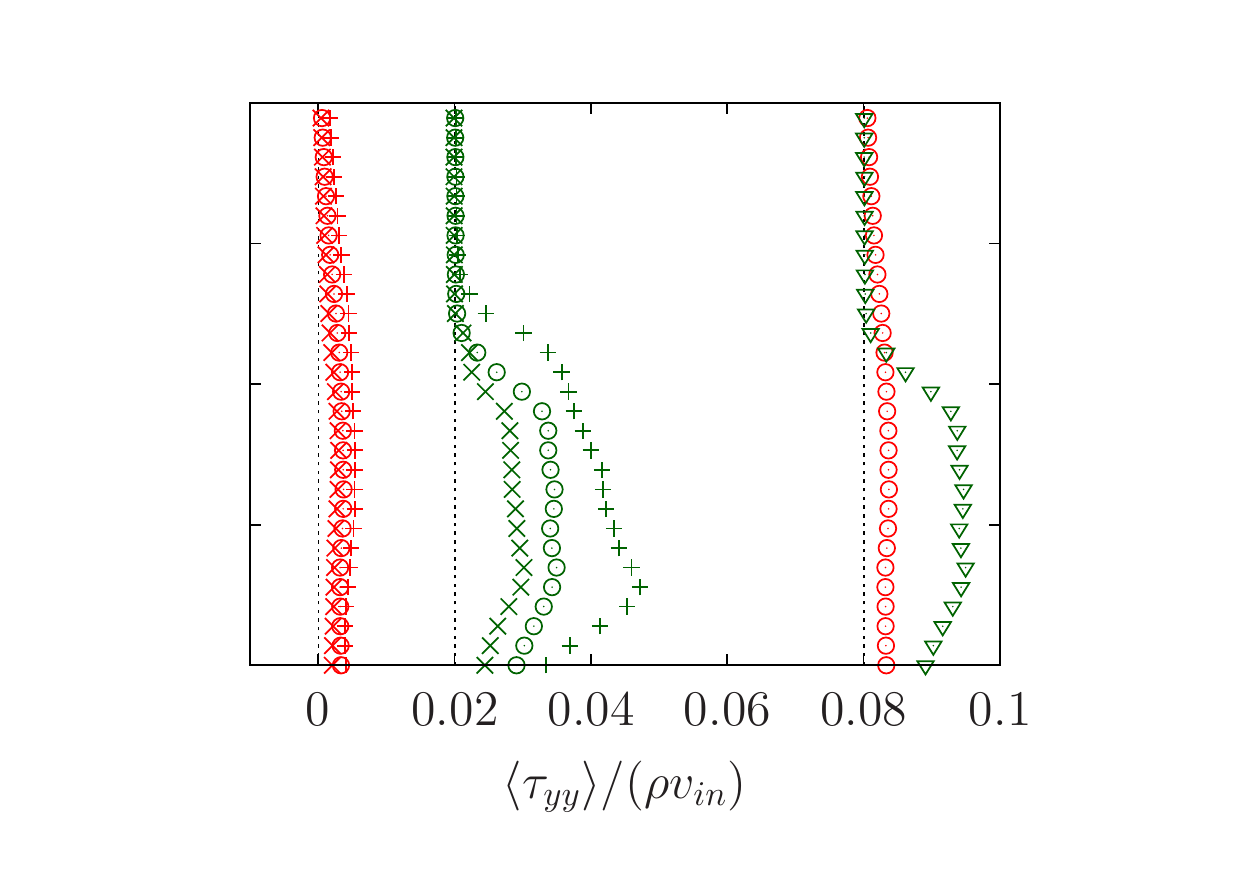}}
\put(-0.05,0.52){\includegraphics[width=0.96\columnwidth, keepaspectratio]{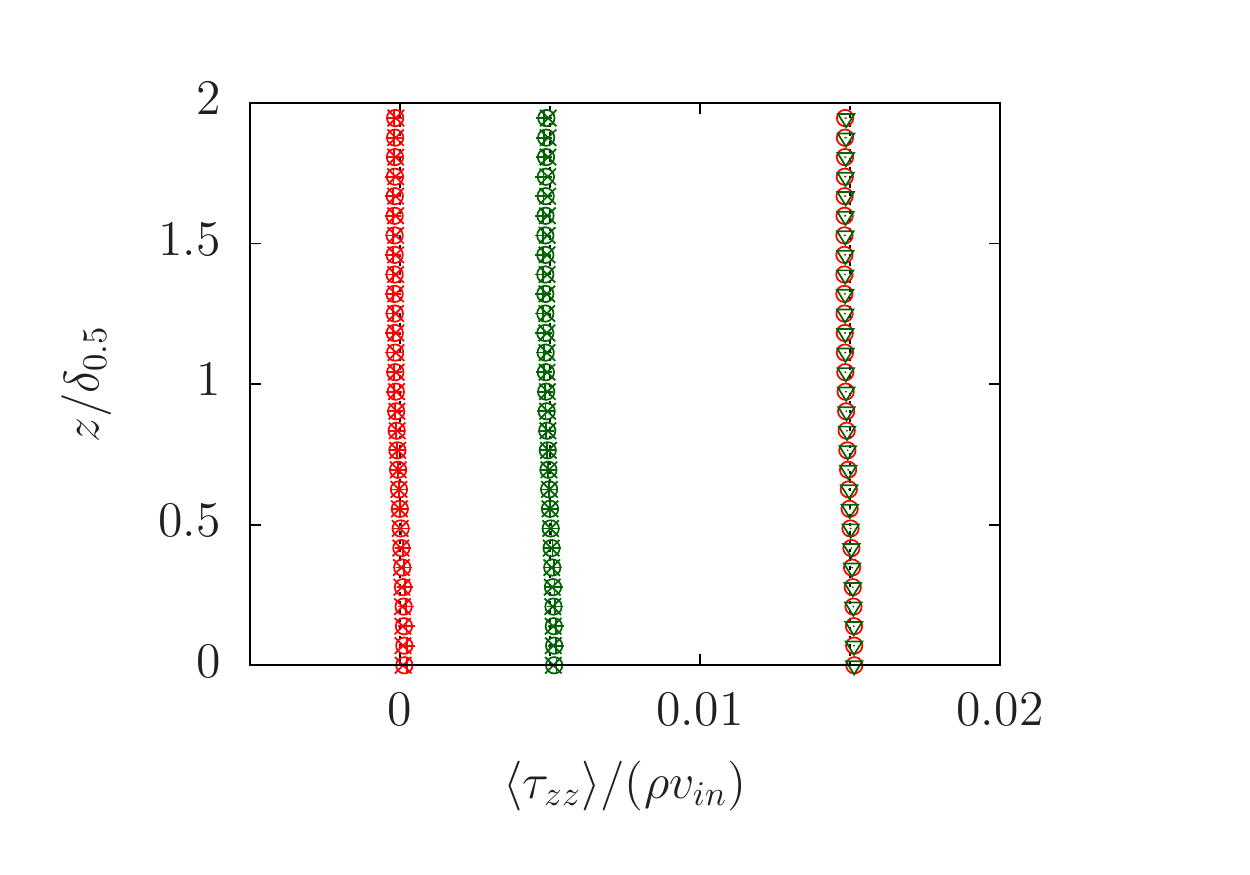}}
\put(0.61,0.52){\includegraphics[width=0.96\columnwidth, keepaspectratio]{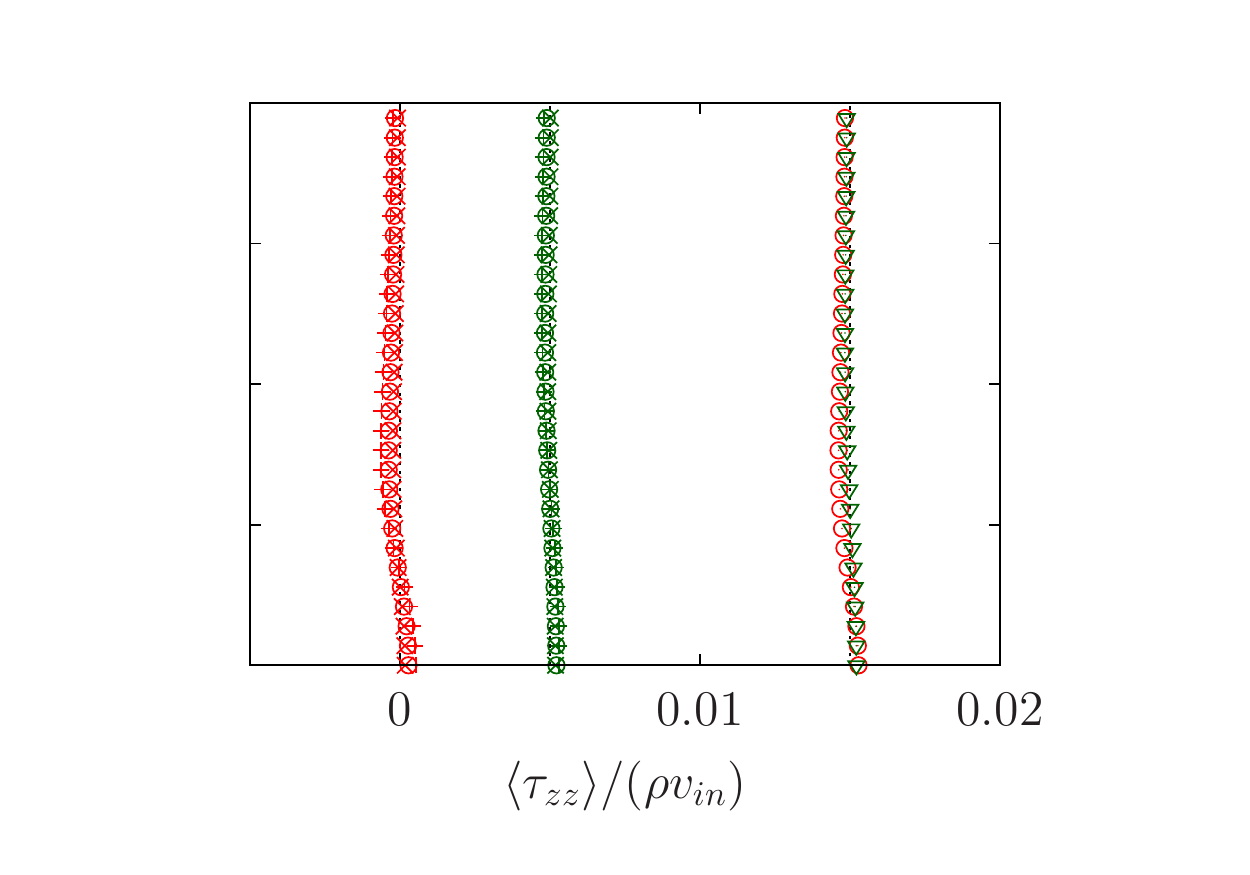}}
\put(1.27,0.52){\includegraphics[width=0.96\columnwidth, keepaspectratio]{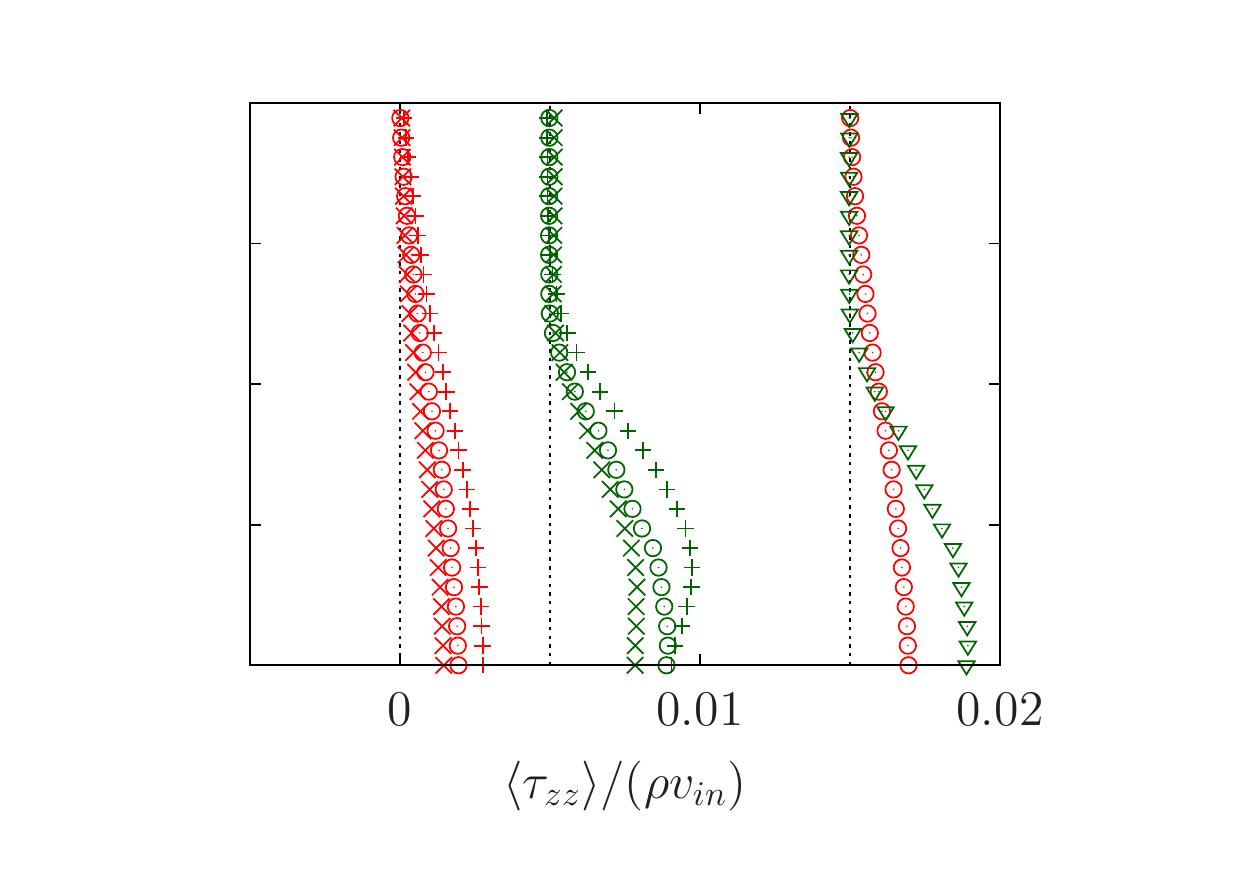}}
\put(-0.05,-0.06){\includegraphics[width=0.96\columnwidth, keepaspectratio]{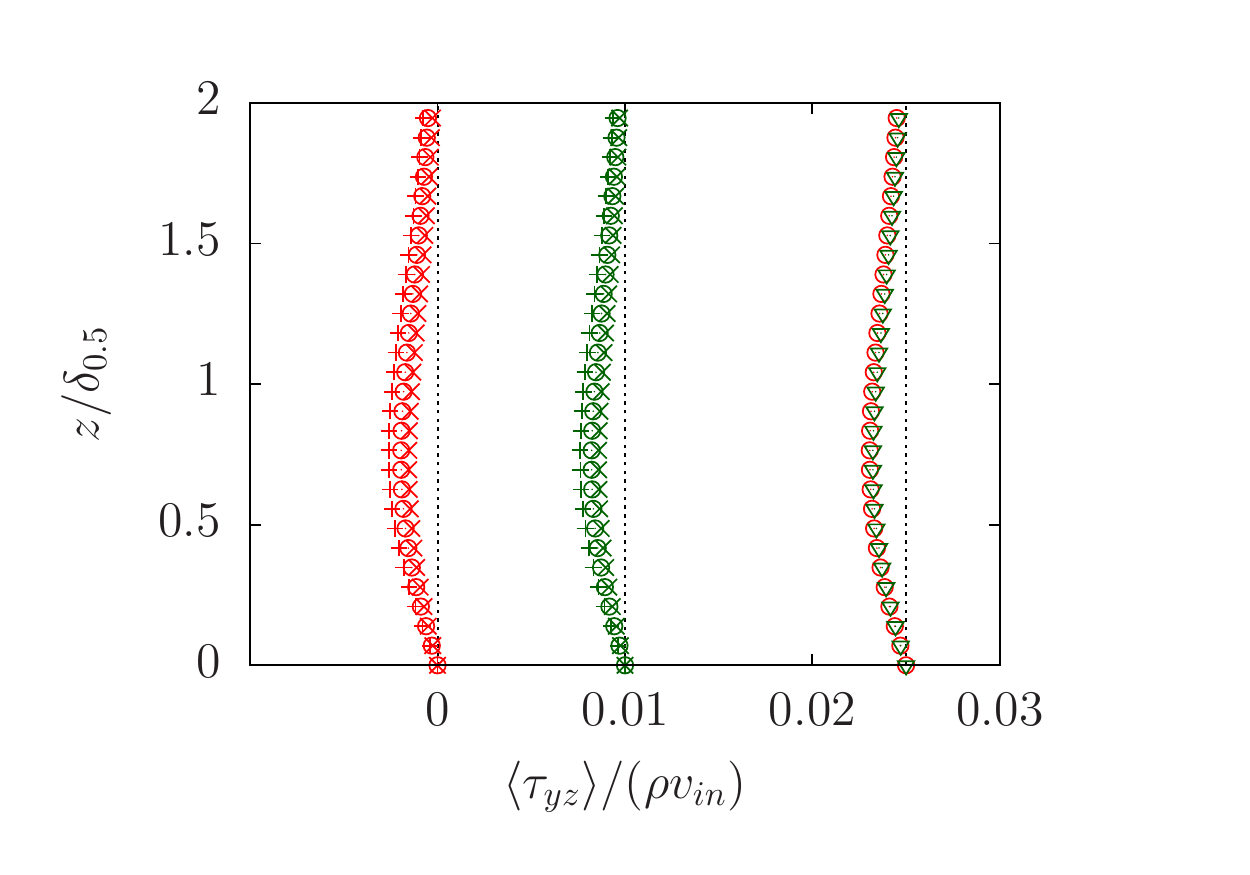}}
\put(0.61,-0.06){\includegraphics[width=0.96\columnwidth, keepaspectratio]{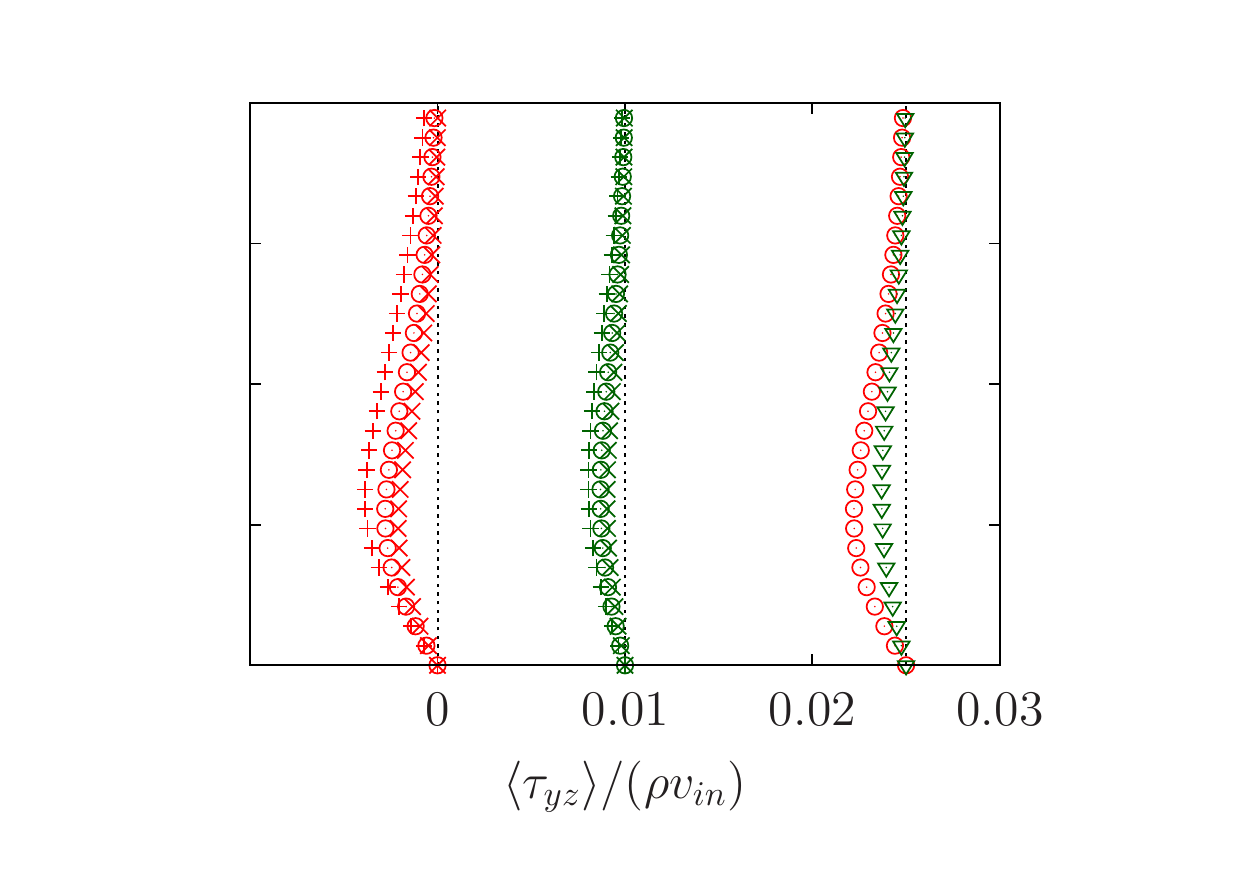}}
\put(1.27,-0.06){\includegraphics[width=0.96\columnwidth, keepaspectratio]{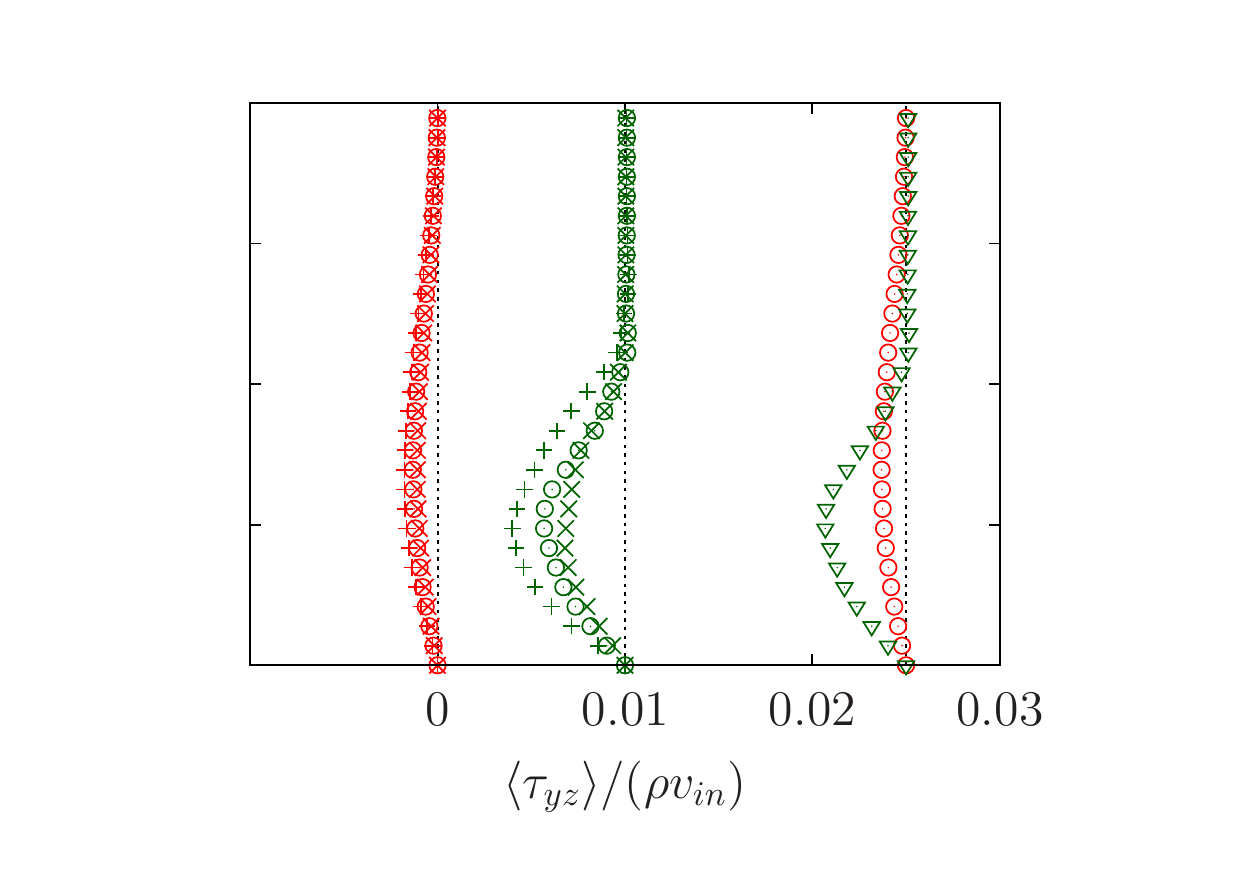}}
\put(0.65,1.64){(a)}
\put(1.31,1.64){(b)}
\put(1.98,1.64){(c)}
\put(0.65,1.06){(d)}
\put(1.31,1.06){(e)}
\put(1.98,1.06){(f)}
\put(0.65,0.48){(g)}
\put(1.31,0.48){(h)}
\put(1.98,0.48){(i)}
\end{picture}
\caption{Profiles of the main components of the non-Newtonian stress tensor: (a)-(c) $\langle \tau_{yy}\rangle$, (d)-(f) $\langle\tau_{zz}\rangle$, (g)-(i) $\langle\tau_{yz}\rangle$ at different stream-wise locations: $+$ at $y=60h$, $\circ$ at $y=80h$ and $\times$ at $y=100h$. Each color represents a different model and for clarity the profiles for the different models are shifted; the dashed vertical lines identify the corresponding zero-levels. The rightmost part of each panel compares the non-Newtonian extra-stresses for different models at the same stream-wise location $y=80h$. The three columns correspond to the different $\gamma$: (a), (d), (g) $\gamma=1$, (b), (e), (h) $\gamma=10$ and (c), (f), (i) $\gamma=100$.} 
\label{fig: tau_ss}
\end{figure*}

We now move to the analysis of the jet statistics in the direction normal to the jet axis, namely the $z$ direction, in order to verify the self-similarity hypotheses used to derive the above scalings. We report data using the self-similar coordinate $z/\delta_{0.5}$ at different stream-wise locations, with the latter chosen in the fully-developed region.  This is done by observing the data from the centerline statistics, which allow us to separate the inlet region from the fully-developed region; in particular for all non-Newtonian fluid models and all values of $\gamma$, the fully-developed region can be safely identified with $40h \le y \le 100h$. 

Figures~\ref{fig: v_ss} and \ref{fig: Restress_ss} show the mean stream-wise velocity and Reynolds stresses at different stream-wise locations normalized by the value of the centerline velocity at the same stream-wise location. In the figures each non-Newtonian fluid model is reported with a different color, while the markers identify the stream-wise location: $y=60h$ ($+$), $y=80h$ ($\circ$) and $y=100h$ ($\times$). Each data series is shifted to the right for ease of reading and a black, dashed line identifies the corresponding zero-level.  On the rightmost side of each panel we compare the three non-Newtonian fluids at the same stream-wise location, $y=80h$.  

At the lowest value of $\gamma$ considered, $\gamma=1$, no difference is observed among the fluid models at all the locations reported. Since the flowing regime at this $\gamma$ is laminar (i.e., the Reynolds stresses are null) and the non-Newtonian contributions are limited,  when the jet-normal coordinate is rescaled by the local jet thickness and the velocity by the local centerline velocity, the velocity profiles exhibit a complete self-similar behaviour. When the time scale ratio is increased to $\gamma=10$, we observe that for the Carreau and the Giesekus fluids the velocity profiles are self-similar at all the different stream-wise locations, while for the Oldroyd-B model we observe a limited deviation beyond $z/\delta_{0.5}=1.5$ at the location closer to the inlet. This discrepancy for the Oldroyd-B fluid can be traced back to the transition from turbulent to laminar regime occurring at about $y\simeq40h$.
At the highest $\gamma=100$, all three non-Newtonian fluid models clearly show turbulent-like motions, with non-zero Reynolds stresses. Both the Carreau and the Giesekus fluids exhibit a quasi self-similar behavior for the stream-wise mean velocity profile, figure~\ref{fig: v_ss}(c).  
The velocity profiles for the Oldroyd-B fluid do not show a clear self-similar behavior: while the bulk statistics, namely centerline velocity and jet thickness, hint at the presence of a fully-developed region over time, the shape of the normalized velocity profile changes when moving downstream. In particular, we observe the formation of a region around the jet centerline with an approximately uniform velocity profiles, resembling a plug-flow. When considering the Reynolds stress profiles, we note that they decay for increasing distances from the inlet for the Giesekus and Carreau fluid, with a higher decay rate for the Carreau fluid. Being the profile normalized by the centerline velocity, this indicates that the velocity fluctuations decay faster than its mean value. On the other hand, the normalized Reynolds stresses are roughly constant for the Oldroyd-B fluid, thus indicating a similar decay rate for the mean and fluctuating part of the velocity.

We now analyze the profiles in the jet-normal direction of the non-Newtonian extra-stress tensor, reported in figure~\ref{fig: tau_ss}. The analysis is limited to the two viscoelastic cases, the  Giesekus and Oldroyd-B fluids.
At the lowest $\gamma=1$, there is no difference among the two non-Newtonian fluid models, nor among the different stream-wise locations. In particular, the normal stresses $\tau_{yy}$ and $\tau_{zz}$ are almost null, while the shear component $\tau_{yz}$ is non-zero and negative, as it originates from the shearing between the jet and the surrounding fluid at rest. At the intermediate $\gamma=10$, $\tau_{yy}$ starts growing, exhibiting a peak at about $z/\delta_{0.5}\approx 0.6$ which is maximum close to the inlet and then rapidly decays. The other normal component of the stress tensor $\tau_{zz}$ instead is positive in the jet core $z/\delta_{0.5} \lesssim 0.5$ and negative elsewhere, with a magnitude always smaller than $\tau_{yy}$. For the Giesekus fluid model the magnitude of the shear component of the non-Newtonian stress $\tau_{yz}$ increases with $\gamma$ between $1$ and $10$, while we observe an almost negligible increase for the Oldroyd-B fluid. Finally, at the highest $\gamma=100$ when the flow is turbulent, both $\tau_{yy}$ and $\tau_{zz}$ are positive; for the Oldroyd-B fluid both normal components of the extra stress tensor substantially grows, while for the Giesekus fluid only $\tau_{zz}$ grows, with $\tau_{yy}$ actually reducing. A similar trend is observed also for the shear component $\tau_{yz}$, with its value growing for the Oldroyd-B fluid and reducing for the Giesekus one, always remaining negative for both the fluids. For all cases, we report a decay of all the components of the extra stress tensor with the stream-wise location.

\begin{figure}
\setlength{\unitlength}{\columnwidth}
\begin{picture}(1,1.20)
\put(0.,0.71){\includegraphics[width=\columnwidth, keepaspectratio]{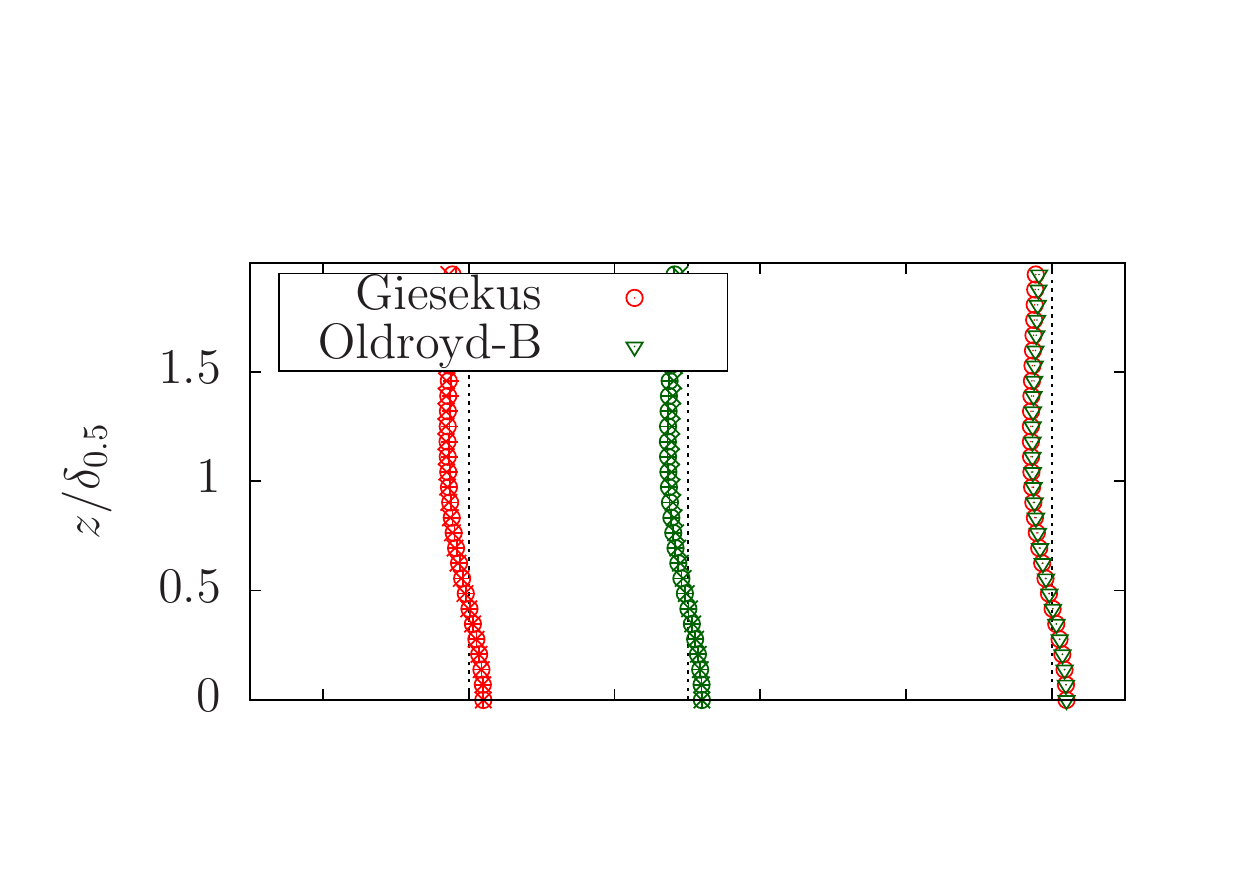}}
\put(0.,0.33){\includegraphics[width=\columnwidth, keepaspectratio]{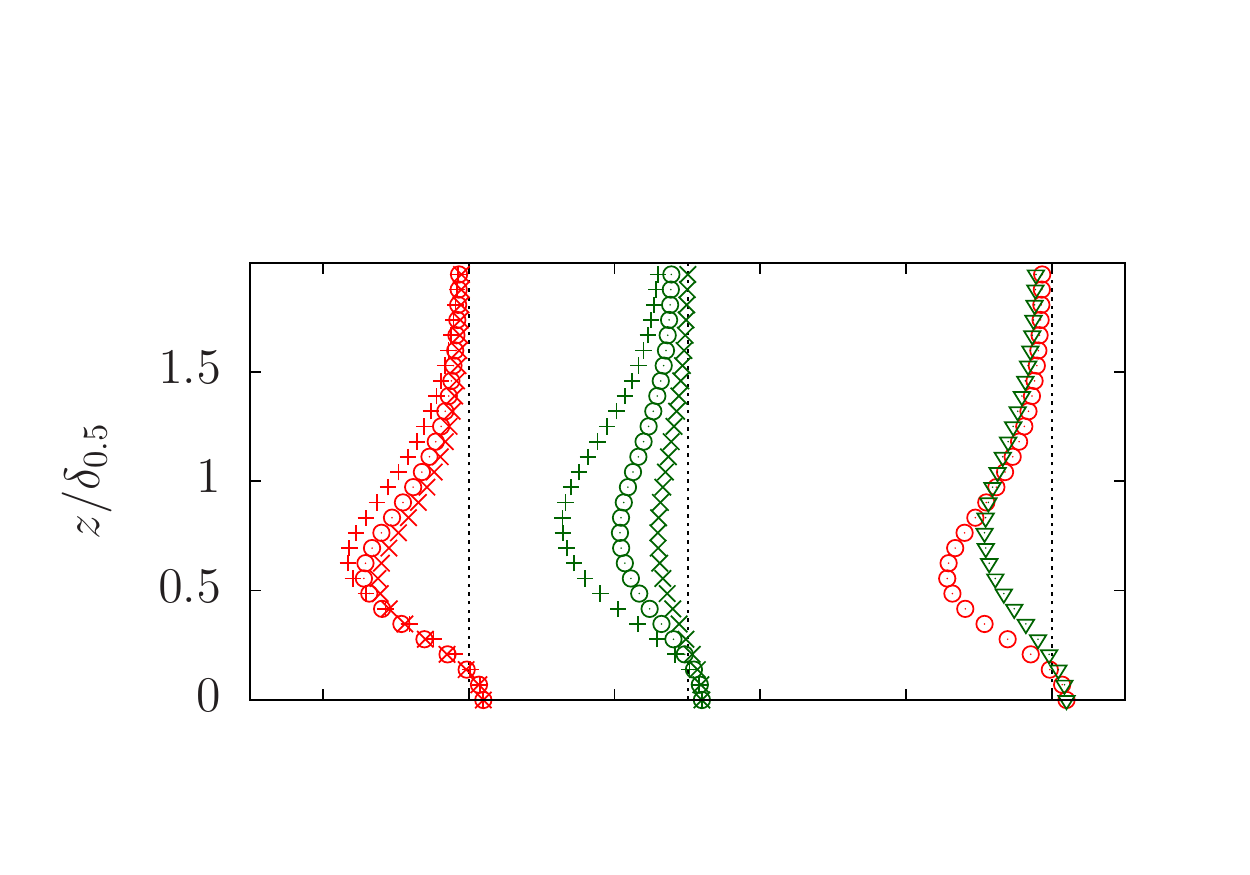}}
\put(0.,-0.05){\includegraphics[width=\columnwidth, keepaspectratio]{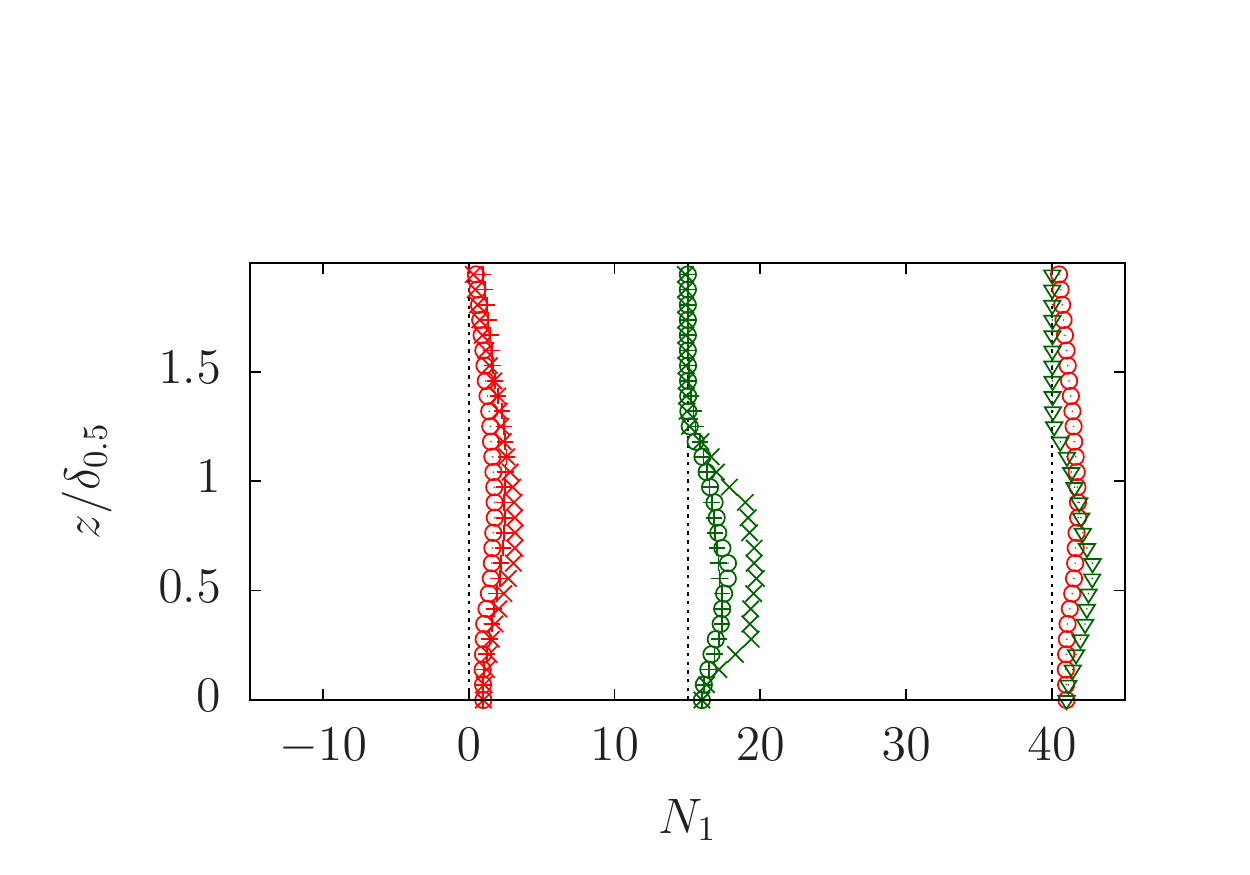}}
\put(0.04,1.16){(a)}
\put(0.04,0.78){(b)}
\put(0.04,0.4){(c)}
\end{picture}
\caption{Profiles of the first normal stress difference $N_1=\langle\tau_{yy}-\tau_{zz} \rangle$ at different stream-wise locations: $+$ at $y=60$, $\circ$ at $y=80$ and $\times$ at $y=100$. Each color represents a different model and for clarity the profiles for the different models are shifted; the dashed vertical lines identify the corresponding zero-levels. The rightmost part of each panel compares the first normal stress difference for different models at the same stream-wise location $y=80h$. The three panels correspond to the different $\gamma$: (a) $\gamma=1$, (b) $\gamma=10$ and (c) $\gamma=100$.}
\label{fig: N1_ss}
\end{figure}
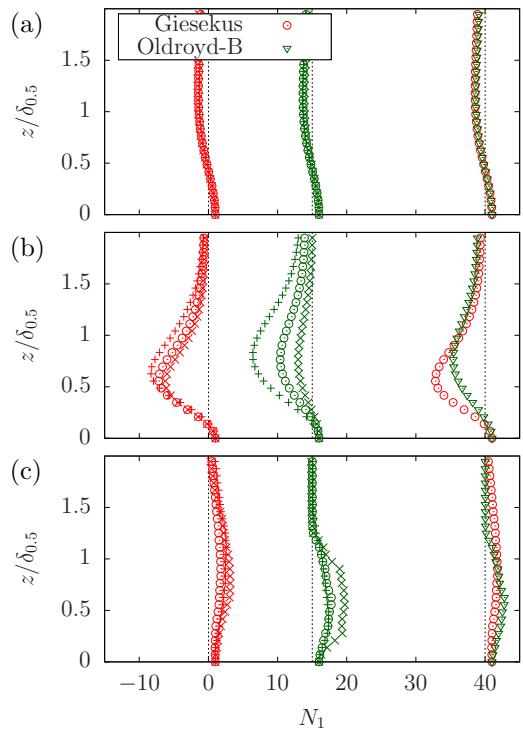

The results above can be better interpreted when considering the first normal stress difference. In particular, from the diagonal components of the non-Newtonian stress tensor, we can compute the first normal stress difference, $N_1=\langle \tau_{yy}-\tau_{zz}\rangle$, shown in figure~\ref{fig: N1_ss}. At low $\gamma=1$, the first normal stress difference is positive at the jet core and changes sign further away from the jet centerline, with no appreciable difference among the fluid models and the  stream-wise locations. The sign of $N_1$ indicates that at the jet core the stream-wise component, $\tau_{yy}$, is larger than the jet-normal, $\tau_{zz}$, sign of an axial stretching; further away from the jet core, the jet-normal component, $\tau_{zz}$, dominates as the ambient fluid is entrained in the jet (jet-normal stretching). When $\gamma$ is increased to $10$, the magnitude of the normal stress difference increases. While the section characterized by positive $N_1$ at the jet core shrinks in size, its value remains unchanged; conversely, much higher values, in magnitude, are observed away from the centerline. A decay of the normal stress difference with the distance from the inlet becomes also clear at the intermediate Weissenberg number. At the highest $\gamma=100$, we do not observe any change in sign in the normal stress difference which is always positive, i.e.,~with the stream-wise component dominating over the jet-normal one. We also note that the magnitude of $N_1$ is large closer to the inlet, $y=60h$, and then stabilizes on a lower value further downstream.

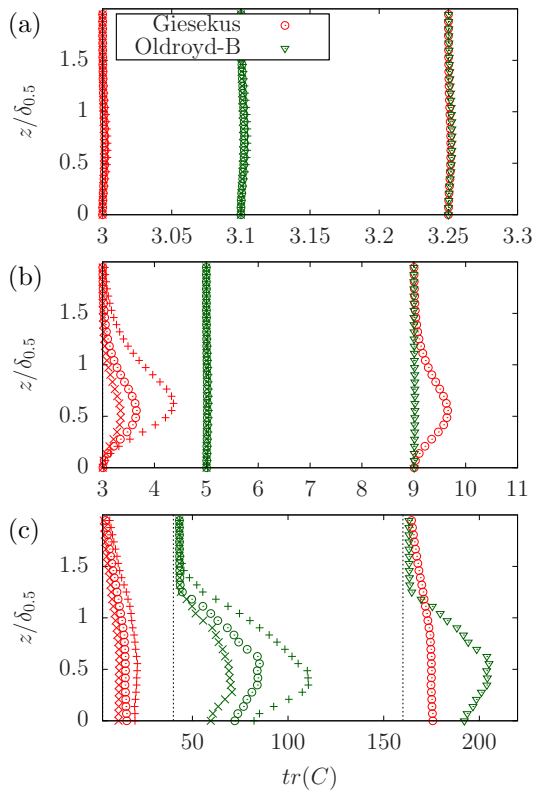
\begin{figure}
\setlength{\unitlength}{\columnwidth}
\begin{picture}(1,1.20)
\put(0.,0.81){\includegraphics[width=\columnwidth, keepaspectratio]{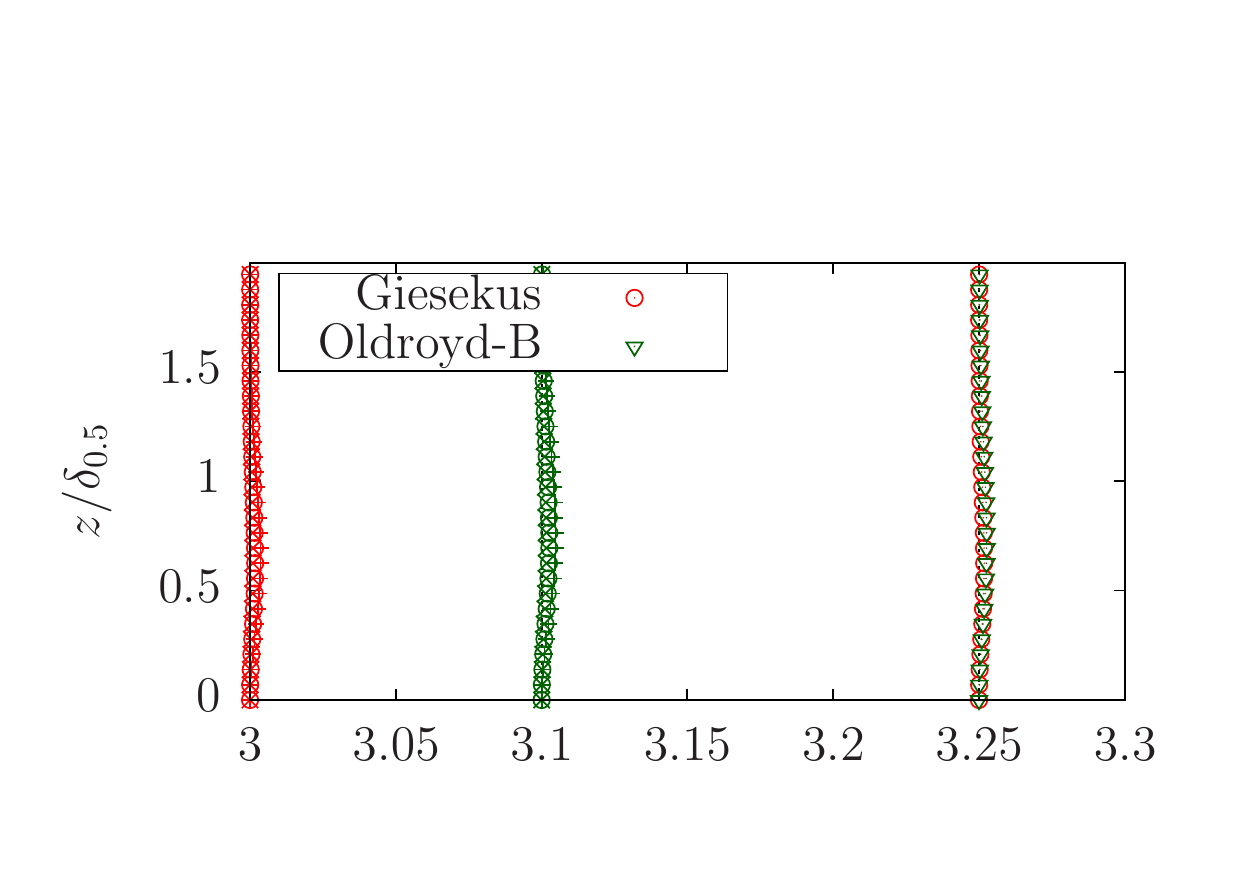}}
\put(0.,0.38){\includegraphics[width=\columnwidth, keepaspectratio]{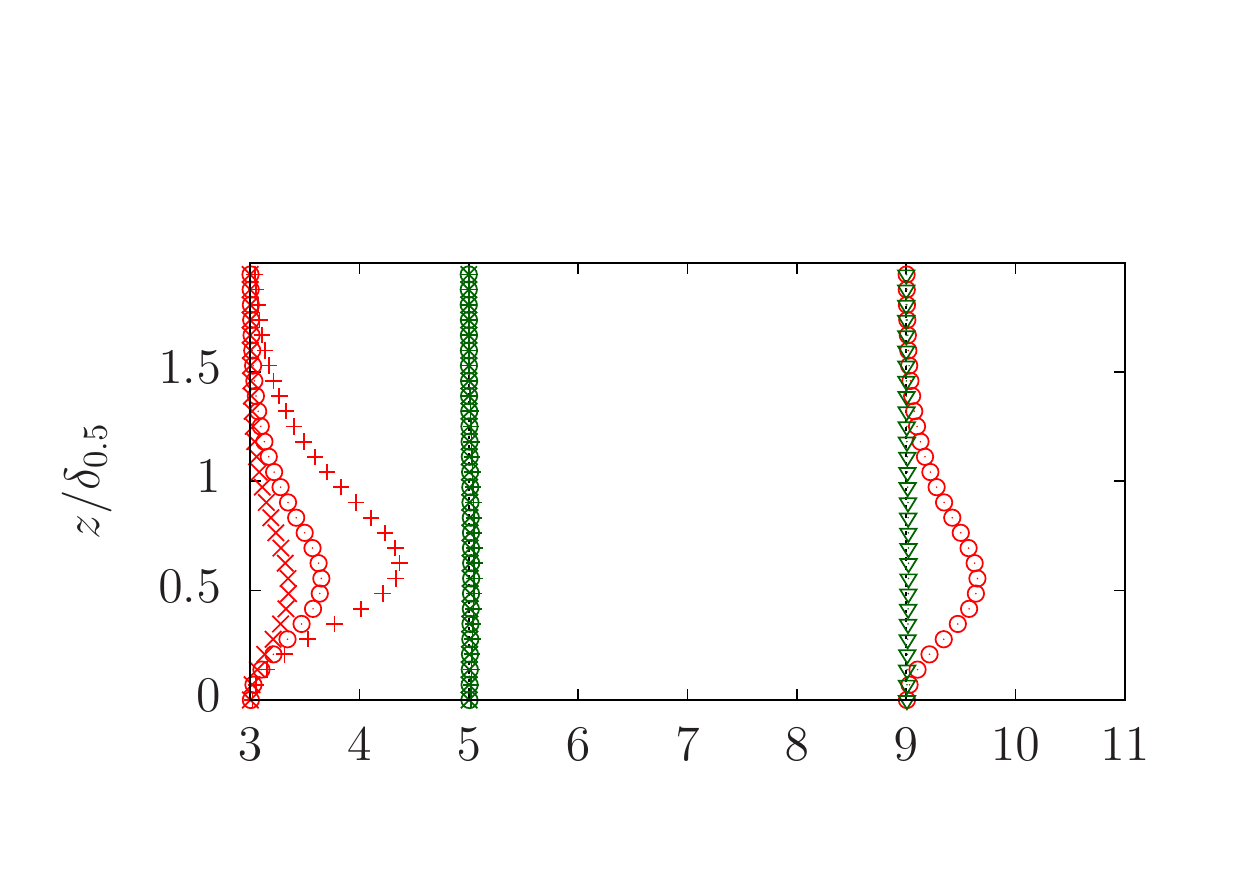}}
\put(0.,-0.05){\includegraphics[width=\columnwidth, keepaspectratio]{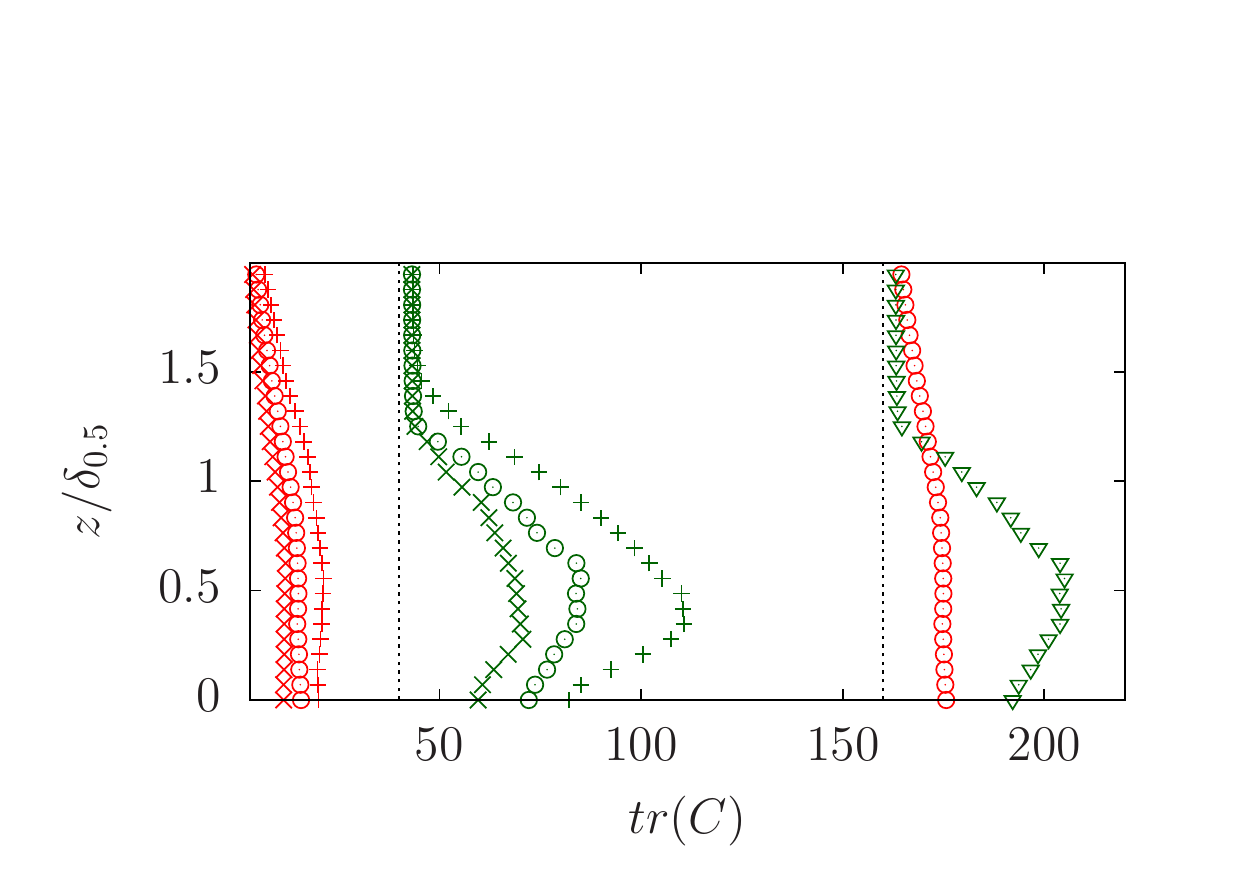}}
\put(0.04,1.26){(a)}
\put(0.04,0.83){(b)}
\put(0.04,0.4){(c)}
\end{picture}
\caption{Profiles of the trace of the conformation tensor $tr (C)$ at different stream-wise locations: $+$ at $y=60$, $\circ$ at $y=80$ and $\times$ at $y=100$. Each color represents a different model and for clarity the profiles for the different models are shifted; the dashed vertical lines identify $tr(C)=3$ for each case. The rightmost part of each panel compares the trace of the conformation tensor for different models at the same stream-wise location $y=80h$. The three panels correspond to the different $\gamma$: (a) $\gamma=1$, (b) $\gamma=10$ and (c) $\gamma=100$.} 
\label{fig: trC_ss}
\end{figure}

Finally, we consider the trace of the conformation tensor, figure~\ref{fig: trC_ss}, $tr(C)=\langle C_{xx}+C_{yy}+C_{zz}\rangle$, which is an indicator of the stretching of the polymers, being equal to $3$ when the polymers are at rest. The trace of the conformation tensor of the Giesekus fluid at the lowest $\gamma$ shows a peak at about $z/\delta_{0.5}=0.6$; when $\gamma$ increases to $10$, the trace of the conformation tensor increases, denoting a higher polymer stretching, while at the highest $\gamma=100$, the peak is not anymore visible, and we observe a flatter profile across the core of the jet. The trend is different for the Oldroyd-B fluid: at low $\gamma$ the polymers are barely stretched, and when $\gamma$ increases from $1$ to $10$, the polymer deformation is only slightly increased, much less than what observed for the Giesekus fluid. This finding is in agreement with the local Deborah number being smaller than one away from the inlet, see figure~\ref{fig: Recent}, and indicating a laminar flowing regime.
Finally, when $\gamma$ is increased to $100$, the polymers become highly stretched for the Oldroyd-B fluid, around $4$ times more than for the Giesekus fluid. Also, for the Oldroyd-B fluid, the trace of the conformation tensor shows a peak away from the centerline of the jet, at about $z/\delta_{0.5}=0.6$, which slightly shifts towards the centerline as $\gamma$ is increased.

We also tested the theory for the far field of viscoelastic jets developed by \citet{guimaraes2020direct}; results are reported in appendix~\ref{sec: appES}.

\subsection{Instability map}

\begin{figure}
\centering
\includegraphics[width=1.0\columnwidth, keepaspectratio]{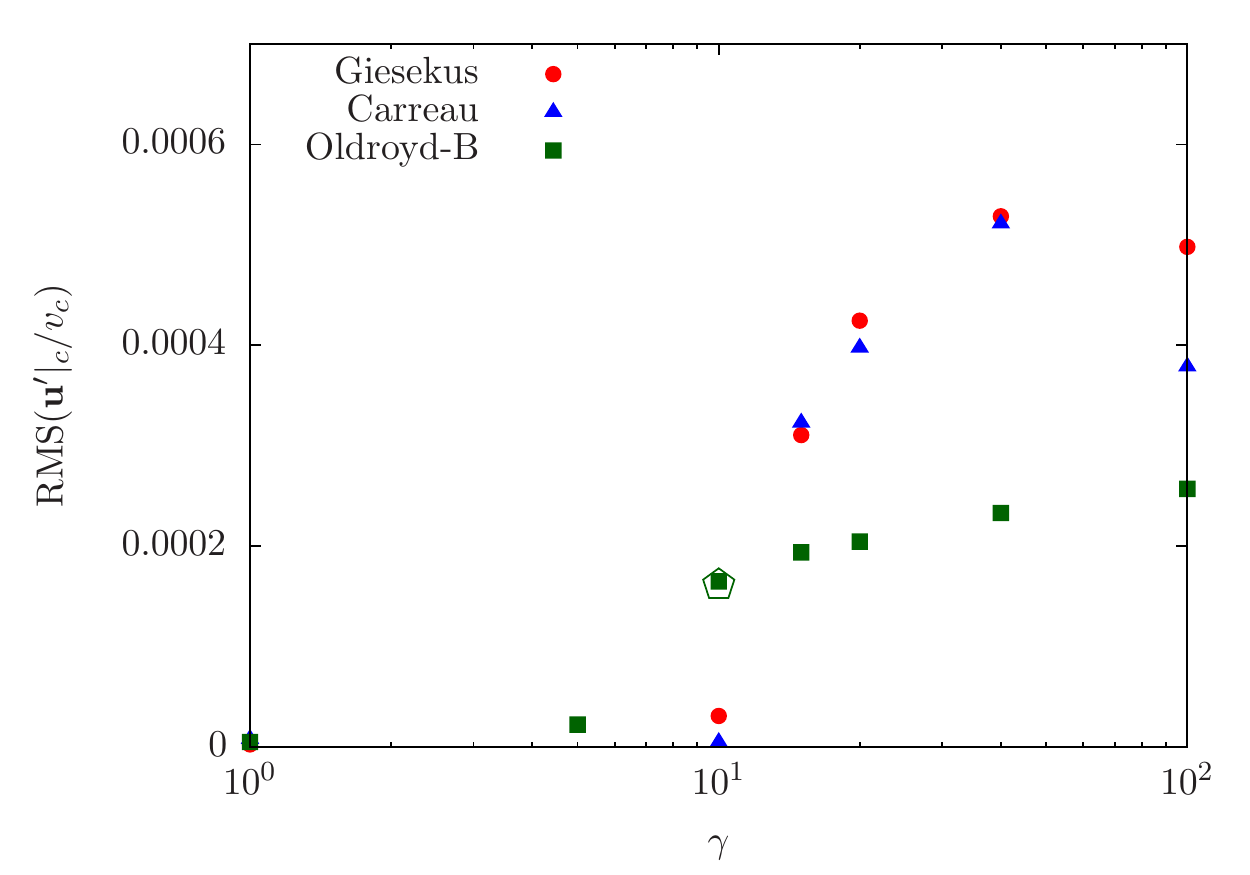}
\caption{Transition map as a function of the time scale ratio $\gamma$ for the three non-Newtonian fluids. Transition from a laminar to a turbulent regime is identified via the value of the root mean square of the fluid velocity fluctuations at the centerline in the self-similar region. The green empty symbol shows the data point obtained with a different inlet velocity profile for the sake of comparison.}
\label{fig: map}
\end{figure}

In this last section we want to investigate more in details the transition from the laminar to the turbulent flow for the three models considered in this study. To do so, we performed additional simulations at intermediate values of $\gamma$ to better capture the transition. As an indicator of the flow behavior, we use the root mean square of the fluid velocity fluctuations at the centerline, normalized by the corresponding mean velocity within the fully-developed region, i.e., RMS$(\mathbf{u}'/v_c)$. Figure~\ref{fig: map} shows this quantity for the three non-Newtonian fluid models: zero values are indicative of a laminar flow regime, whereas non-zeros values highlight the presence of velocity fluctuations and of either an unsteady laminar regime or of a turbulent flow regime. 

From the graph we observe that the Oldroyd-B fluid is the one showing the earliest transition, consistently with what observed previously, see e.g.~figure~\ref{fig: quali}, with a critical value of $\gamma$ of about $\gamma_{cr}\sim5$. The Giesekus fluid instead has a critical value of about $\gamma_{cr}\sim10$, almost double than the one found for the Oldroyd-B fluid. Finally, the Carreau fluid requires an even higher value of the time scale ratio in order to make the flow unstable, with a critical $\gamma$ in the range $10< \gamma_{cr} <15$. These results indicate that while both elasticity and shear-thinning can lead to the transition to a turbulent flow, elasticity alone is the fastest way to transition. Furthermore, when both effects are present at once, an intermediate condition is found, with fluid elasticity and shear-thinning having competing effects. 

Lastly, it is interesting to note that while the Oldroyd-B fluid shows the earliest transition among the three fluids, the magnitude of the velocity fluctuations is lower, about half than what measured for the other two models. This observation can be explained with the large length- and time-scale of velocity fluctuations induced by fluid elasticity, which are thus more correlated (both in time and space) than the one generated by shear-thinning. For the Giesekus fluid, characterized by both fluid elasticity and shear-thinning, these two effects seems to superpose in this regards, leading to the largest values of fluctuations.

\section{Conclusions}
\label{sec: concl}
We perform numerical simulations of planar jets of different non-Newtonian fluids at low Reynolds number, for various values of the time scale ratio $\gamma$. We consider four different fluids, to highlight the roles of elasticity and shear-thinning in the transition to turbulence; in particular, we consider a purely elastic fluid (simulated by the Oldroyd-B model), a purely shear-thinning fluid (simulated by the Carreau model), and an elastic and shear-thinning fluid (simulated by the Giesekus model), in addition to a simple Newtonian fluid for comparison. 
We show the appearing of turbulent motions in low-Reynolds number, non-Newtonian planar jets, for the first time in numerical simulations, in conditions where the corresponding Newtonian jet shows a steady, laminar flow. The non-Newtonian features, namely shear-thinning, fluid elasticity or their combination, are the fundamental factors that trigger and sustain turbulence. 

By varying the value of $\gamma$ we investigate the transition to turbulence for the different fluids. At the lowest value of $\gamma$, all fluid models show a laminar flowing regime and effects from the non-Newtonian feature are weak. As $\gamma$ is increased, we observe a transition to a turbulent-like regime, characterized by disordered fluid motions. This transition occurs first for the Oldroyd-B fluid, followed by the Giesekus fluid and finally by the Carreau fluid. Thus, while both viscoelasticity and shear-thinning are able to trigger the flow instability and to sustain a turbulent-like flow, the first mechanism acts faster. Furthermore, when the two mechanisms are combined, shear-thinning is delaying the transition, thus clarifying that the two effects are actually in competition for the flow destabilization, differently from what usually believed. Finally, at sufficiently high $\gamma$, all the non-Newtonian flows are turbulent.
 
Even when turbulent, the three models provide quantitative and qualitative differences, originated by the different mechanisms involved in the generation of the turbulent fluid motions. These differences can be appreciated in all the flow statistics that we analyzed. When observing the power-spectrum of the turbulent kinetic energy, for the Carreau fluid a $-5/3$ power-law scaling is observed in the inertial range, while for the Giesekus and Oldroyd-B fluids a $-3$ power-law scaling appears. The power-law scaling in the turbulent kinetic energy spectrum is confirmed by the second-order structure function for the Carreau fluid, while for the Giesekus and Oldroyd-B cases the true scaling exponent is masked by the dissipation range scaling. We then investigate the presence of intermittency in the turbulent flow using the extended self-similarity (ESS) and the multifractal spectrum. Both methods yield the same result: intermittency, a feature of inertial Newtonian turbulence, also characterizes elastic turbulence. Indeed the high-order structure function, when plotted using the ESS, deviate from the K41 scaling, and the multifractal spectrum for all cases at the highest $\gamma$ follows the curve measured for high-Reynolds Newtonian turbulence. 

When looking at the bulk properties of the jet, we found that in the laminar regime all non-Newtonian fluids well follow the Newtonian predictions. On the other hand, when the flow is turbulent, the Newtonian prediction are less accurate. Also, for all the non-Newtonian models, the power-law scalings are achieved at much greater distance from the inlet than what observed for a Newtonian fluid. 
This effect results in an initial lower decay rate of the centerline velocity and a lower spreading rate of the jet compared to the Newtonian case, as previously reported by \citet{guimaraes2020direct}. 
By evaluating the local Reynolds and Deborah numbers, we confirmed that the turbulence generated in the presence of shear-thinning alone is mostly inertial, generated in regions with large Reynolds number. On the other hand, that generated by fluid elasticity is triggered when the local Deborah number goes beyond unity. We confirmed that when the two effect are combined, they are in competition: indeed, at intermediate values of $\gamma$, even when the local Deborah number is larger than one, the flow remains laminar if shear-thinning is present. Lastly, we analyze the velocity and polymer statistics along the jet-normal direction and find that self-similarity among different stream-wise positions is observed for the stream-wise velocity, but not among the different models at high values of the time scale ratio $\gamma$. We instead do not observe self-similarity for the Reynolds and extra stress tensors.

In conclusion, we find out that turbulent jets at low Reynolds number exhibit a very complex phenomenology, and that differences in the fluid rheology result in different flows. Elasticity and shear-thinning can coexist in a turbulent flow, with their effects summing up providing higher levels of velocity fluctuations, but they are also in competition regarding the flow transition. Our work, as the first simulations of turbulent jets at low Reynolds number, paves the way for future analysis to better understand the similarities and differences among classical and elastic turbulence. Indeed, while jets are canonical flow configurations for turbulent flows at high Reynolds number, they have received very limited analysis at low Reynolds numbers until recently.

\section*{Acknowledgments}
The research was supported by the Okinawa Institute of Science and Technology Graduate University (OIST) with subsidy funding from the Cabinet Office, Government of Japan. The authors acknowledge the computer time provided by the Scientific Computing Section of Research Support Division at OIST and the computational resources provided by HPCI under the grants hp210246 and hp220099. The authors acknowledge Rahul Kumar Singh (OIST) for the helpful discussion on the derivation of the structure function exponent.

\section*{Declaration of Interests}
The authors report no conflict of interest.

\section*{Appendix}

\subsection{Inlet velocity profile}
\label{sec: inlet}

\begin{figure}
\setlength{\unitlength}{\columnwidth}
\begin{picture}(1,0.80)
\put(0.,0.33){\includegraphics[width=\columnwidth, keepaspectratio]{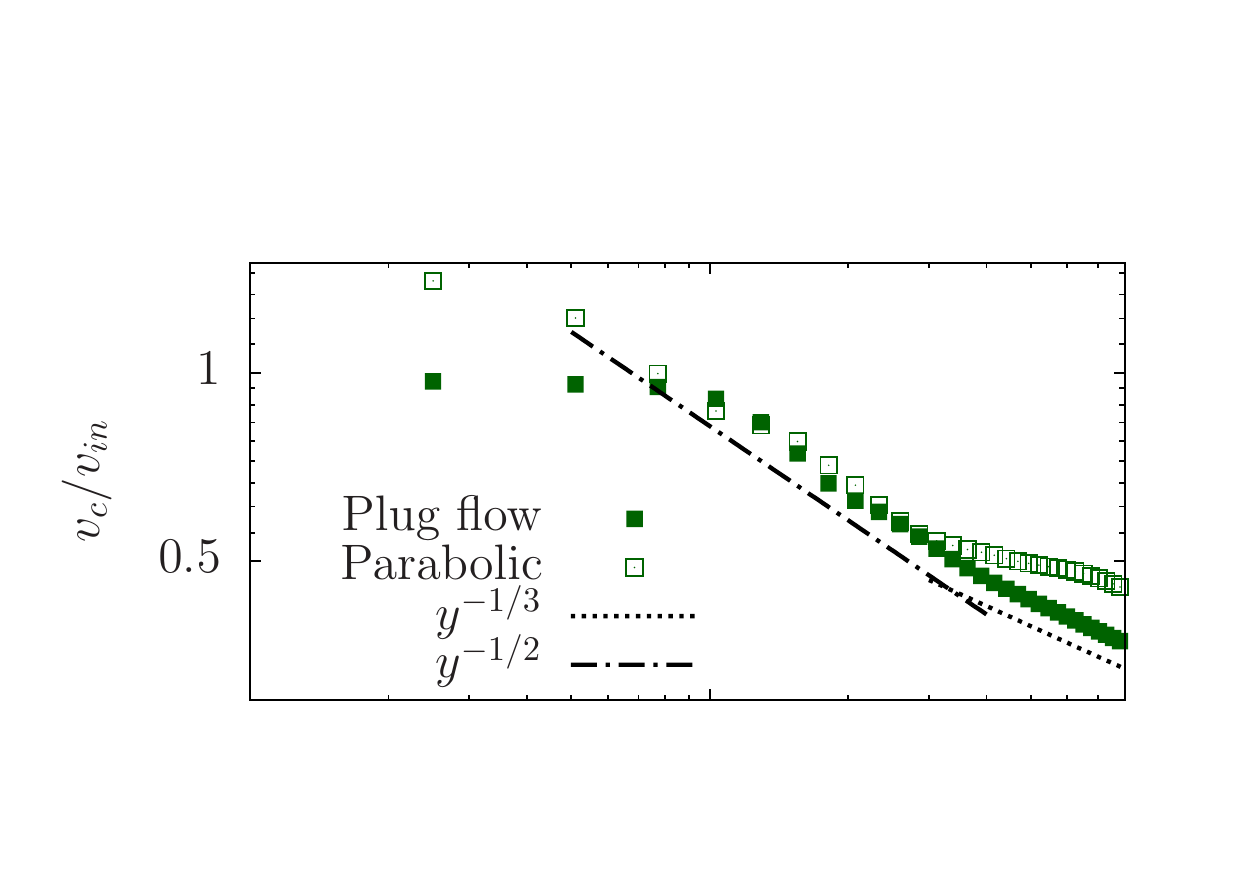}}
\put(0.,-0.05){\includegraphics[width=\columnwidth, keepaspectratio]{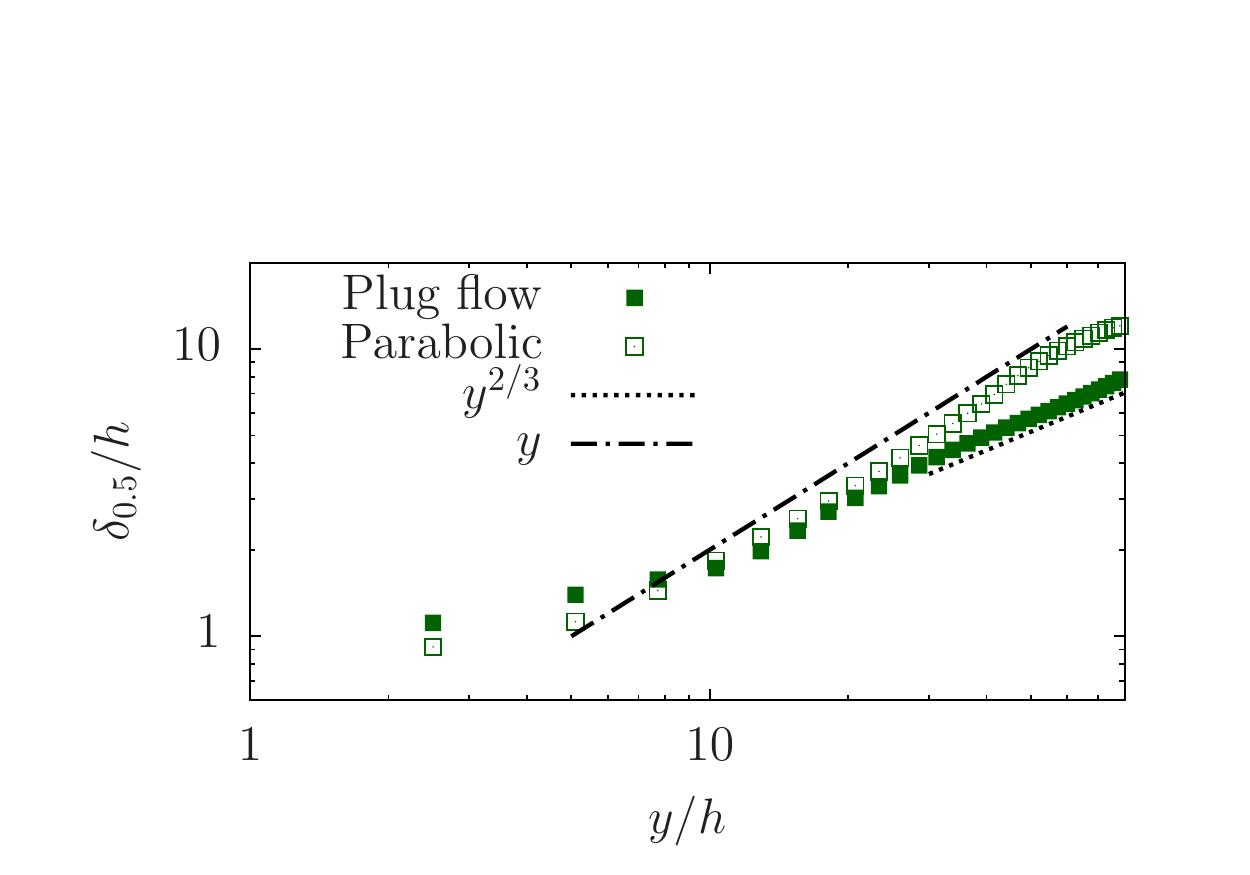}}
\put(0.04,0.78){(a)}
\put(0.04,0.4){(b)}
\end{picture}
\caption{Centerline velocity (panel a) and jet thickness (panel b) for the different inlet velocity profiles at $\gamma=10$. The laminar (dotted line) and turbulent (dash-dotted line) scaling laws are reported for reference.} 
\label{fig: inlet_streamwise}
\end{figure}

We tested different inflow conditions to verify that the plug flow inlet condition used in the main text does not affect the results obtained in our analysis. Thus, we performed additional numerical simulations using a parabolic velocity profile at the inlet (with the same flow-rate) for the Oldroyd-B case at $\gamma=10$. This particular setup was selected to be close to the transition region from laminar to turbulent-like regime to assess the effect of the inlet velocity profile on the transition. The overall transition from laminar to turbulent flow is not particularly affected by the inlet condition (see figure~\ref{fig: map}). The velocity profile at the inlet changes instead the stream-wise location at which turbulence can be no longer sustained and the flow becomes laminar, as shown in figure~\ref{fig: inlet_streamwise}: the transition occurs closer to the inlet when imposing a plug flow velocity profile at the inlet. Our understanding is that the plug flow inlet condition has a stronger shear rate and a lower centerline velocity compared to the plug flow profile, factors that speed up the decay of the jet.

\begin{figure*}
\setlength{\unitlength}{\columnwidth}
\begin{picture}(1,1.5)
\put(-0.05,1.1){\includegraphics[width=0.96\columnwidth, keepaspectratio]{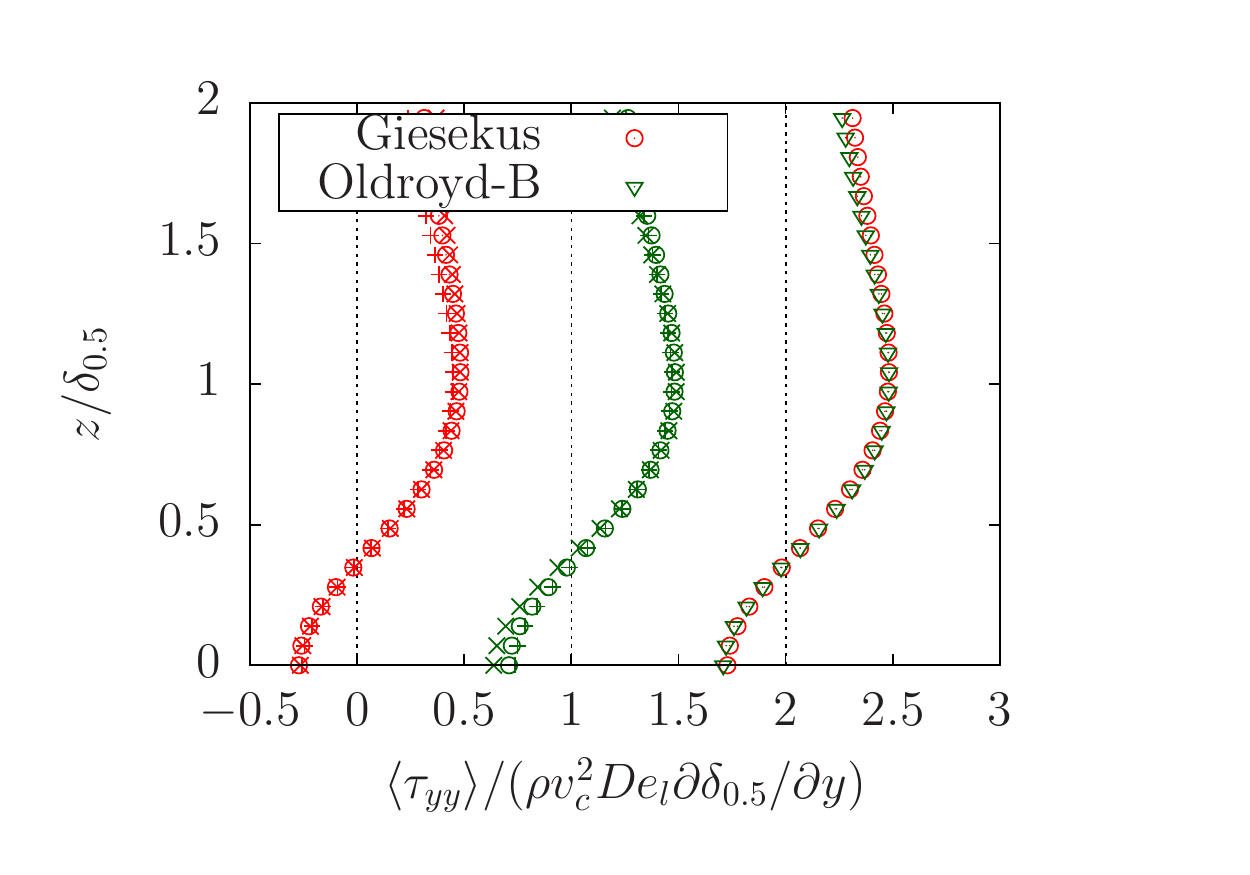}}
\put(0.61,1.1){\includegraphics[width=0.96\columnwidth, keepaspectratio]{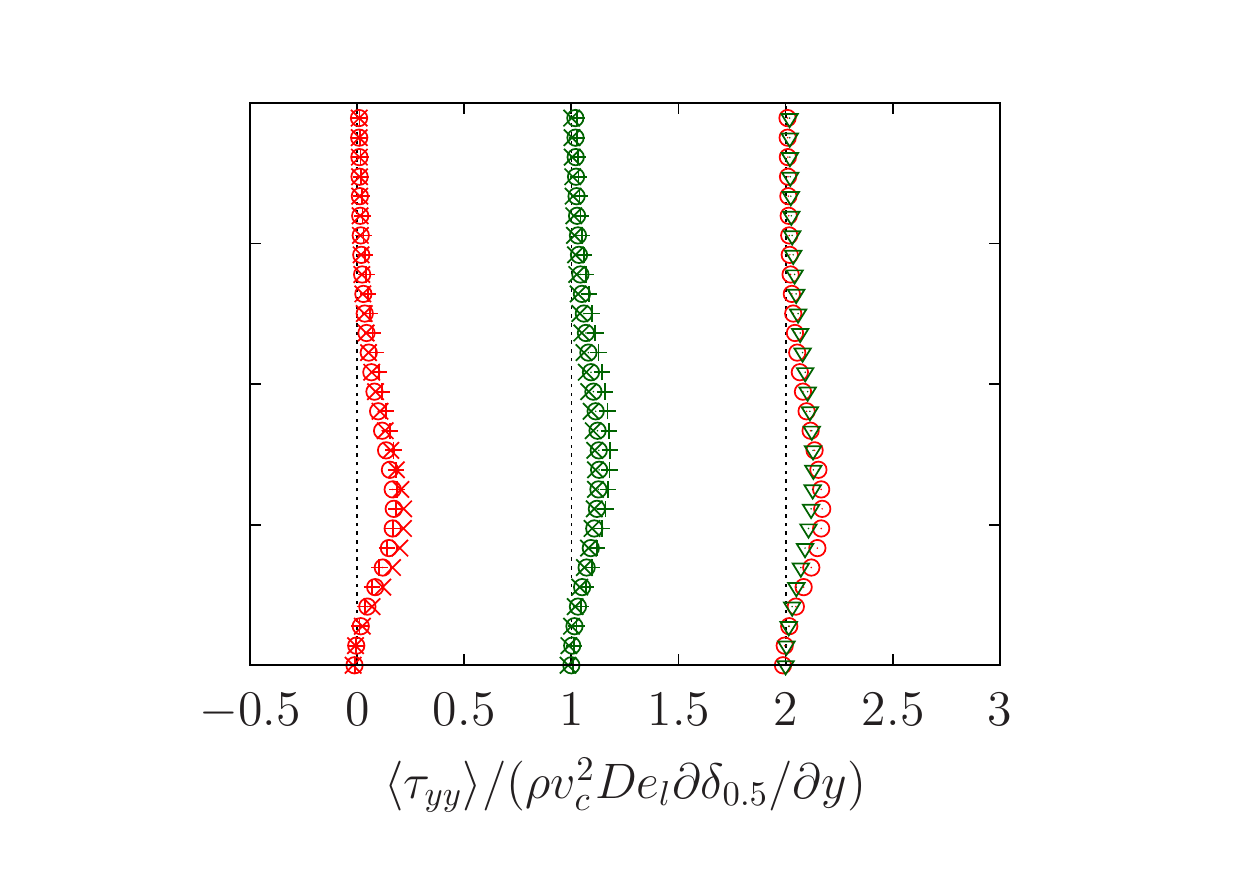}}
\put(1.27,1.1){\includegraphics[width=0.96\columnwidth, keepaspectratio]{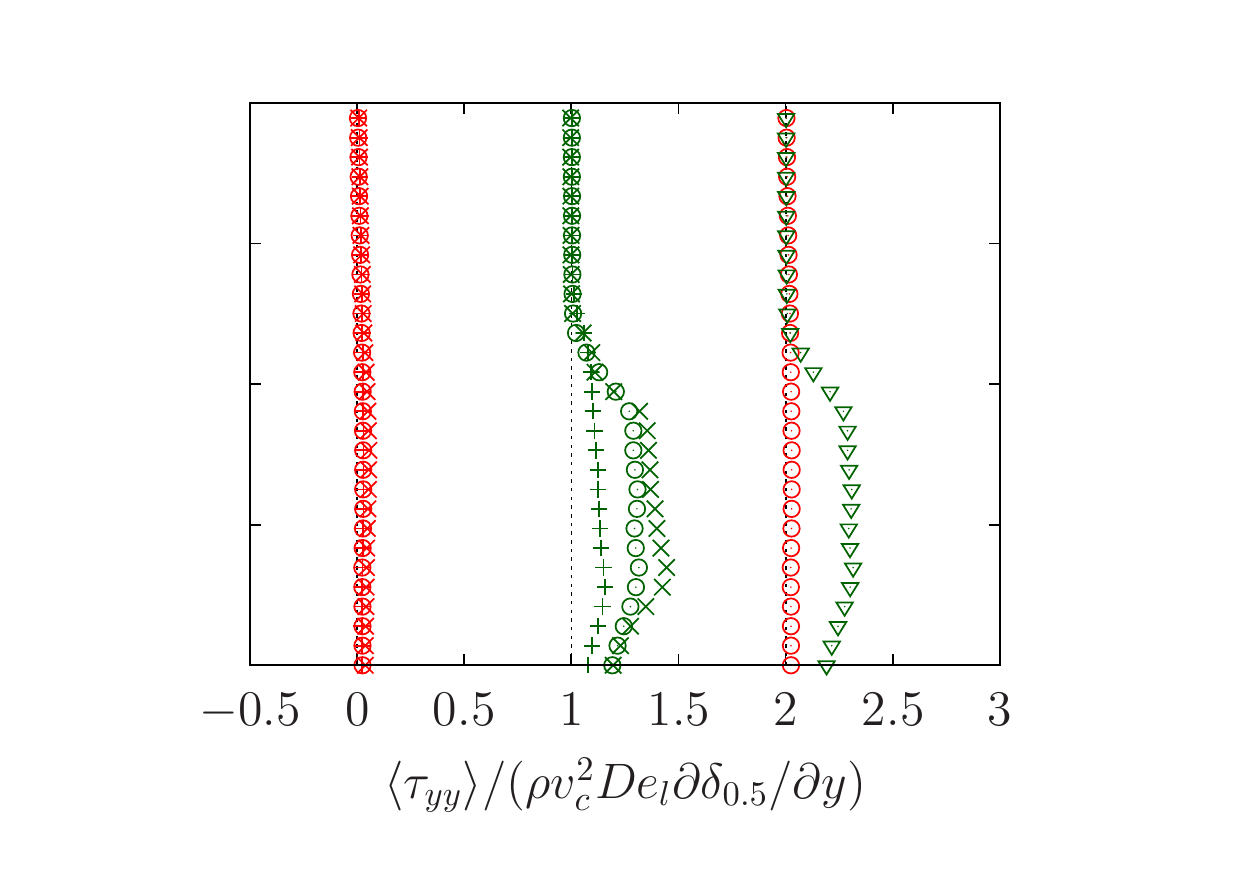}}
\put(-0.05,0.52){\includegraphics[width=0.96\columnwidth, keepaspectratio]{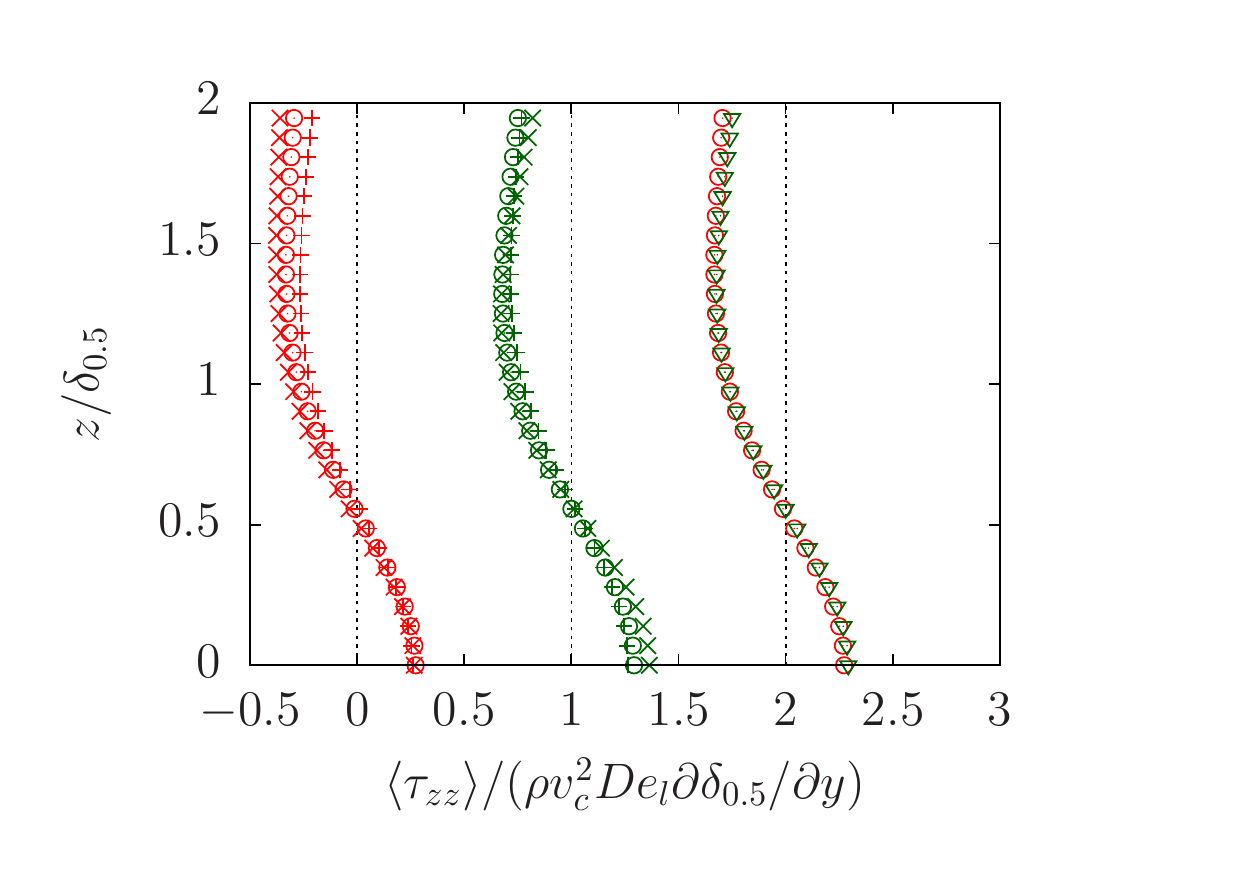}}
\put(0.61,0.52){\includegraphics[width=0.96\columnwidth, keepaspectratio]{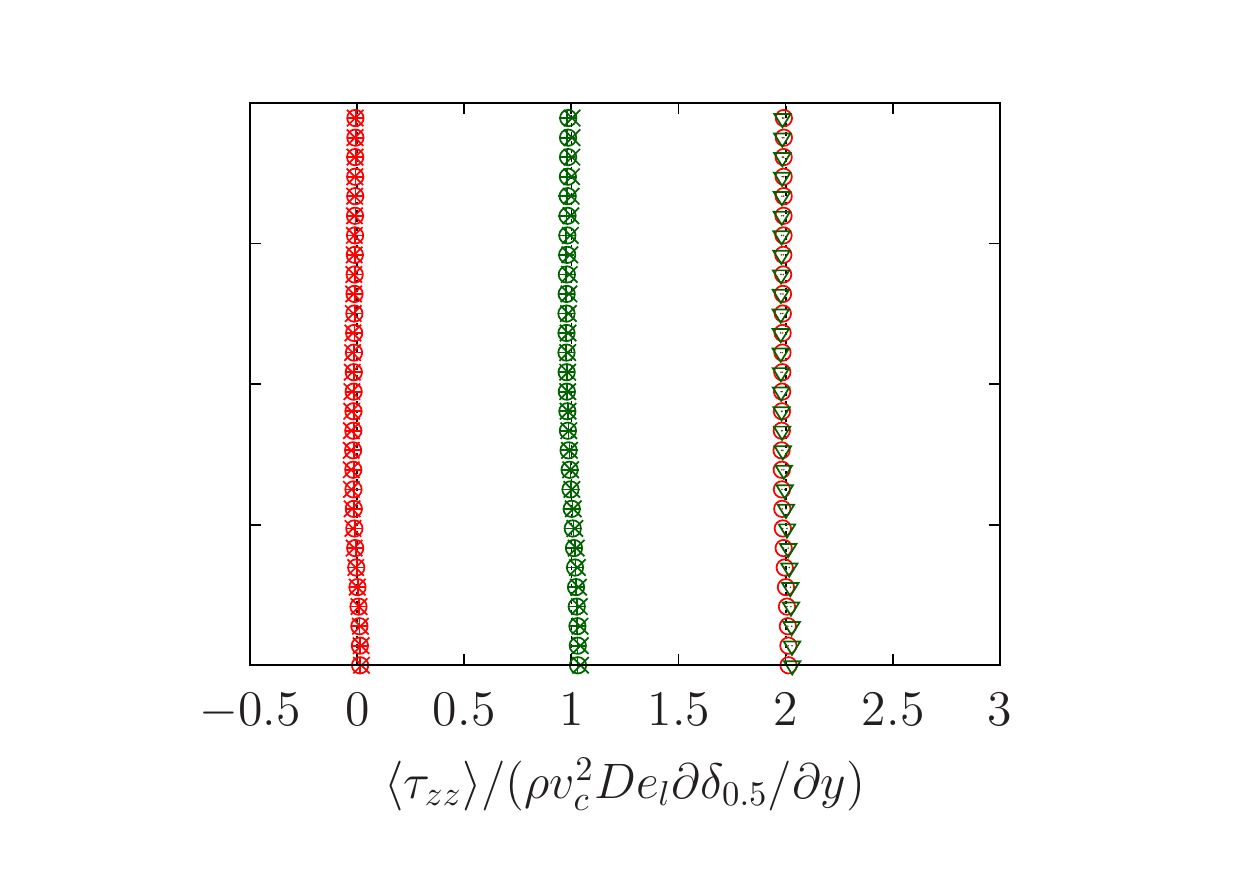}}
\put(1.27,0.52){\includegraphics[width=0.96\columnwidth, keepaspectratio]{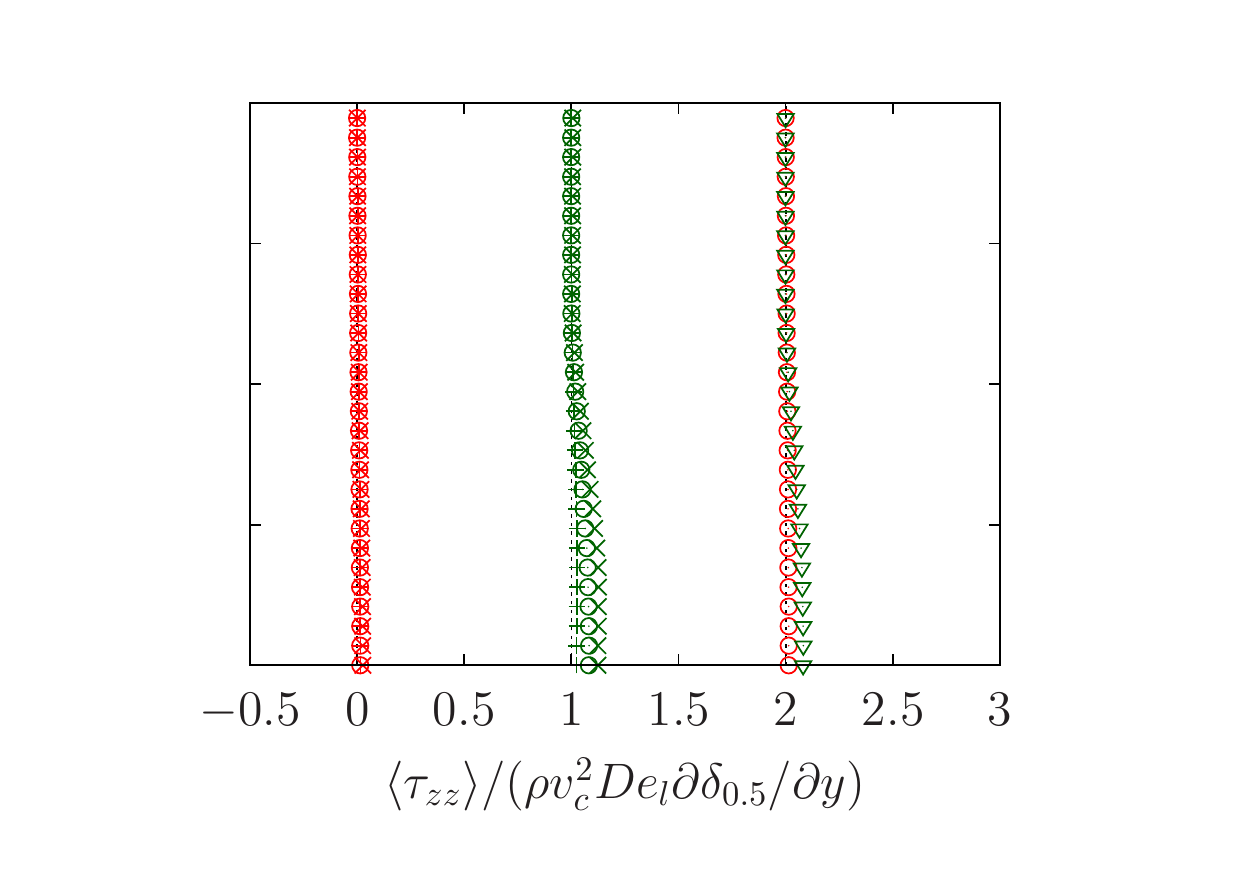}}
\put(-0.05,-0.06){\includegraphics[width=0.96\columnwidth, keepaspectratio]{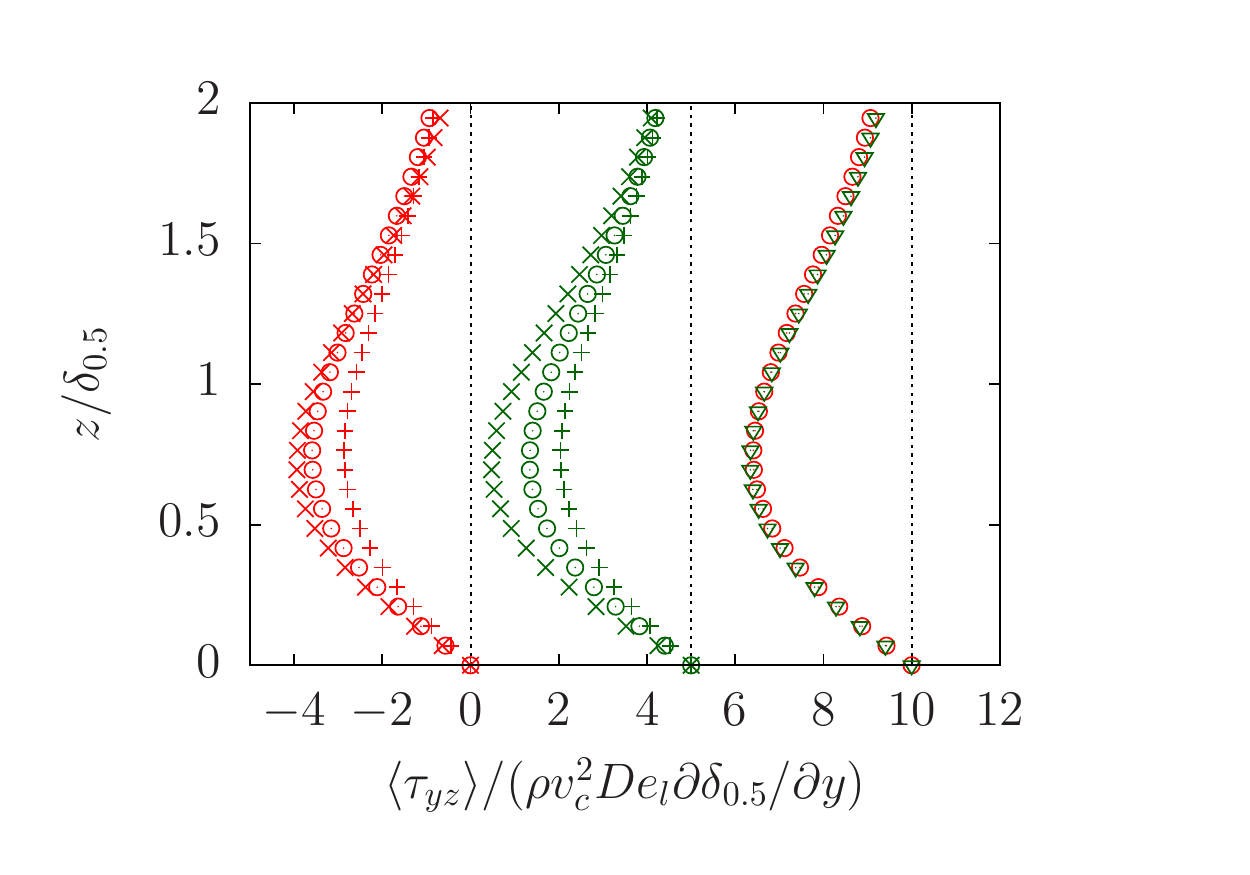}}
\put(0.61,-0.06){\includegraphics[width=0.96\columnwidth, keepaspectratio]{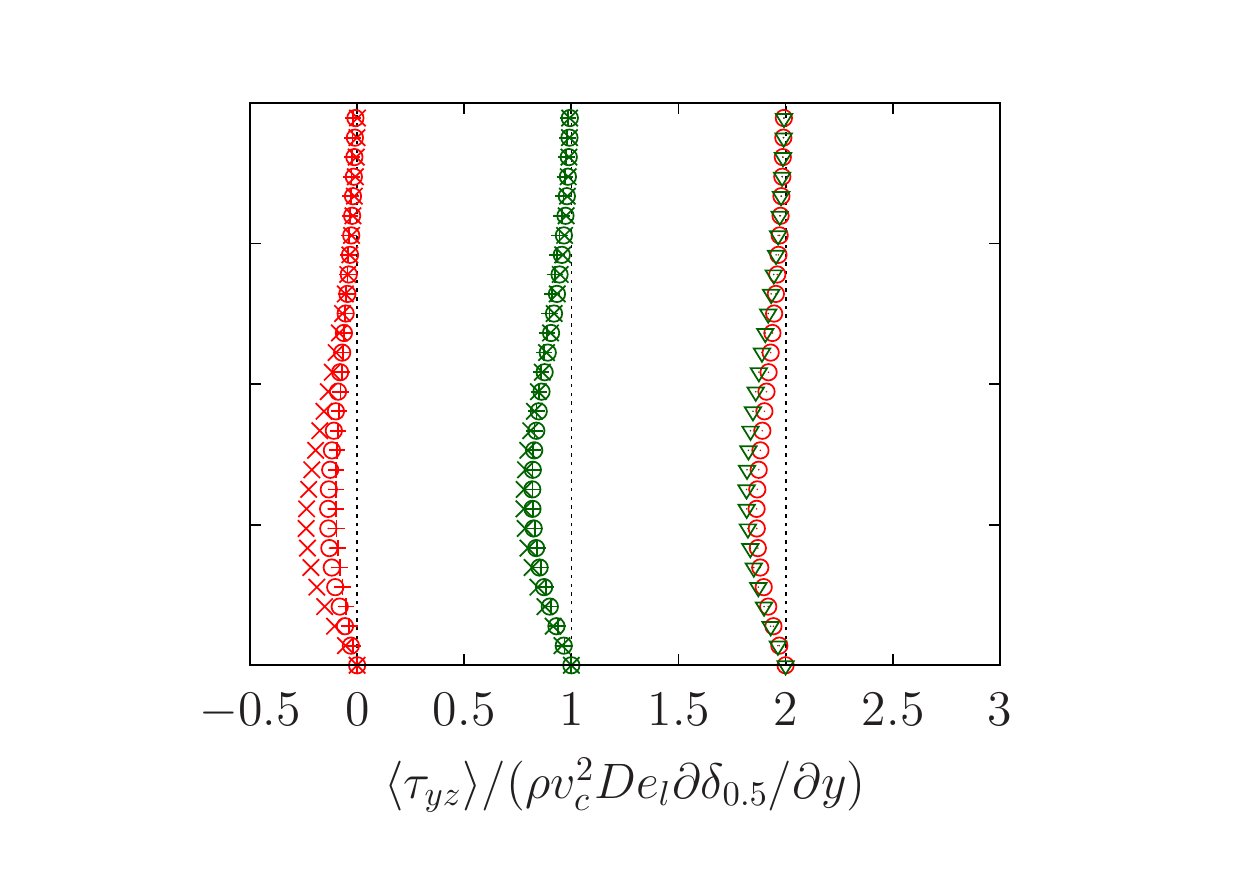}}
\put(1.27,-0.06){\includegraphics[width=0.96\columnwidth, keepaspectratio]{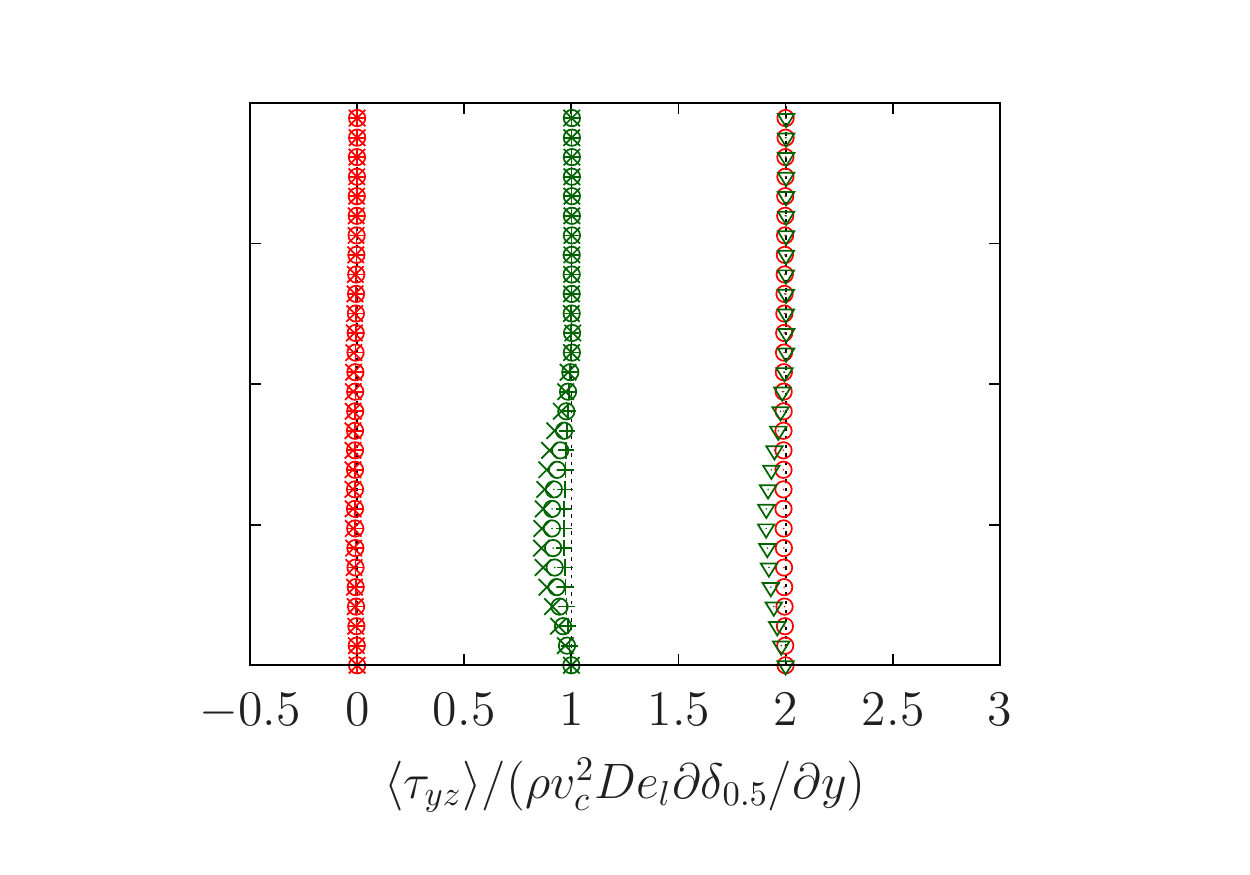}}
\put(0.65,1.64){(a)}
\put(1.31,1.64){(b)}
\put(1.98,1.64){(c)}
\put(0.65,1.06){(d)}
\put(1.31,1.06){(e)}
\put(1.98,1.06){(f)}
\put(0.65,0.48){(g)}
\put(1.31,0.48){(h)}
\put(1.98,0.48){(i)}
\end{picture}
\caption{Profiles of the main components of the time- and space-averaged non-Newtonian stress tensor: (a)-(c) $\langle \tau_{yy}\rangle$, (d)-(f) $\langle\tau_{zz}\rangle$, (g)-(i) $\langle\tau_{yz}\rangle$ at different stream-wise locations: $+$ at $y=60h$, $\circ$ at $y=80h$ and $\times$ at $y=100h$. Each color represents a different model and for clarity the profiles for the different models are shifted; the dashed vertical lines identify the corresponding zero-levels. The rightmost part of each panel compares the non-Newtonian extra-stresses for different models at the same stream-wise location $y=80h$. The three columns correspond to the different $\gamma$: (a), (d), (g) $\gamma=1$, (b), (e), (h) $\gamma=10$ and (c), (f), (i) $\gamma=100$.} 
\label{fig: tau_ssG}
\end{figure*}

\subsection{Rescaling of polymer extra-stresses}
\label{sec: appES}
We also tested the scaling for the extra-stress tensor components proposed by \citet{guimaraes2020direct} in the framework of a new theory for the far field of viscoelastic jets. It was shown that there exists a region where the transverse profiles of the mean polymer shear stresses become self-preserving, provided that the inlet Weissenberg number is large enough. The scaling factor used to rescale the extra-stresses was obtained by balancing the viscoelastic stress power and the turbulence production terms in the turbulent kinetic energy balance equation. The theory relies on the validity of a modified Townsend hypothesis \citep{townsend1980structure} applied to viscoelastic jets, stating that the relative importance of production and viscoelastic stress power is conserved. 
A second assumption involves the choice of the flow time and velocity scales, which are obtained from the inertial-range scaling for the velocity of an eddy of size $l$, $u(l) \approx (\varepsilon_s l)^{1/3}$, with $\varepsilon$ being the local solvent dissipation.

We report in figure~\ref{fig: tau_ssG} the components of the extra-stress tensor normalized by the rescaling factor $\rho v_c^2 De_l \partial \delta_{0.5}/\partial y$ \citep{guimaraes2020direct}. Each different row correspond to a different component of the extra-stress tensor, and each column to a different value of $\gamma$, increasing from left to right. The different colors correspond to different fluid models, red for the Giesekus fluid and green for the Oldroyd-B fluid; different symbols correspond to different stream-wise positions. The profiles for the stream-wise and jet normal components show reasonable self-similarity up to intermediate Weissenberg numbers, both among different models and different stream-wise positions, figure~\ref{fig: tau_ssG}(\textit{a},\textit{b},\textit{d},\textit{e}). Self-similarity is however lost at the highest Weissenberg number.
Conversely, the $yz$ component of the extra-stress tensor does not show any self-similarity among the different stream-wise positions for all Weissenberg numbers considered in this study.
A major source of the discrepancy is that in this work we consider planar jets at a much lower inlet Reynolds number than what done by \citet{guimaraes2020direct}, and turbulent fluctuations are set in motion mainly by fluid elasticity for the Giesekus and Oldroyd-B fluid models, thus the use of the inertial-range scaling for the velocity is not appropriate. In addition, we examine stream-wise positions that are much farther apart than those reported in the work by \citet{guimaraes2020direct}.

\bibliographystyle{jfm}
\bibliography{totalbib.bib}

\end{document}